\documentclass[12pt,notitlepage]{report}
\usepackage{unmthesis}

\usepackage{amsmath}
\usepackage{amssymb}
\usepackage{graphicx}
\usepackage{afterpage}

\renewcommand{\bibname}{References}
\renewcommand{\thesisauthor}{Stephen A. Klinksiek}
\renewcommand{\thesismonth}{December,}
\renewcommand{\thesisyear}{2005}
\renewcommand{\thesistitle}{Finding The Charm In 800 \cmom\\
 \pcu and \pbe Single Muon Spectra}
\renewcommand{\thesistitletypesize}{\Large}
\renewcommand{\thesisauthorpreviousdegrees}{\normalsize B.S.,
 Physics/Mathematics, New Mexico Highlands
 University, 1993 $\; \; \;$\\
 \hspace{15mm}M.S., Physics, University of New Mexico, 1999 $\; \;$}
\renewcommand{\thesissubject}{Physics}
\renewcommand{\thesisgraduateunit}{Department of Physics and Astronomy}

\begin{document}
\pagestyle{empty}
\thesistitlepage
\pagestyle{plain}
\pagenumbering{roman} 
\setcounter{page}{3}
\thesiscopyrightpage
\thesisdedicationpage
\thesisacknowledgments
\newpage
\pagestyle{empty}
\thesisabstracttitlepage
\thesisabstract
\pagestyle{plain}
\pagenumbering{roman} 
\setcounter{page}{7}


Fermilab Experiment 866 took single muon data from 800 \cmom
($\sqrt{s}=38.8$ GeV) \pcu and \pbe interactions in an attempt to
extract the inclusive nuclear open charm/anti-charm
($D/\overline{D}$) differential
cross sections as
a function of \nspt. The muons
were decay products from semi-leptonic
decays of open charm mesons as well as decays from lighter
non-charmed mesons ($\pi$'s and K's). Data were taken simultaneously
from two interaction regions; one of two thin nuclear targets and
a copper beam dump 92 inches
downstream. The open decay length for hadrons
produced in the targets increased the contribution to the muon
spectrum from light hadron decays, relative to those from the dump.
Production cross sections for light hadrons from previous experiments
were used in conjunction with parameterized open charm cross sections
to produce total Monte Carlo single muon spectra that were
subsequently fit to the data.

The sensitivity of this measurement covered an
open charm hadron \pt range of approximately 2 to 7 \nscmom,
center-of-mass rapidity, $y_{cm}$, between 0 and 2, and $x_{F}$
between 0.2 and 0.8. Previous
experimental results for \ppp or \pA open charm production at
comparable energy was limited to $\sqrt{5}$ \nscmom. Three
functions describing the shape of the
open charm/anti-charm cross sections were fit to the data; an
exponential, \nsfuncexp, and two polynomials, \functhree and
\nsfuncfour. The first polynomial was fit with the parameter $n$
as a free parameter, and constant with three integer values, 4,
5 and 6. The second was fit with $n$ held fixed at the constant
integer values only. The best results were with the first polynomial
with $n$ around 6. All three parameterizations resulted in good fits.
Extrapolation of the cross sections to small \pt shows good
agreement with previous experiments. The power $\alpha$
of the nuclear dependency $A^{\alpha(\nspt)}$ was calculated as
a function of \nspt. The result indicates that $\alpha$ is
transverse-momentum dependent, albeit within large errors.

\newpage

\tableofcontents 
\listoffigures
\newpage
\listoftables

\newpage 


\chapter{Introduction}
\pagenumbering{arabic}
\setcounter{page}{1}

Quantum Chromo-Dynamics (QCD), and the Standard Model (SM) as a
whole, have been
remarkably successful in describing the nature of particles and
their interactions. They are widely accepted by the community. The
Standard model is composed of six quarks (referred to as flavors),
six leptons and four
force carriers, neglecting gravity. Quarks and leptons come
in three generations. A quark generation consists of two quarks, one
having charge $+ \, 2 \, / \, 3$ and one having charge
$- \, 1 \, / \, 3$. The quarks also come in three colors,
red, green and blue. Each generation of leptons consists of
a pair with one
neutral and the other having unit charge. Quarks and leptons have
half-integer spin and couple (depending on the interaction) to the
spin 1 force
carriers, gluons, the charged and neutral weak bosons
$ \left( W^{+} \, , \, W^{-} \, \mbox{and} \, Z^{0} \right)$ and
photons. Table \ref{smt1} shows the quarks, leptons and gauge bosons
used in the Standard Model, taken from the Particle Data
Group\cite{pdg}.

\renewcommand{\arraystretch}{1.1}

\begin{table}[ht]
\caption[Quarks, Leptons And Bosons]{Quarks, leptons and vector
bosons used in the Standard Model. Masses (MeV c$^{-2}$) are
given in parentheses \cite{pdg}.}
\label{smt1}
\begin{center}
\begin{tabular}{|c|c|ccc|}
\hline

\multicolumn{5}{|c|}{Quarks and Leptons}\\

\hline
\hline

Type &
 charge &
 $1^{st}$ Generation &
 $2^{nd}$ Generation &
 $3^{rd}$ Generation\\[0.25ex]

\hline

  &
  &
 $u$ &
 $c$ &
 $t$\\

\raisebox{1.5ex}[0pt]{up-type} &
  &
 up &
 charm &
 top\\

\raisebox{1.5ex}[0pt]{quarks} &
 \raisebox{2.5ex}[0pt]{$\displaystyle +\frac{2}{3}$} &
 (1.5 - 4.0) &
 (1150 - 1350) &
 (174000)\\

\hline

  &
  &
 $d$ &
 $s$ &
 $b$\\

\raisebox{1.5ex}[0pt]{down-type} &
  &
 down &
 strange &
 bottom\\

\raisebox{1.5ex}[0pt]{quarks} &
 \raisebox{2.5ex}[0pt]{$\displaystyle - \frac{1}{3}$} &
 (4 - 8) &
 (80 - 130) &
 (4100 - 4400)\\

\hline\hline

  &
  &
 $\nu_{e}$ &
 $\nu_{\mu}$ &
 $\nu_{\tau}$\\

\raisebox{1.5ex}[0pt]{neutral} &
  &
 $e$ neutrino &
 $\mu$ neutrino &
 $\tau$ neutrino\\

\raisebox{1.5ex}[0pt]{leptons} &
 \raisebox{2.5ex}[0pt]{$\displaystyle 0$} &
 ($ < 3 \times 10^{-6} $) &
 ($ < 0.19 $) &
 ($ < 18.2 $ )\\

\hline

  &
  &
 $e$ &
 $\mu$ &
 $\tau$\\

\raisebox{1.5ex}[0pt]{charged} &
  &
 electron &
 muon &
 tau\\

\raisebox{1.5ex}[0pt]{leptons} &
 \raisebox{2.5ex}[0pt]{$ \displaystyle +1 $} &
 (0.5) &
 (105.7) &
 (1777)\\

\hline\hline

\multicolumn{5}{|c|}{Vector Bosons (Force Carriers)}\\

\hline\hline

Type &
 charge &
 Boson &
 \multicolumn{1}{|c|}{mass} &
 \\

\cline{1-4}

neutral &
  &
  &
 \multicolumn{1}{|c|}{} &
 \\

weak &
 \raisebox{1.6ex}[0pt]{$0$} &
 \raisebox{1.6ex}[0pt]{$Z$} &
 \multicolumn{1}{|c|}{\raisebox{1.6ex}[0pt]{(9120)}} &
 \\

\cline{1-4}

charged &
 $ + 1 $ &
 \multicolumn{1}{c|}{$W^{+}$} &
 \multicolumn{1}{c|}{} &
 \\

weak &
 $- \, 1$ &
 \multicolumn{1}{c|}{$W^{-}$} &
 \multicolumn{1}{c|}{\raisebox{1.6ex}[0pt]{(8040)}} &
 \\

\cline{1-4}

electro- &
  &
 \multicolumn{1}{c|}{$\gamma$} &
 \multicolumn{1}{c|}{} &
 \\

magnetic &
 \raisebox{1.6ex}[0pt]{$0$} &
 \multicolumn{1}{c|}{photon} &
 \multicolumn{1}{c|}{\raisebox{1.6ex}[0pt]{($<6\times10^{-23}$)}} &
 \\

\cline{1-4}

  &
  &
 \multicolumn{1}{c|}{$g$} &
 \multicolumn{1}{c|}{} &
 \\

\raisebox{1.6ex}[0pt]{strong} &
 \raisebox{1.6ex}[0pt]{$0$} &
 \multicolumn{1}{c|}{gluon} &
 \multicolumn{1}{c|}{\raisebox{1.6ex}[0pt]{($0$)}} &
 \\

\hline

\end{tabular}
\end{center}
\end{table}

\renewcommand{\arraystretch}{1.0}

Hadrons composed of three
quarks, such as the proton ($u,u,d$) and neutron ($u,d,d$) are
referred to as baryons, and hadrons made of one quark and one
anti-quark are called
mesons. The ($u,u,d$) quarks in the proton are referred to as
valence quarks. Hadrons have virtual $\qqbar \,$ pairs as well,
which are referred to as \textit{sea quarks}.
Flavor is conserved in strong and electromagnetic interactions,
so net 'new' flavor in strong or electromagnetic
interactions must be 0. Charm, bottom and top
quarks are often referred to as heavy quarks. Mesons containing
charm come in two varieties, those having one charm or anti-charm
quark, referred to as open charm, such as the $D^{+}$ which has
one charm and one anti-down quark
$ \left( c \,\overline{d} \right) $, and those containing one charm
and one anti-charm quark, referred to as hidden charm, such as the
$J \, / \, \Psi \; \left( c \, \overline{c} \right)$.

\begin{center}
\begin{figure}[!ht]
\resizebox{5.8in}{3.1in}
{\includegraphics[7,413][598,728]{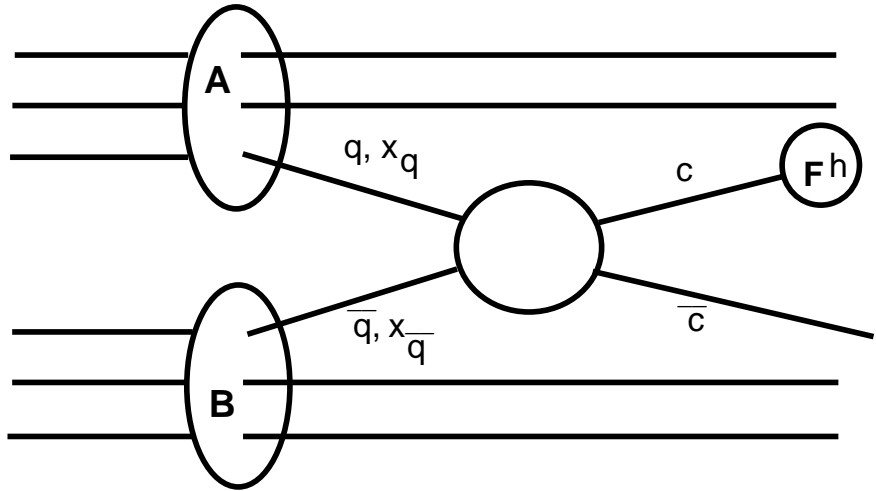}}
\caption[Hadron-Hadron Scattering]{Representation of
$\qqbar \rightarrow c \overline{c} $ from the interaction of a quark
$q$ in hadron $A$ with momentum fraction $x_{q}$ and a sea
anti-quark $\overline{q}$ from hadron $B$ having momentum fraction
$x_{\overline{q}}$. The charm quark subsequently hadronizes to
a meson at the process labelled $F^{h}$.}
\label{had-scatt}
\end{figure}
\end{center}

Study of the production of hadrons containing charm or heavier
quarks contributes to the understanding of the theory of Quantum
Chromo-Dynamics (QCD). QCD uses factorization theory to describe
the production and hadronization of heavy quarks into hadrons
seen in the lab. Figure \ref{had-scatt} shows a parton-parton
(partons are quarks or gluons) interaction in a collision of
two baryons, $A$ and $B$. The interaction produces one charm and
one anti-charm quark, $c$ and $\overline{c}$. The charm quark
then hadronizes into a charm meson, shown as $F^{h}$ in the
Figure.

A shorthand method used to describe what is shown in
Figure \ref{had-scatt} is:
$$ A + B \rightarrow \nsplusd + \nsminusd + X $$ where the charm
quark hadronized into a \nsplusd, the anti-charm quark hadronized
into a \minusd and $X$ is used to denote 'anything else', which must
include two baryons.

All mesons are unstable and decay. The short hand notation for one
of the ways a \plusd meson may decay, called a hadronic mode, is:
$$ \nsplusd \rightarrow \minusk \pluspi \nspluspi $$ An example
of a decay mode referred to as a semi-leptonic mode is
$$ \nsplusd \rightarrow \muplus \nu_{\mu} \, \nsminusk$$ Experiments
studying the production
of open charm usually 'tag' events by looking for the decay
products from various decay modes. For the hadronic mode above, the
event would be tagged if a \minusk and 2 \pluspi tracked back to
a common vertex, suggesting they were the decay products from the
\nsplusd. The invariant mass is then calculated, and compared to
the mass of the \nsplusd. If the event passes all cuts used to
eliminate bad events, it is used in the analysis of the production
characteristics of the \plusd for their experiment.

This analysis used data from 800 \cmom \pcu and \pbe
interactions to determine the production of hadrons containing
open charm as a function of the hadron \nspt. The data consisted of
single muon events. The E866 spectrometer was designed to study
di-muon events, where the two muons tracked back to a decay vertex.
Single muon events are events where a hadron decays semi-leptonically
to one muon, plus anything else, as shown above. The E866
spectrometer could not track any remaining secondary particles from
the decay, so no decay vertex was available. The single muon data
was taken from two production regions simultaneously,
one a thin target of either copper or beryllium, the other a solid
copper beam dump. The targets had an open decay length of 92 inches
before hadrons (and remaining proton beam) interacted in the
dump. Hadrons containing open charm decay
roughly $10^{4}$ times sooner than light hadrons, so the ratio of
muons from open charm to light hadrons is significantly enhanced
in the data taken from the dump, relative to the data taken from
the targets. Use of the difference between muon spectra from the
two production regions allowed extraction of the \textit{inclusive}
open charm/anti-charm differential cross sections as a function
of the hadron transverse momentum between 2.25 and 10.0 \nscmom.
Mesons containing open charm are shown in Table \ref{ochads}.

\renewcommand{\arraystretch}{1.5}

\begin{table}[ht]
\caption[Mesons Containing Open Charm]{Mesons containing open charm. Mass
(MeV/c$^{2}$) is rounded off to the nearest MeV/c$^{2}$.}
\label{ochads}
\begin{center}
\begin{tabular}{|c|c|c|c|c|c|c|}
\hline

\multicolumn{7}{|c|}{Pseudoscalar Mesons (Spin 0)}\\

\hline

  &
 $D^{+}$ &
 $D^{0}$ &
 $D^{+}_{s}$ &
 $D^{-}$ &
 $\overline{D}^{0}$ &
 $D^{-}_{s}$\\

\hline

Quarks &
 $c,\overline{d}$ &
 $c,\overline{u}$ &
 $c,\overline{s}$ &
 $\overline{c},d$ &
 $\overline{c},u$ &
 $\overline{c},s$\\

\hline

Mass &
 1869 &
 1865 &
 1969 &
 1869 &
 1865 &
 1969\\

\hline

\multicolumn{5}{|c|}{Vector Mesons (Spin 1)} &
 \multicolumn{2}{c}{}\\

\cline{1-5}

  &
 $D^{* \, +}$ &
 $D^{* \, 0}$ &
 $D^{* \, -}$ &
 $\overline{D}^{* \, 0}$ &
 \multicolumn{1}{|c}{}\\

\cline{1-5}

Quarks &
 $c,\overline{d}$ &
 $c,\overline{u}$ &
 $\overline{c},d$ &
 $\overline{c},u$ &
  \multicolumn{2}{|c}{}\\

\cline{1-5}

Mass &
 2010 &
 2007 &
 2010 &
 2007 &
 \multicolumn{2}{|c}{}\\

\cline{1-5}

\end{tabular}
\end{center}
\end{table}

\renewcommand{\arraystretch}{1.0}

The term inclusive (versus exclusive) is used to classify the
production of open charm by the final state of the reaction
$$ p + N \rightarrow D + X $$ where $p$ is a proton interacting
with a nucleon $N$ in the
material, $D$ is a hadron containing open charm or open
anti-charm and $X$ is anything else.\footnote{To conserve baryon
number, $X$ must be composed of at least two baryons and the
other charm quark must be included in either the baryons or
as another meson.} This analysis could not
identify the parent hadron, so \textit{any} hadron produced
that decayed to an open charm or anti-charm hadron is included
in the measured cross sections as if it were produced as an
open charm or anti-charm hadron, such as

$$ p + N \rightarrow B + X \quad \mbox{where} \quad
B \rightarrow D + \pi $$ Exclusive cross sections measure the
strength of a reaction to a
specific set of particles such as:

$$ p + N \rightarrow \nspluspi + \nsminuspi + p + N $$

Additional information can be derived as well, such as the
production dependency on $A$.\footnote{$A$ is the atomic
weight of a material used as a target to produce hadrons in
interactions. The scaling of the production to the atomic weight
is commonly referred to as the nuclear dependency which is
discussed in Chapter \ref{mc} for the production of light hadrons,
as well as in Chapter \ref{results} regarding the values for open
charm production as determined from this analysis.} Prior
experimental results from meson interactions have shown an
enhancement of the production of open charm hadrons containing one
of the valence quarks of the incident meson, called the leading
particle effect. This enhancement has not been observed in proton
interactions. If this effect were present in proton-nucleon
interactions, it would be seen as an enhacement of hadrons relative
to anti-hadrons. The hadron/anti-hadron ratio determined by this
analysis is presented as well.

\afterpage{\clearpage}

\section{Heavy Hadron Production}

The theoretical description of heavy hadron production in pQCD
(perturbative Quantum Chromodynamics) is done in two parts, referred
to as factorization;
production of the heavy quarks using partonic cross sections, and
the process of hadronization, where the bare quarks are transformed
into hadrons seen in the lab.

A cross section, $\sigma$, is used to measure the effectiveness of
an interaction such as
$$ a + b \rightarrow c + d $$ where
$a$ and $b$ are the interacting (incident)
particles and $c$ and $d$ are particles produced
in the interaction which may be different from either $a$ or
$b$. The cross section has units of area, typically given as
barns (1 barn = $10^{-24}$ cm$^{2}$).

Partonic cross sections in QCD are modeled from Deep Inelastic
Scattering cross sections. The cross sections for deep inelastic
scattering on unpolarized nucleons can be written, generically,
in terms of structure functions:

\begin{equation}
\frac{d^{2}\sigma^{i}}{dx \, dy} =
 {\displaystyle \frac{4 \pi \alpha^{2}}{x y Q^{2}}}
 \, \eta^{i} \, \left[ \left( 1 - y -
 {\displaystyle \frac{x^{2} y^{2} M^{2}c^{4}}{Q^{2}}} \right)
F^{i}_{2} + y^{2} x F^{i}_{1} \mp \left( y -
 {\displaystyle \frac{y^{2}}{2}} \right) x F^{i}_{3} \right]
\label{psigma}
\end{equation} where $i=NC,CC$ is for neutral or charged
current scattering and $F^{i}_{1}$, $F^{i}_{2}$ and $F^{i}_{3}$
are structure functions. $Q$ is the four-momentum transferred
in the interaction, and $\alpha$ (called
the fine structure constant) is defined as
$\alpha=e/\hbar c$. (The units for $\alpha^{2}$ are
(GeV$^{2}$ \, cm$^{2}$).) In the quark-parton model, contributions
to the structure functions are expressed in terms of quark
distribution functions of the proton. The quark distribution
function, $q(x,Q^{2})$, is the number of quarks,
or anti-quarks, of the designated flavor ($q=u,\overline{u},d,
\overline{d},s,\cdots$) that carry a momentum fraction between
$x$ and $x+dx$ of the protons momentum where the protons momentum
is large. For the charged current reaction $ e^{-} p \rightarrow
\nu X $
$$ F^{W^{-}}_{2} = 2 x \left( u + \overline{d} + \overline{s} + c
+ \cdots \right) $$

\begin{center}
\begin{figure}[!ht]
\resizebox{4.87in}{2.78in}
{\includegraphics[102,171][453,372]{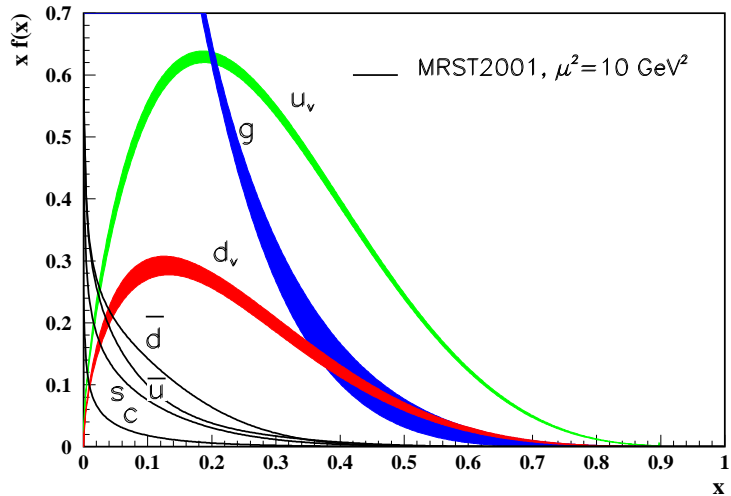}}
\caption[Parton Distributions]{Distributions of $x$ times the
unpolarized parton distributions $f(x)$ (where $f=u_{v},d_{v},
\overline{u},\overline{d},s,c,g$) using the MRST2001
parameterizations
\cite{eur-phys-j-c-28-455,eur-phys-j-c-23-73} (with uncertainties
in $u_{v},d_{v}$ and
$g$) at a scale $\mu^{2}=10$ (GeV$^{2}$/c$^{2}$). Figure is taken
from \cite{pdg}.}
\label{pdfs}
\end{figure}
\end{center}

One prediction of the quark-parton model is that the structure
functions scale in the Bjorken limit that $Q^{2}$ and
$\nu \rightarrow \infty$ with $x$ fixed. Scaling implies
$$F^{i}(x,Q^{2}) \rightarrow F^{i}(x)$$ QCD uses scale dependent
parton distribution functions (PDFs), $f(x,\mu^{2})$, to describe
the process above. Here, $f=g \, \mbox{or} \, q$ and $\mu$ is
typically the scale of the probe, $Q$. At a given $x$, these
correspond to the density of the partons in the proton integrated
over transverse momentum $k_{t}$ up to $\mu / c$. The fine structure
constant $\alpha$ used in Equation \ref{psigma} is redefined to
be scale dependent $\alpha_{s}(\mu^{2}) = \frac{g^{2}_{s}(\mu^{2})}
{\hbar c}$ where $g_{s}$ is the SU$_{c}$(3) coupling constant.
Parton distributions have been measured by many experiments. Figure
\ref{pdfs} shows the
distributions $x$ times the unpolarized parton distribution
functions $f(x)$ at a scale $\mu^{2}=10$ GeV$^{2}$/c$^{-2}$.

Production of heavy quarks in hadron-hadron interactions at leading
order (LO) in perturbation theory is the result of two parton-parton
interactions
$$ \qqbar \rightarrow \QQbar \quad \mbox{and} \quad
g \, g \rightarrow \QQbar$$ where $\qqbar \,$ (often referred to as
quark-quark annihilation) represents a quark from one hadron
interacting with an anti-quark from the other hadron, $g \, g$ is
used to represent the interaction between a gluon in one hadron
interacting with a gluon in the other, and $\QQbar \,$ is the
produced heavy quark and anti-quark (they must be produced as a
pair). In \ppp or \pA collisions, $\qqbar \,$ interactions are between
a valence quark in one hadron, and a sea anti-quark in the other.

\begin{center}
\begin{figure}[ht]
\resizebox{4.5in}{1.76in}
{\includegraphics[55,530][377,672]{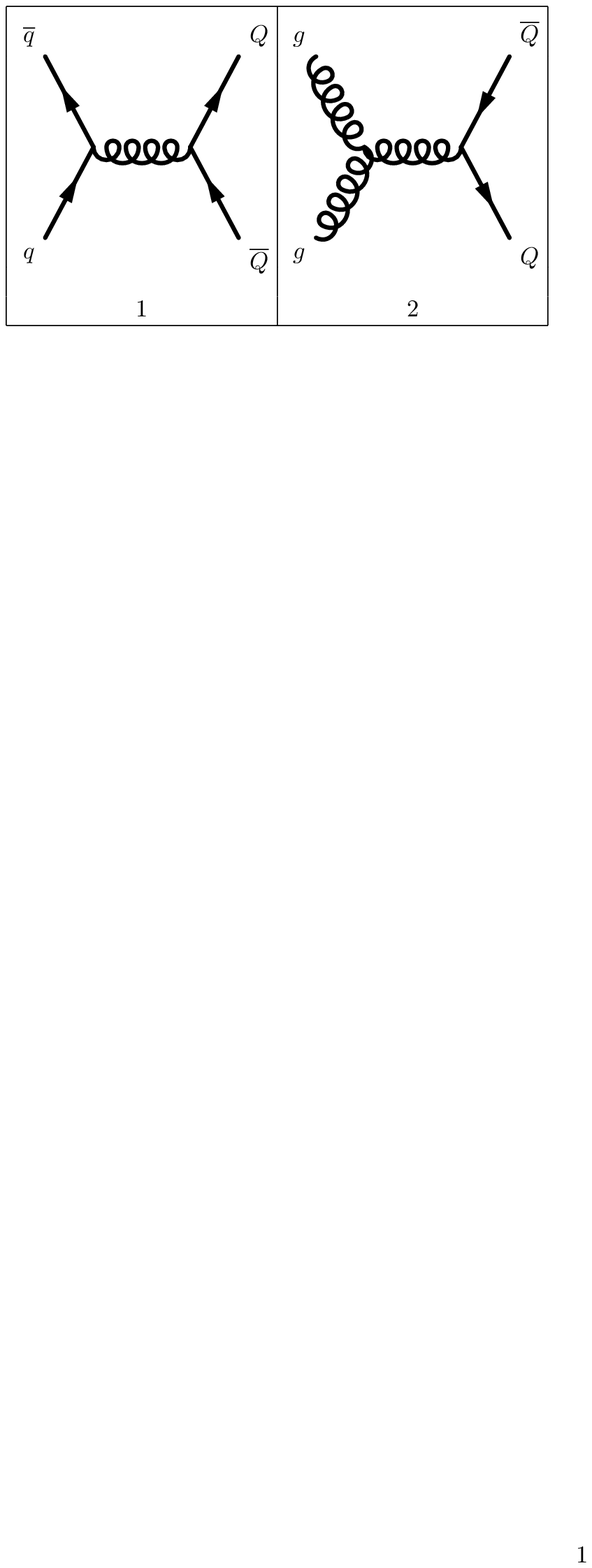}}
\caption[Leading Order Diagrams]{The Leading
Order Feynman diagrams $q \, \overline{q} \rightarrow
Q \, \overline{Q}$ (1) and $g \, g \rightarrow Q \,
\overline{Q}$ (2).}
\label{feyn-lo}
\end{figure}
\end{center}

Feynman diagrams are used to describe
the process as well as to calculate the amplitudes of the
interaction. The Feynman
diagrams of the leading order (LO) processes for the production
of charm are shown in Figure
\ref{feyn-lo}. The total partonic cross section from
hadro-production is proportional to the sum of all the combinations
of the quarks in one hadron with anti-quarks in the other (and
conversely, the sea anti-quarks in the first with quarks in the
other) plus the contribution from gluons in both.

Hadronization is usually described in terms of
fragmentation functions, $F(x,s)$. The Particle Data Group
\cite{pdg} define the fragmentation function for a hadron of type
$h$ at c.m. energy $\sqrt{s}$ represented as a sum of contributions
from the different parton types $i=u,d,s,\overline{u},\overline{d},
\overline{s},\cdots,g$ as
\begin{equation}
F^{h}(x,s) =\sum_{i} \int^{1}_{x} \frac{dz}{z} \,
C_{i}(s;z,\alpha_{s}) \, D^{h}_{i}(x/z,s)
\end{equation} where $D^{h}_{i}$ are the parton fragmentation
functions analogous to the parton distribution functions above,
$x=2 \, E_{h} / s \le 1$, $z=x_{h}/x_{i}$ and $C_{i}$ are
coefficient functions of the partons $i$.

At lowest order in $\alpha_{s}$, the coefficient function $C_{g}$
for gluons is 0, but for quarks, $C_{i} = g_{i}(s) \,
\delta \left( 1 - z \right)$. At higher orders of $\alpha_{s}$ the
coefficient and parton fragmentation functions are factorization
scheme dependent. Measured fragmentation
functions need to be parameterized at some initial scale, $t_{0}$,
typically 2 GeV$^{2}$ for light quarks and gluons. A general
parameterization is:

$$ D_{p \rightarrow h} = N \, x^{\alpha} \,
\left( 1 - x \right)^{\beta} \, \left( 1 +
{\displaystyle \frac{\gamma}{x}} \right) $$ where the normalization
$N$, $\alpha$, $\beta$ and $\gamma$ are usually dependent on the
scale $t_{0}$ as well as the type of parton, $p$ and hadron $h$.

\newpage

There have been numerous studies of the production of open
charm using nuclear targets. The majority of fixed target studies
have been done using charged meson beams on
nuclear targets. A reasonably complete compilation of
experimental results is found in \cite{hep-ph-9702287}. Here, and
in most literature, the term open charm production is used as
a 'catch all' phrase for all hadrons containing one charm or
anti-charm quark. Results are usually presented for the total
open charm production, or one or more sub-categories of these
hadrons, such as hadrons containing one charm quark
($D$) or one anti-charm quark
($\overline{D}$) or those with charge $\pm$1 or neutral.

\section{Results From Prior Experiments}

\begin{center}
\begin{figure}[!ht]
\resizebox{4.45in}{3.78in}
{\includegraphics[259,287][521,506]{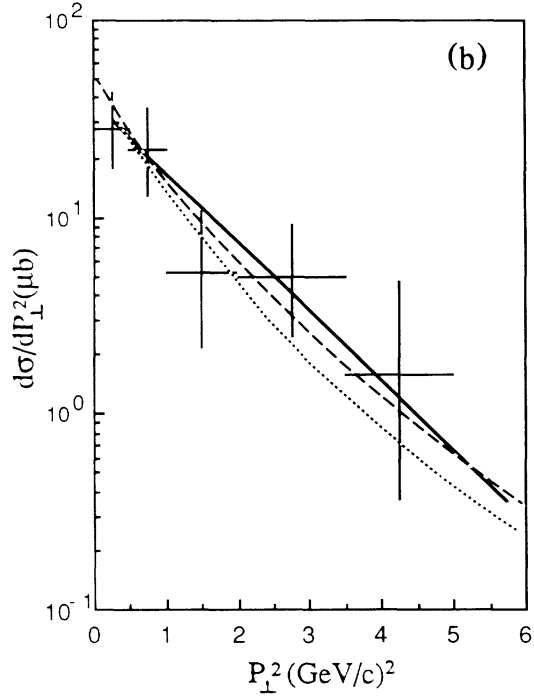}}
\caption[Data From LEBC-MPS]{Total inclusive open charm
differential cross section
$ d \sigma ( D + \overline{D} ) \, / \, d p^{2}_{T} \, $
($\mu$b c$^{2}$ GeV$^{-2}$) measured by the LEBMC-MPS Collaboration
(Fermilab Experiment 743)\cite{prl-61-2185}. Solid curve shows the
results of a fit to the empirical form $ \left( 1 - | x_{F} |
\right)^{n} \, exp \left( - a \, \nspt^{2} \right)$.}
\label{prl-61-2185-pt2}
\end{figure}
\end{center}

Discussion concerning results from fixed target \ppp and \pA open
charm production often cite publications from four
experiments; Fermilab Experiment 743 (the LEBC-MPS Collaboration)
\cite{prl-61-2185}, Fermilab Experiment 653
\cite{phys-lett-b-263-573}, Fermilab Experiment 789
\cite{prl-72-2542} and Fermilab Experiment 769
\cite{prl-77-2392,prl-77-2388}. Figure \ref{prl-61-2185-pt2} shows
the total open charm differential cross section from
800 \cmom \ppp interactions as measured by the LEBC-MPS
Collaboration \cite{prl-61-2185}.

\begin{center}
\begin{figure}[!ht]
\resizebox{4.7in}{2.46in}
{\includegraphics[247,137][540,289]{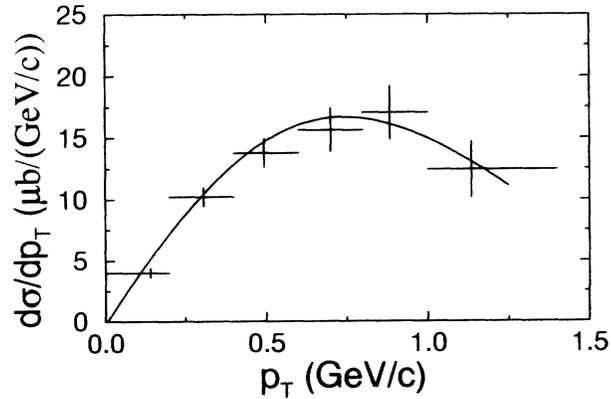}}
\caption[Data From E789]
{Differential cross section per nucleon,
$ d \sigma (\nsdzero) \, / \, d \, \nspt +
d \sigma (\nsadzero) \, / \, d \, \nspt $ versus \nspt.
Uncertainties shown are statistical only and do not inclue an
additional systematic uncertainty of 12.8 percent. Figure is from
Fermilab Experiment 789\cite{prl-72-2542}.}
\label{prl-72-2542-pt}
\end{figure}
\end{center}

\begin{center}
\begin{figure}[!ht]
\resizebox{5.7in}{2.7in}
{\includegraphics[43,97][572,353]{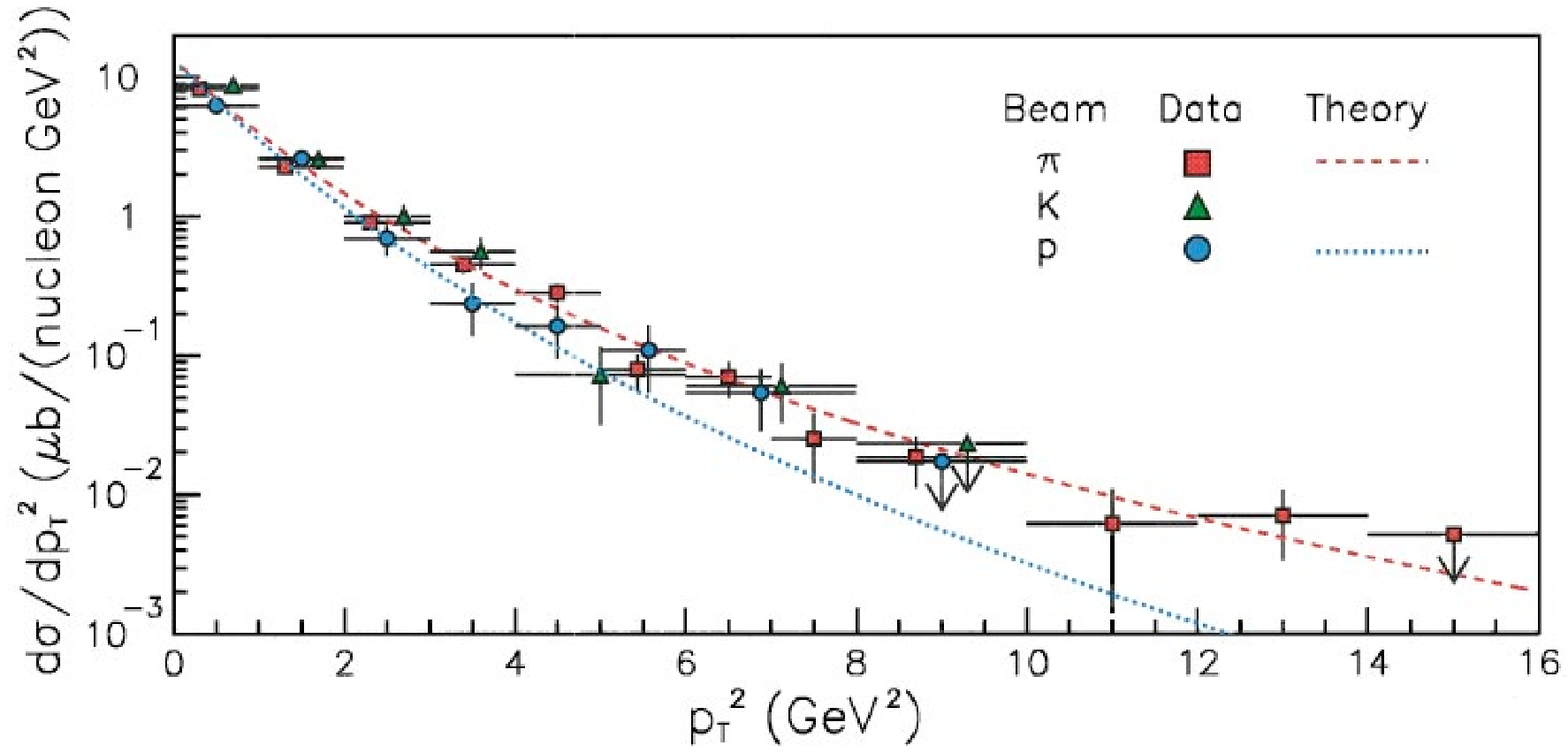}}
\caption[Data From E769]
{Measured $D$ meson
$ ( \nsplusd, \, \nsminusd, \, \nsdzero, \,
\nsadzero, \, D^{+}_{s} \, \mbox{and} \, D^{-}_{s} ) \quad
d \sigma \, / \, d p^{2}_{T} \, $
$ \left( \mu\mbox{b} \, \mbox{c}^2 \, \mbox{nucleon}^{-1} \,
\mbox{GeV}^{-2} \right) \,$ $ \left( x_{F} > 0 \right) \,$ for
production induced by $\pi$, $K$ and $p$ beams and NLO QCD
predictions for charm quarks \cite{nucl-phys-b-405-507} ($\pi$ and
$p$ beams). In addition to the statistical errors shown, there are
overall normalization errors of about 6\%, 6\% and 9\% for $\pi$,
$K$ and $p$ results respectively. Figure is taken from
\cite{prl-77-2392}.}
\label{prl-77-2392-pt2}
\end{figure}
\end{center}

Fermilab Experiment
789 measured neutral open charm production for 800 \cmom protons
incident on beryllium and gold targets \cite{prl-72-2542}. Their
results are shown in Figure \ref{prl-72-2542-pt}.
The Fermilab E769 Collaboration studied open charm production using
secondary 250 \cmom $\pi^{\pm}$, $K^{\pm}$ and $p$ 
beams on a multifoil target of beryllium, copper, aluminum and
tungsten \cite{prl-77-2392,prl-77-2388}. The results from
\cite{prl-77-2392} are presented in Figure \ref{prl-77-2392-pt2}.

The production of open charm using charged meson beams shows
significant differences from the production of open charm using
proton beams on nuclear targets. The differences are seen in the
ratio of charged to neutral charm production and
the ratio of hadron to anti-hadron production. The second is
referred to as the leading particle effect. Results to date
from either meson or proton induced production reveal that
open charm meson production is shaped more or less the same as the
next-to-leading order (NLO) predictions for the production of the
quarks themselves (see Figure \ref{prl-77-2392-pt2}).

Theory suggests that the ratio of charged to neutral open charm
production should be approximately 0.32
\cite{hep-ph-9702287,hep-ph-0306212}, and results from open charm
production using charged meson beams shows rough agreement. The
results from fixed target \ppp and \pA experiments, until recently,
have shown this ratio to be more or less unity. A recent result
published by the HERA-B Collaboration reports the ratio
$\sigma(\nsplusd) \, / \, \sigma(\nsdzero) = 0.54 \pm 0.11 \pm 0.13$
from 920 \cmom \pA induced production \cite{hep-ex-0408110}. The
STAR Collaboration reported the same ratio as 0.40 $\pm$ 0.09 $\pm$
0.13 using data from $\sqrt{s_{NN}}=200$ GeV \ppp and
\textit{d-Au} collisions \cite{nucl-ex-0410038}. The leading particle
effect reported by meson-nucleon experiments is contrary to
predictions as well. Proton-nucleon experiments have not seen this
effect. It is generally thought that some momentum is lost during
the hadronization process. Results from a variety of experiments,
such as E769 above, do not show this effect.

\afterpage{\clearpage}

\chapter{Apparatus}
\label{apparatus}

\section{General Description}

Fermilab Experiment 866 (E866/NuSeA) was designed for
collecting and analyzing di-muon events from 800 \cmom protons
incident on various nuclear targets. E866 was a continuation of
several Fermilab Experiments including 772 and 789, where the
detector had several major
improvements to increase the accuracy of the measured trajectory of
the muons
as well as increased data taking capabilities through improvements
in the data aquisition system and trigger configurations. The
experiment has resulted in five previous Doctoral
theses \cite{hawker} \-- \cite{jwebb} and several published
results \cite{prl-80-3715}$ \, - \, $\cite{hep-ex-032019}.  

\begin{center}
\begin{figure}[ht]
\resizebox{5.8in}{3.0in}
{\includegraphics[79,220][535,453]{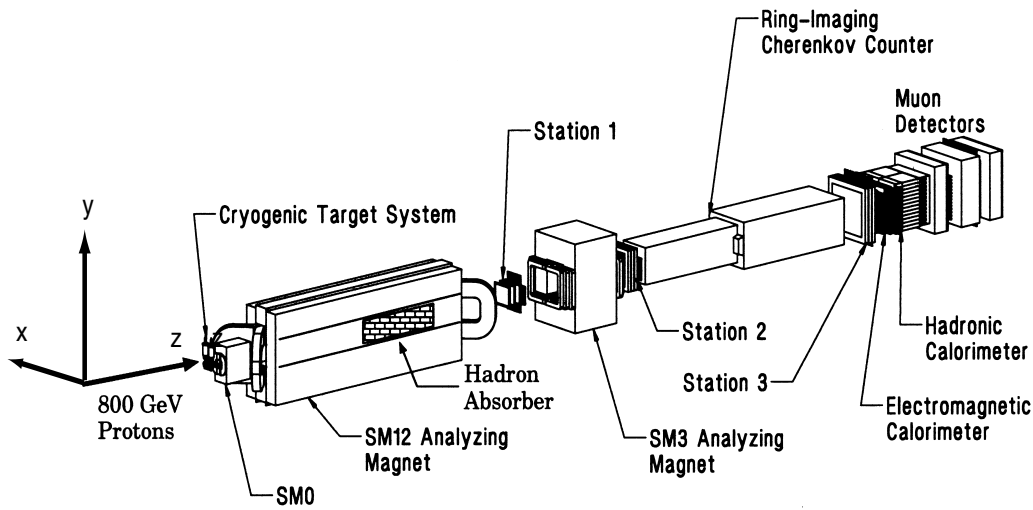}}
\caption[E866 Spectrometer]{FNAL E866/NuSeA Spectrometer. For this
analysis the spectrometer had no cryogenic target system, but a
target wheel instead, missing is the dump shown in figure
\ref{sm12}. The ring-imaging cherenkov counter was inactive,
and the muon detectors are referred to as station 4 in the
text.}
\label{spectrometer}
\end{figure}
\end{center}

Figure \ref{spectrometer} shows the FNAL E866/NuSeA spectrometer
for the original configuration of the experiment. There were
modifications to the original configuration to take data for
this analysis that will be pointed out and explained as
necessary.

The E866 spectrometer was located in the East hall of the Meson
Experimental Area, and used the 800 \cmom proton beam extracted
from the accelerator ring for a period of approximately twenty
seconds out of each minute. Each spill contained "buckets"
with a 19 $ns$ time structure based on the radio accelerator
RF of 53 MHz. 

\section{Beam Monitors}

Beam monitoring was accomplished with various detectors
for size, position and intensity. The beam size and position was
measured using RF cavities and segmented wire ion chambers (SWIC).
The beam was last monitored approximately 70 inches upstream of the
targets. This was done using a movable SWIC having a wire spacing
of 2 mm horizontally and 0.5 mm vertically. There were several 
intensity monitors, but these were unreliable at the low intensity
(approximately $10^{10}$ protons per spill) used for the single
muon data.

\section{Targets}
\label{targets}

The data was taken with a rotating target wheel instead
of the cryogenic target system shown in Figure
\ref{spectrometer}. The experiment designated the center of the
opening of
the upstream face of spectrometer magnet SM12 as the origin. The
target wheel was located
at $z=-24$ inches. The wheel had three used positions; one empty
target frame, one beryllium target of thickness 2.036
inches  (referred to as the beryllium target), and another copper
target  of thickness 1.004 inches (referred to as the copper
target). The proton interaction lengths of these targets, calculated
with  $\rho_{Cu}=8.96$ \density and $\lambda_{I,Cu}=134.9$ \ilength
for copper  and $\rho_{Be}=1.848$ \density and $\lambda_{I,Be}=75.2$
\ilength for beryllium,  were 0.127 and 0.169 interaction
lengths respectively. The wheel was rotated during the 40 second
period between spills according to two rotation schedules.

\section{Spectrometer}
\label{e866spectrometer}

The spectrometer had three dipole magnets with fields in the 
$x$ direction as defined in Figure \ref{spectrometer}. For this
analysis data was taken with currents in magnets SM12 and SM3 only.
The currents were configured both parallel and anti-parallel during
data taking, although results are only shown for the parallel
configuration. The magnet currents were set to 1420 amps in SM12
and 4200 amps in SM3. Only the results from the parallel magnetic
field configuration are given in this analysis, because the
analysis lost the computer during the analysis of the opposite
polarity data. The magnetic configuration would only allow for
a separate set of data to fit, since combining the two results
was not possible. The acceptances for both magnetic field
configurations was very similar, so no new information was
lost.

\begin{center}
\begin{figure}[ht]
\resizebox{5.6in}{5.6in}
{\includegraphics[130,225][500,575]{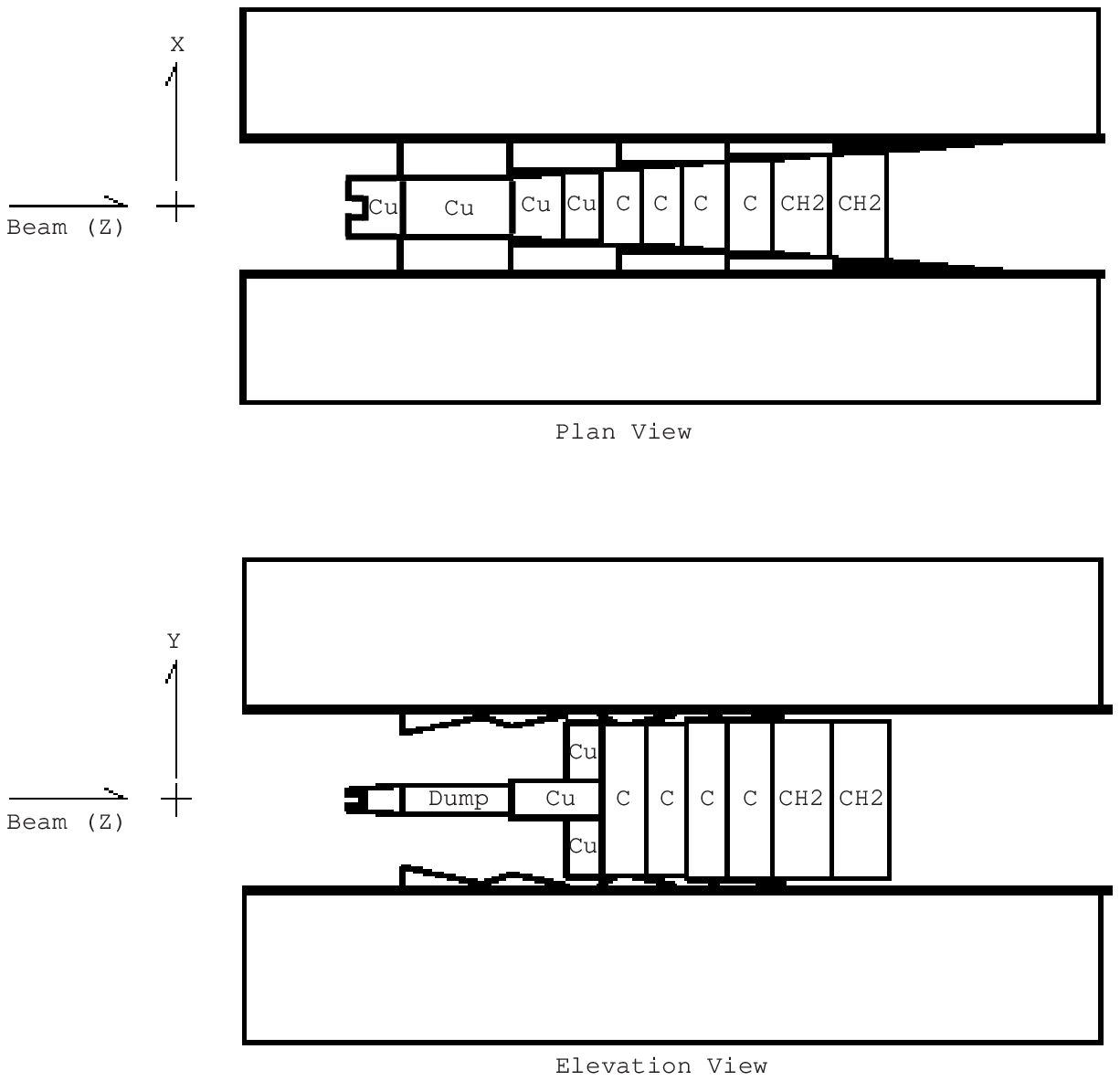}}
\caption[Spectrometer Magnet SM12 Detail]{Spectrometer magnet
SM12 as configured for this analysis showing the copper beam dump
and hadron absorbers in plan view (top) and elevated view (bottom).
Cross indicates the location of the targets relative to the
spectrometer magnet and dump.}
\label{sm12}
\end{figure}
\end{center}

Particles created in the targets, and any remaining proton
beam, entered magnet SM12. The configuration of this magnet when
the data for this analysis was taken is shown in detail in
Figure \ref{sm12}. The magnet was 567 inches long and provided a
momentum deflection of 7 \cmom when operated at its maximum current
of 4000 amps and provided a momentum deflection of approximately
2.4 \cmom for this data.  The copper beam dump was 168 inches long
beginning
68 inches inside the volume of the magnet. The dump spanned the
entire volume in $x$, and was approximately 8 inches tall for the
first 80 inches, and 10 inches tall thereafter. The face of the
dump had a rectangular hole 12 inches deep by 2 inches square to
reduce back-scattered particles. The dump provided a second target 
having 26 proton interaction lengths, or approximately 220
radiation lengths. 

Behind the dump was the hadron absorber that filled the entire
volume in $x$ and $y$ consisting of 24 inches of copper, 4 layers
of carbon 27 inches thick each and two 36 inch layers of borated
polyethylene. The absorber wall, having over thirteen interaction
or sixty radiation lengths, effectively allowed only muons to
pass downstream. The remainder of the inside of the magnet was
filled with a helium bag.

Between magnets SM12 and SM3 is the first of three similar tracking
stations. Stations 1, 2 and 3 had multiple pairs of drift chambers
and one or more layers of hodoscopes. Station 1 had three pairs of
drift chambers, U1, U1$^{\prime}$, Y1, Y1$^{\prime}$ and V1 and
V1$^{\prime}$. Planes U and U$^{\prime}$ had sense wires oriented
+14$^{\circ}$ $\left[tan(\theta)=0.25\right]$ in the $x-y$ plane, Y
and Y$^{\prime}$ were horizontal and V and V$^{\prime}$ were
oriented -14$^{\circ}$. The primed planes were offset in the
direction perpendicular to the sense wires by half of one cell to
help remove ambiguities in the drift direction. Table
\ref{wireplanes} (page \pageref{wireplanes}) gives a summary of the
physical characteristics of the drift chambers for each station.

Hodoscopes at each station allowed fast track evaluations or 'roads'
used to trigger valid events to be taken to tape.
Horizontally aligned planes of scintillators determined rough $y$
positions, while vertically aligned planes gave rough $x$ positions.
Each plane was optically split to provide quadrants
named up left (UL), up right (UR), down left (DL) and down right
(DR). Each plane was designated by orientation and station number.
Gaps created in the splitting resulted in small dead spots for muon
tracks at or near $x=0$ for $y$ hodoscope planes, or $y=0$ for
$x$ hodoscope planes.

\begin{center}
\begin{figure}[ht]
\resizebox{5.6in}{5.6in}
{\includegraphics[130,225][500,575]{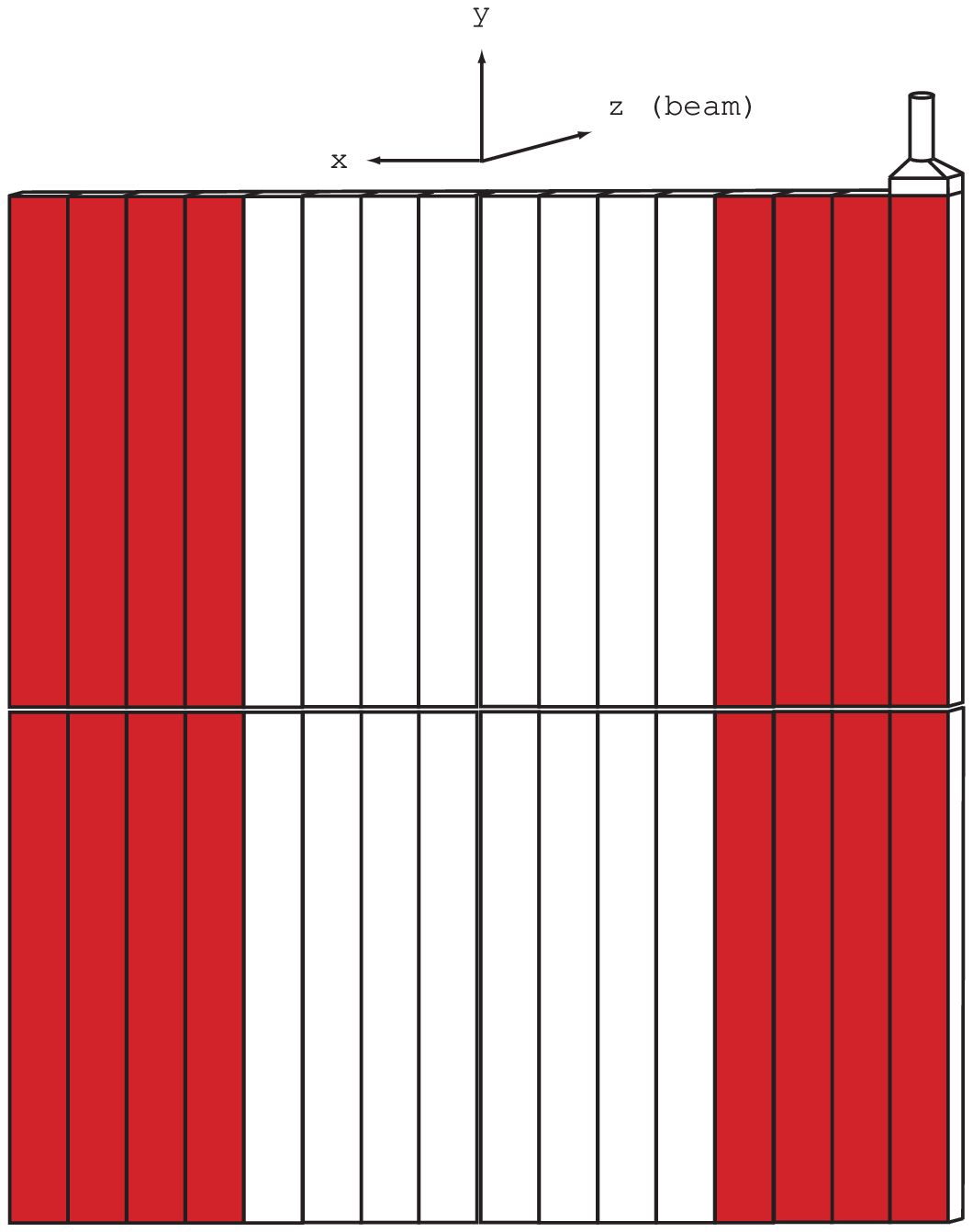}}
\caption[X Plane Of Hodoscopes]{View of an $x$ plane of hodoscopes.
Gaps between quadrants are exaggerated. Only one phototube is shown.
Shaded region represents hodoscopes having high voltage supplied to
the attached phototubes. $y$ hodoscopes are similar except being
rotated 90 degrees and all scintillators have high voltage
supplied to the phototubes.}
\label{hodos}
\end{figure}
\end{center}

Specific to this analysis, data were taken with the middle half of
each $x$ measuring hodoscope plane turned off to reduce trigger
rates that would have been unacceptabes had they been left on. This
configuration was
necessary to reduce the event rate, even at the low intensity
requested. This configuration produced a loss in acceptance for
muons having $\nsmupt \le 1.75 \nscmom$, and created two inner
acceptance edges which further reduced the acceptable minimum
muon transverse momentum as described in \ref{xhodos} and
\ref{traceback}. Figure \ref{hodos} shows an $x$ hodoscope plane,
and Table \ref{hodoplanes} gives the specifications of all
hodoscope planes as used for taking data for this analysis.

Particles then entered the second magnet, SM3, which was used to
determine the momentum of the particle in the $y-z$ plane. At its
normal current setting of 4230 Amps, the magnet provided a
transverse momentum deflection of 0.91 \nscmom. The direction of
the deflection relative to the magnetic field direction determined
the charge of the particle. Muons then passed Station 2, a tracking
station similar to Station 1 except only one plane of $y$ measuring
hodoscopes, Y2, is present.

Muons then traversed a non-operational Ring Imaging Cherenkov
counter (RICH). This detector had been used in previous experiments
for particle identification, but was not used for this analysis.
The interior was helium filled to reduce multiple scattering.
Downstream of the RICH detector was Station 3, a larger version of
the prior stations. Station 3 had both $x$ and $y$ measuring
hodoscope planes.

Both calorimeters following Station 3 shown in Figure
\ref{spectrometer} were inactive, but had been left in place to
provide additional hadron absorbing material. Shown in Figure
\ref{spectrometer} are large amounts of zinc and concrete placed
between three layers of proportional tubes, PTY1, PTY2 and PTX in
Station 4, labeled as 'Muon Detectors' in figure \ref{spectrometer}.
Station 4 had two hodoscope layers for measuring both
$x$ and $y$, called Y4
and X4.

All wire planes and proportional tubes used the same gas mixture,
50\% argon, 50\% ethane and a small amount of ethanol alcohol added
by bubbling the gas through ethanol which was kept at a constant
25$^{\circ}$ $F$.

\afterpage{\clearpage}

\section{Special X Hodoscope Setting}
\label{xhodos}

\begin{center}
\begin{figure}[ht]
\resizebox{5.5in}{6.8in}
{\includegraphics[75,90][540,675]{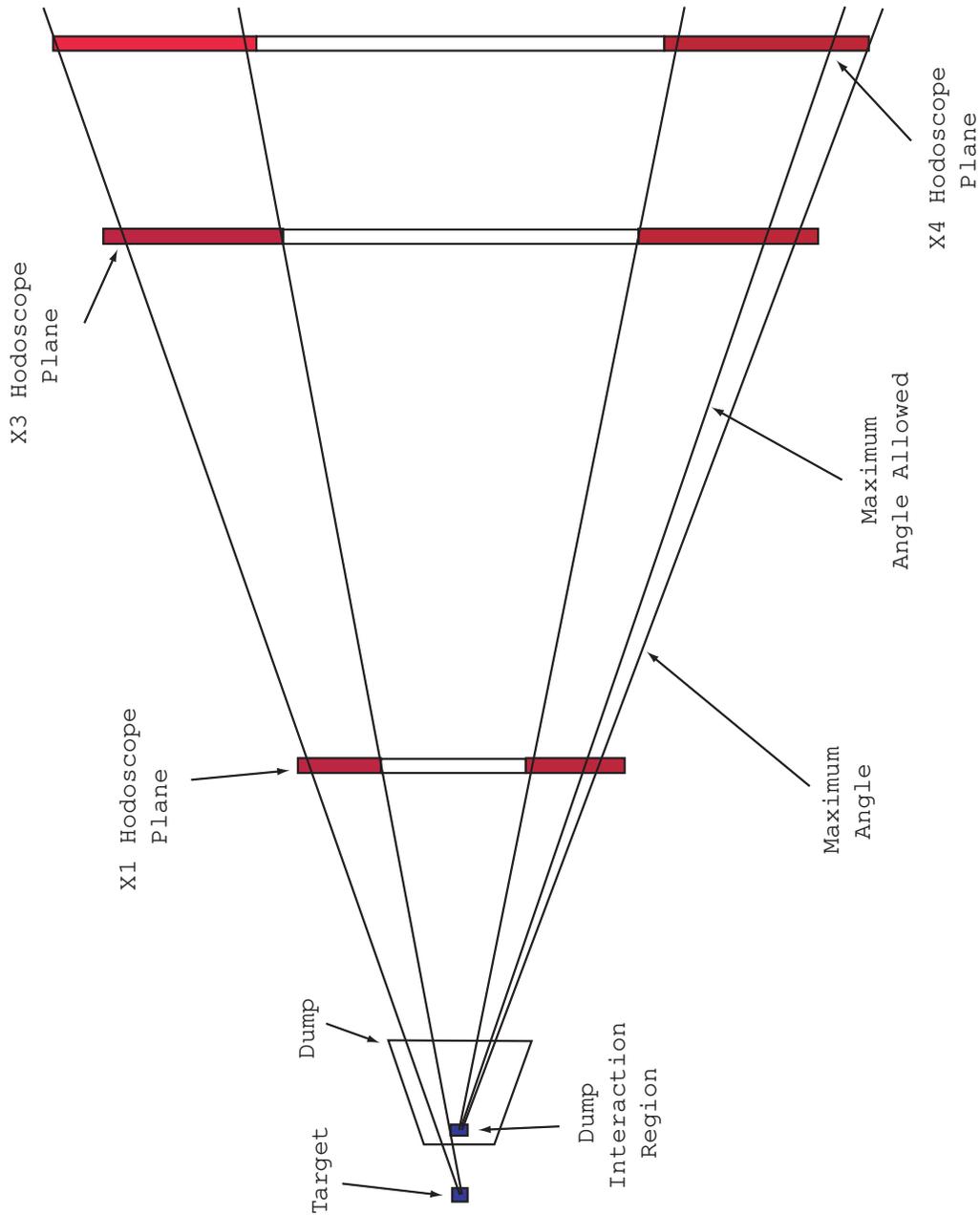}}
\caption[Modified $x$ Acceptances]{Drawing of change in acceptance of
the spectrometer in the $x-z$ plane, as well as the minimum and
maximum opening angles caused by disconnecting the high voltage
supplies to the middle half of the hodoscopes in all $x$ hodoscope
planes. Note the extreme shortening of the $z$ axis, though
relative distance from targets to hodoscope planes is preserved.
Maximum angle refers to the maximum physical opening, while
maximum angle allowed is the angle in the final cuts to insure
all events passed through a hodoscope in all three planes.}
\label{xplane}
\end{figure}
\end{center}

The single muon analysis used a special configuration of the
$x$ measuring hodoscopes. The high voltage supply for the middle
half of all $x$ hodoscope layers was turned off (referred to as
being pulled), shown in figure \ref{hodos}. This reduced the event
rate for single muon events as well as providing a way to
distinguish between events from the target and dump. This was a
significant change from the normal operation of the apparatus.
Single muon
events required hits on the same side in all three $x$ hodoscope
layers. This $x$ hodoscope configuration plus the required trigger,
described in \ref{triggers}, limited the transverse momentum
of accepted events, because the muon must have a minimum absolute
value for the slope in the $x-z$ plane,
$ | \nstantx | = | \nsxmom/\nszmom | $. This
minimum slope is shown in figure \ref{xplane}. The figure also
shows how the loss in acceptance allowed separation of target and
dump events for single muons. The minimum and maximum
$ | \nstantx | $ for muons to be accepted, based on the
spectrometer survey are listed in Table \ref{hodoplanes}.

\afterpage{\clearpage}

\section{Event Triggers and Readout}
\label{triggers}

The FNAL E866/NuSeA trigger has been described in detail in
several references \cite{nim-a-418-322-1998} \cite{kaplan}.
The data taken for the single muon analysis contained both single
and dimuon events, requiring dimuon as well as single muon
triggers. Dimuon events were used to calibrate the analysis
routine as described in Chapter \ref{analysis}.
The calibrations found from the di-muon events were then used to
analyze the single muon events.

Signals from the hodoscopes were first sent to LeCroy 4416
discriminators whose signals were reshaped and synchronized to the
accelerator RF clock. All triggers are hit correlations from
some or all hodoscope layers and are referred to as Physics A,
Physics B and Diagnostics triggers. The single muon data was
taken using 5 single and 5 dimuon physics triggers in two
separate but similar configurations. The trigger system for the
left half of the spectrometer is shown in figure \ref{trig_nim}
\cite{hawker}.

The trigger system had a total propagation delay from inputs
from hit hodoscopes to output of the event to memory storage of
approximately 20 nsec, providing single bucket resolution. This
was accomplished by comparing hodoscope hits in the seven layers
to hardwired and software defined matrices in modules called
Track Correlators (TCs) or Matrix Modules (MMs). The four Trigger
Matrix modules (called MUL, MUR, MDL and MDR) were lookup tables
loaded onto a set of ECL SRAM chips. These modules used
correlations in the Y1, Y2 and Y4 hodoscopes only. A track of
interest would define a 'road' in the Y view under the
deflection of the magnetic fields in SM12 and SM3. Hits on Y2 and
Y4 were correlated against predicted hits on Y1 and if the desired
correlation existed, output for the event was generated. These
were used only for dimuon events. Four S4XY Track Correlators,
also used for di-muon events only, used signals from the X4 and
Y4 hodoscopes only. The correlations required for a di-muon event
to be accepted could be changed by software.

The three $x$ hodoscope layers were hardwired into an and for each
side, called X134L and X134R. Any muon which traversed an energized
hodoscope paddle in all three $x$ layers on one side satisfied the
X134 trigger.

These were passed to the two main physics Track
Correlators (along with the other information from the Matrix
Modules and S4XY Track Correlators) called Physics Triggers TC and
Diagnostic Triggers TC. These two main correlators determined the
validity of an event and if it was to be taken to tape, then
informed the Master Trigger to set the busy signal and stream the
event to a large memory buffer, and subsequently written to 8mm
tape.

The di-muon triggers were labeled Physics TcA1 through
4, Diagnostic 3 and all four S4XY triggers. All dimuon triggers
were not prescaled due to the low beam intensity.

One example of an opposite sign dimuon pair physics trigger was a
coincidence of a track through the upper and lower left (or
alternatively right) quadrants (one being the \muplus and the other
the \nsmuminus), and an example for a like sign dimuon event where
one muon goes through the upper right quadrant in coincidence with
a muon passing through the upper left quadrant, all using the four
Matrix Modules MUL, MUR, MDL and MDR. The Diagnostics trigger
required a left-right coincidence in five of the seven layers of
hodoscopes on both sides of the spectrometer.

The X134L/R three-fold coincidence trigger, called PhysB1, was
the exclusive single muon trigger used for analysis. For diagnostics
purposes there were 4 other single muon triggers, PhysB2, Diag1,
Diag2 and Diag4. The diagnostic triggers were useful for studying
edge effects of the pulled hodoscopes as well as
hodoscope efficiencies for two of the three X layers.

The first trigger, PhysB1, required a hit in all three X hodoscope
banks on either side. PhysB2 was a prescaled trigger to take 1 of
2000 events where there was a track through a quadrant (any one
quadrant of the four) using the MUR/L MDR/L matrix modules, thus
avoiding the large opening angle required for trigger PhysB1.

The main single muon event trigger was PhysB1, called the
X134L/R trigger referring to the three fold coincidence of the
three X hodoscope planes on either side, left or right. This
trigger is often referred to by the single muon analysis as the
'trigger bit 5' trigger since that bit was set if the event had
that coincidence. All single muon events were required to have a
valid PhysB1 trigger set.

All events satisfying one or more triggers were written to tape. This
was accomplished using Nevis Transport electronics \cite{kaplan},
and VME as well as CAMAC modules. All detector subsystems fed data
onto the Transport bus, which was then transferred to the memory
buffer. The memory buffer, which was VME based, formatted the event
data and transferred it to 8mm tape. Since the typical spill lasted
20 seconds out of every minute, the memory buffering allowed
higher event rates with less deadtime. Each event written to tape
contained the values from the coincidence registers from the
hodoscopes and proportional tubes and readouts from the time to
digital converters (TDC's) from signals in the drift chambers, as
well as beam position, size and intensity, target position,
voltages and information from various monitors.

\section{The Single Muon Data}

\renewcommand{\arraystretch}{1.5}

\begin{table}[ht]
\caption[Single Muon Data]{Data taken for this analysis. All data
were taken with a common trigger configuration. Currents in SM12
and SM3 set to -1400 and -4200 amps, respectively. All data was
taken using the same trigger configuration file, labeled jpsism2.
The triggers for this configuration are given in the text.}
\label{data1}
\begin{center}
\begin{tabular}
[c]{|lccrc|}
\hline

  &
  &
  &
 Events & \\

\raisebox{2.0ex}[0pt]{Run} &
 \raisebox{2.0ex}[0pt]{Date} &
 \raisebox{2.0ex}[0pt]{Time} &
 $\times \, 10^{6}$ &
 \raisebox{2.0ex}[0pt]{Trigger}\\

\hline\hline

2748 &
 07/10/97 &
 10:14 &
 5.778 &
 jpsism2\\

2749 &
 07/10/97 &
 14:06 &
 6.047 &
 ''\\

2750 &
 07/10/97 &
 15:41 &
 6.067 &
 ''\\

2751 &
 07/10/97 &
 17:16 &
 5.918 &
 ''\\

2752 &
 07/10/97 &
 19:40 &
 5.942 &
 ''\\

2753 &
 07/10/97 &
 21:42 &
 5.750 &
 ''\\

2754 &
 07/10/97 &
 23:30 &
 7.995 &
 ''\\

2755 &
 07/11/97 &
 01:38 &
 6.172 &
 ''\\

2756 &
 07/11/97 &
 03:41 &
 5.944 &
 ''\\

2757 &
 07/11/97 &
 06:06 &
 3.150 &
 ''\\

\hline

\end{tabular}
\end{center}
\end{table}

\renewcommand{\arraystretch}{1.0}

\afterpage{\clearpage}

\chapter{Analysis}
\label{analysis}

The 10 runs on 10 tapes listed in
Table \ref{data1} (page \pageref{data1}) contained the single muon
data used in this analysis. The E866 collaboration analyzed data
for dimuon events with an analysis routine first developed for the
E605 experiment. The dimuon code was modified to analyze both
single and dimuon events for this analysis. All modifications to
the original routines are presented in detail. Analysis of the
single muon data was performed in four stages;
\begin{enumerate}

\item Separate single from dimuon events, separate target events
  from the dump events, then place the separated raw data into Data
  Summary Tapes, or DSTs, for further analysis.

\item Iterative analyses of the dimuon DSTs were then performed
  to determine the correct values, or calibrations, needed to
  initialize the code. No further analysis of dimuon data was
  performed.

\item Multiple analyses of the single muon data was performed using
  the single muon DSTs, placing the fully analyzed events into
  large arrays called n-tuples.\footnote{The n-tuple as
  well as PAW which was used to do the final analysis of the data
  were produced by CERN.}

\item Output final data (the single muon spectra) to arrays for use
  in the fitting routines to determine the open charm cross
  sections.

\end{enumerate}

\begin{center}
\begin{figure}[ht]
\resizebox{5.5in}{6.7in}
{\includegraphics[15,60][590,765]{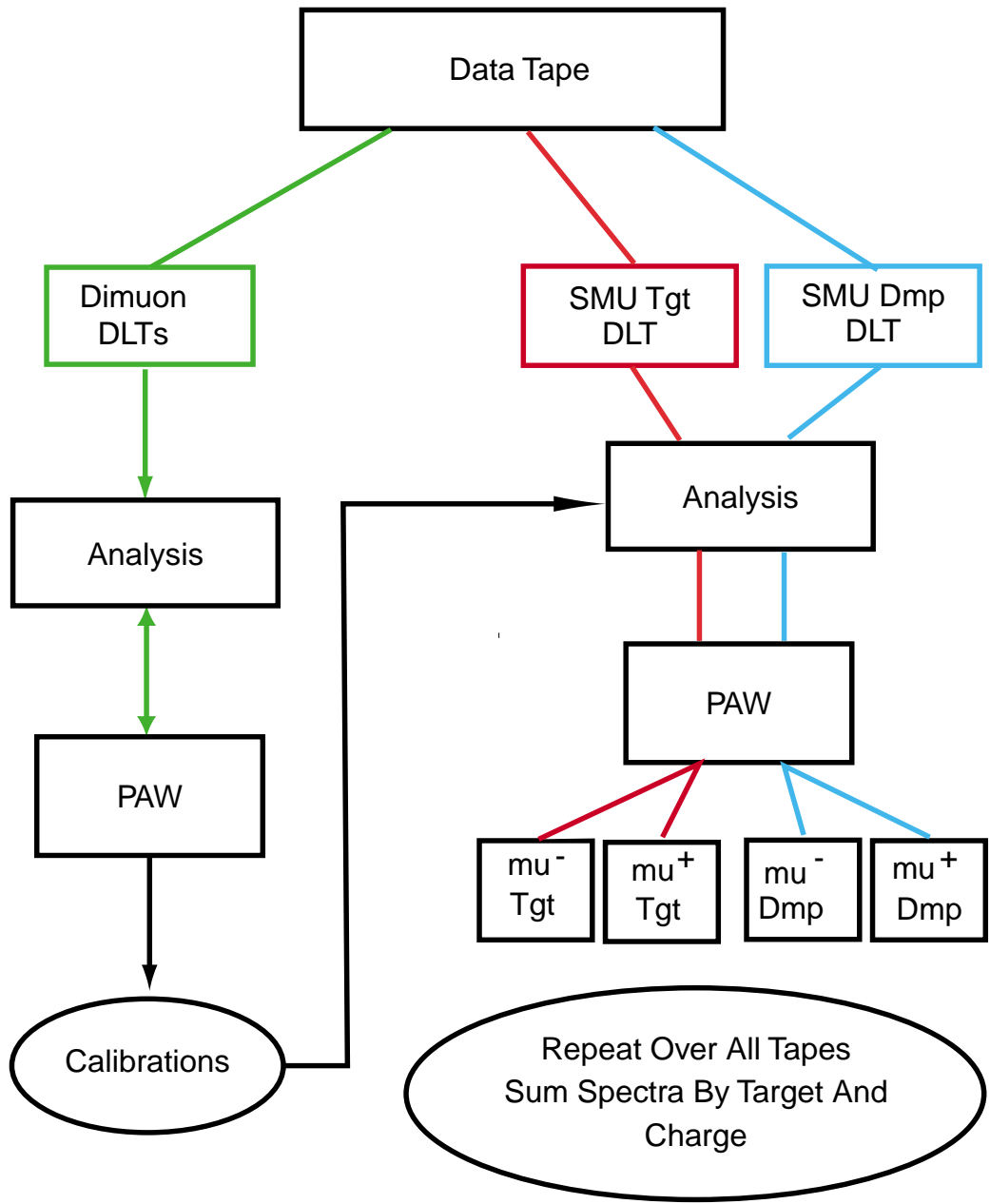}}
\caption[Flow Chart Of The Analysis]{Flow chart of the process
used to analyze the single muon data. The data was histogrammed
as a function of the muon \pt and placed into arrays for use
in the least-squares minimization routines described in Chapter
\ref{fitting}. Each target resulted in four spectra, one each for
\muplus and \muminus from the target and one each
from the dump.}
\label{data-to-array}
\end{figure}
\end{center}

Figure \ref{data-to-array} shows a flow chart of the method used to
analyze the single muon data, and may be useful for the following
discussions. All analysis of the raw data
as well as generation and analysis of Monte Carlo events was
performed on a DEC Alpha 500 workstation.

\afterpage{\clearpage}

\section{Initialization and Unpacking}

The first step in the analysis code was to initialize for the
specific run. This initialization included setting the magnetic
fields in SM12 and SM3 to the current direction and value at the
start of the run, trigger information and input spectrometer
calibrations. The code identified the number of the run from the
data tape and used an internal look up table to set the magnetic
field configurations whereas the trigger information and
spectrometer calibrations needed were read from specified trigger
or data tables.

The code then unpacked and processed the events. Unpacking an event
was a process that read the output DAQ information contained on the
tape for that event and allocated the information to several large
arrays which were designed to speed up the analysis of each track.

\section{Tracking}
\label{tracking}

Tracking began with searching for hit clusters in all the wire
planes in Stations 2 and 3 (See Figure \ref{spectrometer}, page
\pageref{spectrometer}). This required hits in 4 of the 6 planes
in each station to qualify as a potential hit cluster. Those
clusters having hits in 2 of the 3 planes were classified as
doublets, those having all three planes registering
a hit were classified as triplets. All clusters required at least
one hit in the Y view plane. Once a list of doublets and triplets
had been made, combinations of hits in the two stations were linked
to produce track candidates. The linkage was a combination of hit
clusters that loosely pointed back to the target region.

The next tracking process was to extend the candidate tracks to hit
clusters in the Station 1 chambers. Each candidate track was
projected upstream through SM3 in the \textit{x-z} plane to a
vertical
band in Station 1. Only clusters found within this band were
considered for that track. If a cluster was found in the band,
the entire track from Station 1 to Station 3 was refit using all
18 wire planes in the three stations. The $z$ coordinate in SM3,
$Z_{SM3}$ was allowed to vary. All tracks required at least 14
hits in the 18 planes. The final fit at this point in the tracking
routine gave the position of the track in $x$, $y$ and
$z$ (called $X_{SM3}$, $Y_{SM3}$ and $Z_{SM3}$ respectively) at
the SM3 bend plane as well as the slopes \nstantx(SM3) and
\nstanty(SM3), where \tantx and \tanty are the ratios of the
$x$ and $y$ momenta to the $z$ momentum using the sign convention
that \tantx (or \tanty) is positive if \xmom or \ymom is in
the positive $\hat{x} \, \mbox{or} \, \hat{y}$ direction.

The difference
$ \mbox{\nstanty}^{U}(SM3) - \mbox{\tanty}^{D}(SM3) $, where U and D
refer to upstream or downstream of the SM3 bend plane, with the
field map of SM3 determined the \textit{y-z} projection of the
momentum. Combined with \nstantx(SM3), the four momentum
$\left( \frac{E}{c}, \, \overrightarrow{p} \right)$ at
Station 1 was fully determined. A final cut on the 18 plane fit
requiring the track to have \pdfchi less than 5 was made.

The full reconstruction of the track downstream of Station 1 was
then completed. Each track found above making all cuts was projected
back to Station 4. Since there was considerable hadron absorber
material throughout this projection, the projected intercept at
Station 4 was compared to a
'window' of possible hits in the 3 proportional tube layers as well
as the hodoscopes. These windows were wide enough to allow for
multiple scattering within 5 standard deviations at each detector
plane around the projected straight line intercept. High
momentum tracks had windows of less than a few cells. The detectors
were then scanned for hits within the windows. A track was further
considered if it had 3 of the 5 planes (3 proportional tube planes
and 2 hodoscope planes) registering a hit.

This procedure was done for all candidate tracks found from the
event on tape, and the resulting number of such tracks as well as
numbers of hits, clusters and fitting information was passed to
arrays for further use.

\section{Trace-back Through SM12}
\label{traceback}

Once all candidate tracks for an event were found, the event was
passed to a section of code that would 'trace-back' each candidate
track to the assumed place of origin on the $z$ axis, $Z_{tgt}$. For
this Chapter, 'target' (such as $tgt$ in $Z_{tgt}$) is used to
denote \textit{either} the actual targets in the target wheel (target
analysis), or the dump (dump analysis), unless it is necessary to
distinguish between the two. This was done sequentially to each
track in the order they were
found in the tracking section of the code. Since there were no
detectors upstream of Station 1, this was done using the
magnetic field map of magnet SM12 as well as experimentally
determined energy losses
for the absorber materials \cite{nim-a-251-21}, to 'project'
the most probable path of
the muon back toward its assumed origination. Once this projection
was calculated, the effects of multiple scattering were added by
use of a single bend plane approximation, resulting in the final
calculated momenta for the muon at its assumed point of origin.

The trace-back section of the analysis code is described here as
it was originally written. The
single muon analysis made several changes to the procedure which will
be described in Section \ref{smuchanges}, however, the basic
trace-back was preserved and valid for either dimuon or single muon
events. Trace-back was done in three stages for each track found.

The first stage entailed adding back in lost energy
due to interactions within materials as well as changes in
trajectory due to the magnetic field within magnet SM12 beginning
at $Z_{SM3}$, the location of the bend plane in SM3. The trace-back
procedure 'swam' the particle back to a fixed point along $z$,
called $Z_{tgt}$, which was either the location of the thin targets
($z=-24.0$ inches) for target events, or a fixed $z$ location inside
the dump for dump events ($Z_{tgt}$ for dump events was
set to the distance at which 1/2 of the protons intering the
dump would have interacted, $Z_{tgt}=85.1$ inches).

The code began by adding in the beam offsets $X_{off}$ and $Y_{off}$,
and then trace the muon upstream until it reached either the
beginning of the copper dump at $z=68$ inches for target events,
or until the next incremental distance in $z$ (referred to as
sections which were 2 inches in length) upstream of $Z_{tgt}$
where events from the dump were forced to originate from.

Candidate tracks must pass cuts requiring the track to actually
remain within the volume of the magnet during the trace-back. The
\textit{x-y} plane at $Z_{tgt}$ was called the analysis plane. The
placement of the forced points of origination are discussed in
detail in Section \ref{calibrations}. 

Corrections for energy loss used the mean energy loss
for a muon in 2 inches of the material being
traversed \cite{nim-a-251-21}.\footnote{The energy loss code
calculated the mean energy loss for a muon of given momentum while
travering the entire amount of absorber. The loss was then
averaged over the 2 inch increments.} The materials, in the order
taken by the trace-back procedure,
were 72 inches polyethylene, 108 inches carbon, 24 inches copper
(called the copper wall) and either up to 168 inches of copper for
target events or about 151 inches of copper for dump events. The
amount of the copper dump traversed was stored for use during the
second stage of the trace-back.

The second stage of the trace-back used the amount of dump material
the track had traversed to determine the location
along $z$ at which correction for multiple scattering should
be done. The analysis code used a single bend-plane approximation
to correct for multiple scattering that occured while the muon
traversed the dump and hadron absorber materials in magnet SM12. The
single bend-plane approximation was based on determining a point
along the $z$ axis called $Z_{scat}$, where the effects of the
multiple scattering could be approximated in a single large scatter,
or bend (see for example \cite{prd-43-2815}). The location of
$Z_{scat}$ was calculated using
$$ Z_{scat} = a \, l_{dmp} + min $$ where $a$ is a percentage of the
amount of dump material traversed and $min$ is the minimum distance
along $z$ at which the plane could be set. Since all single muon
events required full traversal of the dump, which restricted all
hadrons to have a set open decay length, $a$ was set to 0.

The multiple scattering correction required retracing the muon from
downstream of magnet SM12 to $Z_{tgt}$ (at either the targets or
in the dump), storing the 3 momenta
$\overrightarrow{p}_{scat}$ and position
$(X_{scat},Y_{scat},Z_{scat})$ found at $Z_{scat}$ during the
trace-back, as well as the position $(X_{tgt},Y_{tgt},Z_{tgt})$ at
$Z_{tgt}$. $X_{tgt}$ and $Y_{tgt}$ are commonly referred
to as the uniterated $x$ and $y$ intercepts at the analysis plane.
The difference between the uniterated intercepts and the beam
centroids, $\Delta X$ and $\Delta Y$ , as well as an angular
difference or correction to the track, $\Delta\theta_{x}$ and
$\Delta\theta_{y}$ were calculated using
\begin{align}
\left(
\begin{array}
[c]{l}%
\Delta X\\
\Delta Y
\end{array}
\right)   &  =\left(
\begin{array}
[c]{l}%
X_{tgt}-X_{off}\\
Y_{tgt}-Y_{off}%
\end{array}
\right) \\
\left(
\begin{array}
[c]{l}%
\Delta\theta_{x}\\
\Delta\theta_{y}%
\end{array}
\right)   &  =\frac{1}{Z_{tgt}-Z_{scat}}\left(
\begin{array}
[c]{l}%
\Delta X\\
\Delta Y
\end{array}
\right)
\end{align}

The third stage was an iterative process of tracing the muon back
to the analysis plane from the scattering plane and testing the
iterated $x$ and $y$ intercepts at the analysis plane against a
distribution of acceptable values for $X_{tgt}$ and $Y_{tgt}$. If
the current trace-back failed, new values for the angles
$\Delta\theta_{x}$ and $\Delta\theta_{y}$ were calculated and the
process was repeated. The kinematical variables (all referred to as
iterated variables) were calculated after this final
step in the retracing of the muon.

\section{Spectrometer Calibrations}

The analysis code required several input parameters to
be set during initialization. The calibrations required
were a scalar multiple for the magnetic field strength of SM12,
called \textit{TWEE}, the beam offsets $X_{off}$ and $Y_{off}$, used
to center the beam at $Z_{tgt}$, the beam angles \textit{XSLP} and
\textit{YSLP} and the position at which to place the scattering
plane, $Z_{scat}$, called \textit{ZSCPLN}.

Since the targets were relatively thin, it was assumed that the
most accurate calibration for the magnetic field strength scalar,
\textit{TWEE}, would be found by fitting the dimuon data to the
$J/\Psi$ mass (3.097 \nscmass). The
calibrations for dump analyses were then found using this field
scaler, with a small variation (less than 2 percent) allowed for
uncertainties in the field map as well as the effect of the
thickness of the target and location of the plane of analysis used
for dump events. All dimuon target analyses used data taken with
the beam incident to all three targets while analyses for determining
the dump calibrations only used data taken with the target wheel at
the empty position.

The calibrations were determined using an iterative process,
changing the value for one of the variables and performing a new
analysis of the Data Summary Tape. Results were plotted
and subsequent changes were made to optimize the variable for that
run. After all variables were determined for both target and dump
the code was calibrated with the final settings and the
initial reductions for single muons from that run were performed.
Calibrations for this analysis are given in Table
\ref{lscalibrations} page \pageref{lscalibrations}.

\subsection{Cuts Used For Dimuon Events}
\label{dimucuts}

The cuts on dimuon events used for determining the optimal
spectrometer calibrations were:

\begin{enumerate}
\item  Valid PhysA1, 2, 3 or 4 trigger must be set.

\item  Event must have only two candidate tracks.

\item  Event must have two valid tracks after tracing section of
  code.

\item  Event must have oppositely charged muons. (Opposite sign
  pair.)

\item  The estimated $z$ vertex, \textit{ZUNIN}, must be within
  50 inches of $Z_{tgt}$.

\item $ \mbox{\tanty} \leq 0.030 $ for both \muplus and \nsmuminus.

\item $ \mbox{\tantx} \leq 0.028$ for both \muplus and \nsmuminus.

\item $ 2.1<m_{\mu\mu}<4.1 $ (\nscmass).
\end{enumerate}

The cut on \tanty was to insure both
muons from the event traversed the entire length of the dump as
explained in Section \ref{smucuts}. The cut on \tantx was used to
reduce
the number of events where either muon may have scattered back
into the acceptance after interacting in the walls of SM12, and
$m_{\mu\mu}$ is the reconstructed mass of the dimuon pair in
\nscmass.

To select higher quality events, the E866 analysis
code calculated an estimate of the position along the $z$ axis
where the parent hadron would have been created for the dimuon
pair being analyzed. This calculation was based on information
found as the two muons were traced back to the analysis plane at
$Z_{tgt}$ during the second stage of the trace-back. The second
trace-back section of code determined the four intercepts,
$X_{tgt}^{\pm}$, and $Y_{tgt}^{\pm}$ as well as the two four
momenta, $(\frac{E^{\pm}}{c},\overrightarrow{p}^{\pm})$ at
$Z_{tgt}$ where $\pm$ refers to the two muons, \muplus and
\nsmuminus, of the dimuon pair. The distance in $y$ between the pair
was calculated by $Y_{tgt}^{-}-Y_{tgt}^{+}$, and the estimated
$z$ vertex of the parent hadron, \textit{ZUNIN}, was then
calculated using:
\begin{equation}
ZUNIN = - \left[ \frac{Y_{tgt}^{-}-Y_{tgt}^{+}}
{ ( \mbox{\nstanty}^{-} - \mbox{\nstanty}^{+} ) } \right] \quad
\mbox{(inches)}
\end{equation} The smaller the absolute value of \textit{ZUNIN}
calculated, the more likely the parent hadron was produced at or
near $Z_{tgt}$.

The magnetic field strengths used in magnets SM12 and SM3 while
taking the single muon data, plus the restriction that all events
must have all muons trace back through the entire length of the
dump, made the cut on the estimated $z$ vertex of the dimuon
pair, \textit{ZUNIN}, insufficient by itself to isolate events
originating from the target. Limiting contamination of dump events
in a target analysis was
accomplished using the cut on \textit{ZUNIN} listed above plus the
reconstructed mass cut. The effect of the mass cut on \textit{ZUNIN}
for the target analysis of run 2753 is shown in Figure
\ref{tgtamzunin}. Left is \textit{ZUNIN} for all events with the
beam incident on any of the three targets, and right is the same
data with just the mass cut applied. Figure \ref{tgtairzunin} shows
the reduction of dump events contaminating a target analysis using
this cut by looking at data taken with the target position empty.
Left is a plot of \textit{ZUNIN} for a target analysis of run 2753
for all spills where the target position was empty.
Right is the same data after the mass cut was applied.

\begin{center}
\begin{figure}[!pt]
\resizebox{5.6in}{2.1in}
{\includegraphics[25,495][530,675]{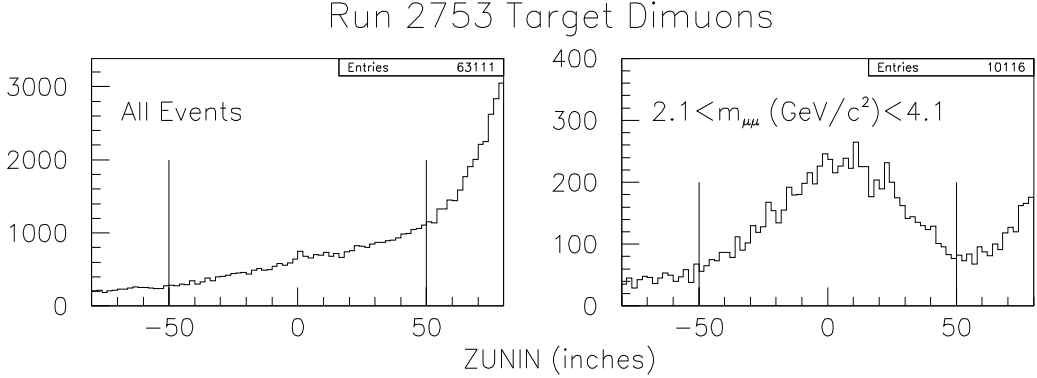}}
\caption[Estimated $z$ Vertex \textit{ZUNIN}.]
{Estimated $z$ vertex, \textit{ZUNIN}, for a target analysis of
dimuon events from Run 2753 with any of the targets presented to the
beam. Left is all events passing the dimuon cuts except for the
\textit{ZUNIN} and mass cuts, plotted as a function of
\textit{ZUNIN}. Right is same data with the mass cut,
$2.1<m_{\mu\mu}<4.1$ \cmass, applied. Vertical lines
indicate the limits on \textit{ZUNIN} used in calibrating the
spectrometer. Compare with figure \ref{tgtairzunin} where no
target was present.}
\label{tgtamzunin}
\end{figure}

\begin{figure}[!pt]
\resizebox{5.6in}{2.1in}
{\includegraphics[25,495][530,675]{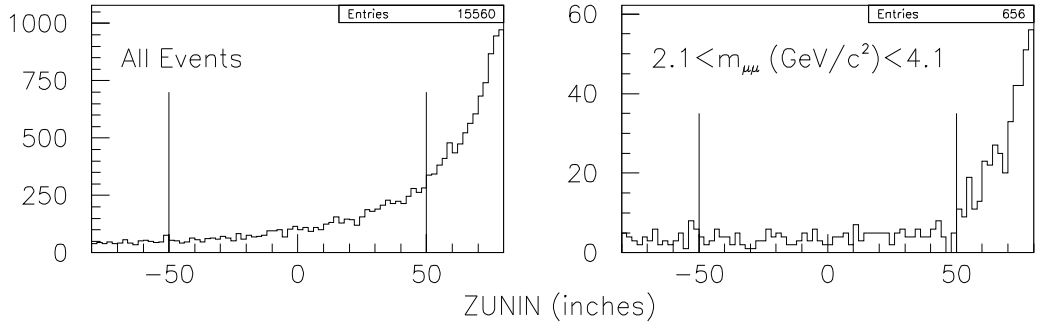}}
\caption[\textit{ZUNIN} For Target Analysis With Target Position Empty.]
{Plots of the estimated $z$ vertex, \textit{ZUNIN},
for a target analysis of Run 2753 where the target wheel was
in the empty position. Left is the dimuon events passing all cuts
except the mass and \textit{ZUNIN} cuts. Right is the same
data after the mass cut, $2.1<m_{\mu\mu}<4.1$ \cmass, was applied.
Vertical lines indicate the limits imposed on \textit{ZUNIN} while
determining the calibrations for the spectrometer. COmpare to figure
\ref{tgtamzunin}.}
\label{tgtairzunin}
\end{figure}
\end{center}

\afterpage{\clearpage}

The absolute calibration of the luminosity monitor IC3 was
unknown for these low intensities, but it was assumed that the
total recorded by the monitor
when the signal not busy was set (IC3SB) could be used to estimate
the number of dimuon events in a target analysis that originated in
the dump by comparing the appearant number of events in a target
analysis when there was no target present. The number of events in
a target analysis (which was the
sum of the events from all the targets) can be estimated from:
$$ N_{T}(D) = \frac{N_{0}}{IC3SB_{0}} \,
\sum^{3}_{i=1} \left( F_{i} \, IC3SB_{i} \right) $$ where
$N_{T}(D)$ is the number of events in a target analysis that
originated in the dump, $N_{0}$ is the number of events in a target
analysis performed for the empty target position, $IC3SB_{i}$ was
the total $IC3SB$ when target $i$ was presented to the beam (the
targets were labelled 0 through 3 for empty, thin copper,
beryllium and thick copper respectively) and $F_{i}$ was the
fraction of the beam incident on the dump:
$$ F{i} = exp \, \left( - l_{i} \,
\frac{\rho_{i}}{\lambda_{I,i}} \right) $$ Here $l_{i}$ is the
thickness of target $i$ (cm), $\rho_{i}$ the density of the material
in the target (\nsdensity) and $\lambda_{I,i}$ is the nuclear
interaction length of the material of target $i$ (\nsilength). An
example of the estimated dump contamination for Run 2751 is
presented in Table \ref{destimate} (page \pageref{destimate}).

\afterpage{\clearpage}

\subsection{Calibrations}
\label{calibrations}

\begin{center}
\begin{figure}[!ph]
\resizebox{5.6in}{5.7in}
{\includegraphics[35,155][530,670]{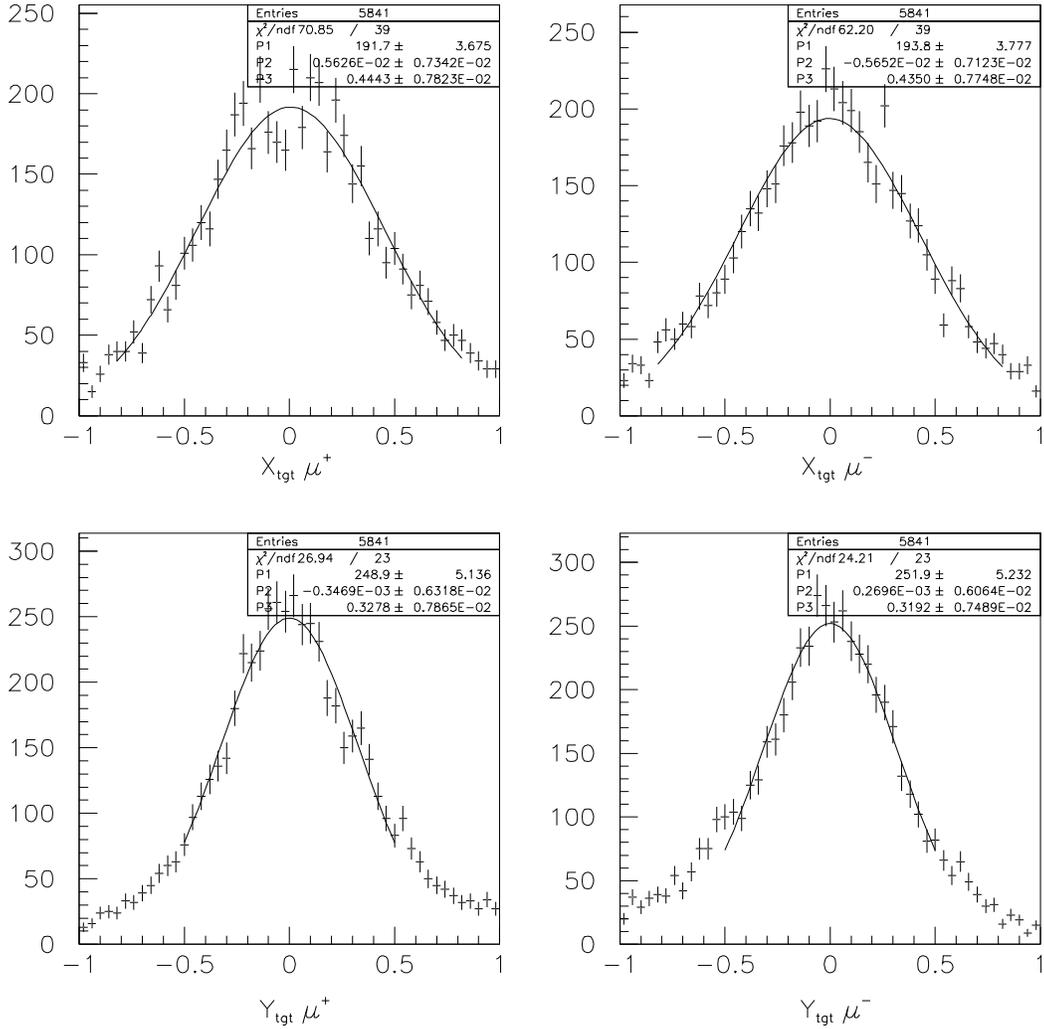}}
\caption[Plots of $X_{tgt}$ and $Y_{tgt}$.]
{Plots of $X_{tgt}$ and $Y_{tgt}$ used in determining the beam
offsets $X_{off}$ and $Y_{off}$. Figure is the final settings
used for Run 2752. Vertical error bars are statistical only.}
\label{xysettings}
\end{figure}
\end{center}

The optimum settings for the beam offsets, $X_{off}$ and $Y_{off}$,
as well as the magnetic field strength scalar $TWEE$ were found
using plots of the $X_{tgt}^{\pm}$ and $Y_{tgt}^{\pm}$ intercepts
at $Z_{tgt}$ for the \bothmu individually, as well as the $J/\Psi$.
The analysis plane,
$Z_{tgt}$, was set to the known location of the targets,
$Z_{tgt}=-24.0$ inches. $TWEE$ and $Y_{off}$ were changed until the
centroids of $Y_{tgt}^{\pm}$ were centered about $Y_{tgt}=0$. Due to
tracking limitations in the $x$ direction, $X_{off}$ was varied
until the centroids of $X_{tgt}^{\pm}$ were equidistant from
$X_{tgt}=0$. Figure \ref{xysettings} shows typical plots of
$X_{tgt}^{\pm}$ and $Y_{tgt}^{\pm}$ for the target dimuon analysis
of run 2751.

Beam angle corrections $XSLP$ and $YSLP$ (referred to as
$\theta_{x}^{\prime}$ and $\theta_{y}^{\prime}$ below) were then
determined for both target and dump analyses since the magnetic
field in SM12 deflected the proton beam, using the two functions
\begin{equation}
\frac{ \Sigma p_{x} }{ \Sigma p_{z} } =
\frac{ \xmag (\mu^{+}) + \xmag (\mu^{-}) }
 { \zmag (\mu^{+}) + \zmag (\mu^{-}) }
\label{xtheta}
\end{equation}
\begin{equation}
\frac{ \Sigma p_{y} }{ \Sigma p_{z} } = 
\frac{ \ymag (\mu^{+}) + \ymag (\mu^{-}) }
 { \zmag (\mu^{+}) + \zmag (\mu^{-}) }
\label{ytheta}
\end{equation}

\begin{center}
\begin{figure}[!ph]
\resizebox{5.7in}{3.0in}
{\includegraphics[35,410][530,670]{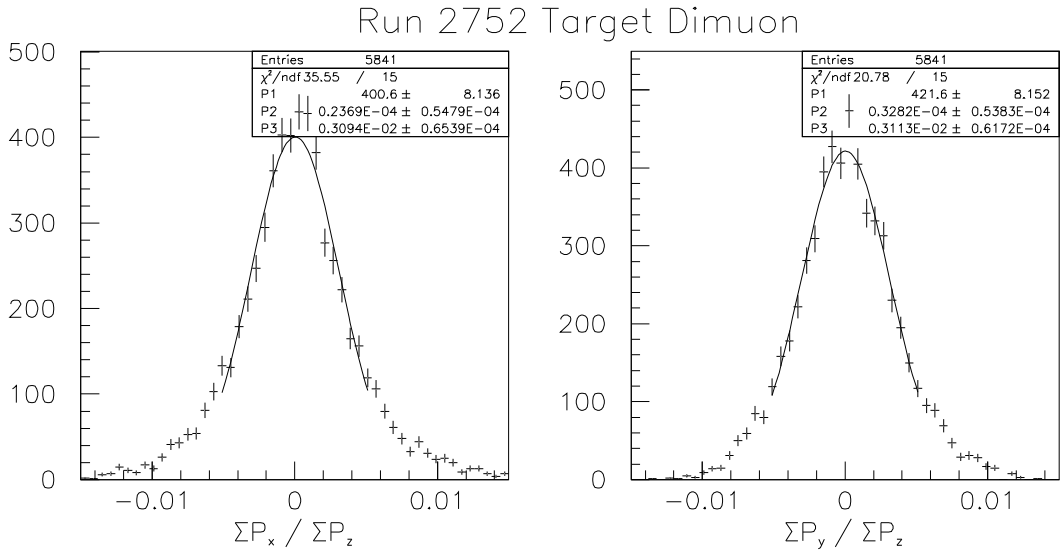}}
\caption[Plots Of The functions
$\Sigma p_{x}$ / $\Sigma p_{z}$ and
$\Sigma p_{y}$ / $\Sigma p_{z}$.]{Plots of functions \ref{xtheta}
and \ref{ytheta} used to calibrate the beam angles for
Run 2752. Vertical error bars are statistical only.}
\label{deltathetas}
\end{figure}
\end{center}

\begin{center}
\begin{figure}[!ph]
\resizebox{5.7in}{3.0in}
{\includegraphics[30,410][530,675]{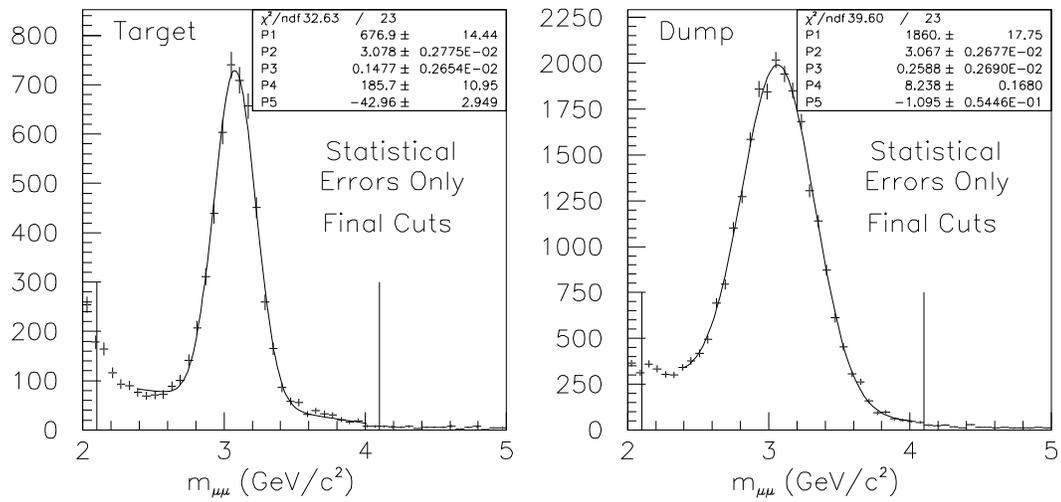}}
\caption[Spectra Of $m_{\mu\mu}$ For Run 2752.]{Reconstructed
dimuon mass, $m_{\mu\mu}$ \cmass for the target (left) and dump
(right) of run 2752. Fits are to the $J/\Psi$ ($m_{J/\Psi} =
3.097$ \nscmass). Errors are statistical only. Lines indicate
the mass cut used in the dimuon analysis.}
\label{1702masses}
\end{figure}
\end{center}

The beam angle corrections were varied until the centroids of the
two plotted variables were centered about 0. Figure
\ref{deltathetas} shows these two variables plotted during the
dimuon target analysis of run 2752. Figure \ref{1702masses} shows
the reconstructed mass spectra for dimuon pairs calculated from
target (left) and dump (right) analyses of run 2752 using the
spectrometer settings used to create Figures \ref{xysettings} and
\ref{deltathetas}. Table \ref{lscalibrations} (page
\pageref{lscalibrations}) give the calibrations found for the runs
used in this analysis.

\afterpage{\clearpage}

\begin{center}
\begin{figure}[!ph]
\resizebox{5.6in}{5.7in}
{\includegraphics[20,155][535,680]{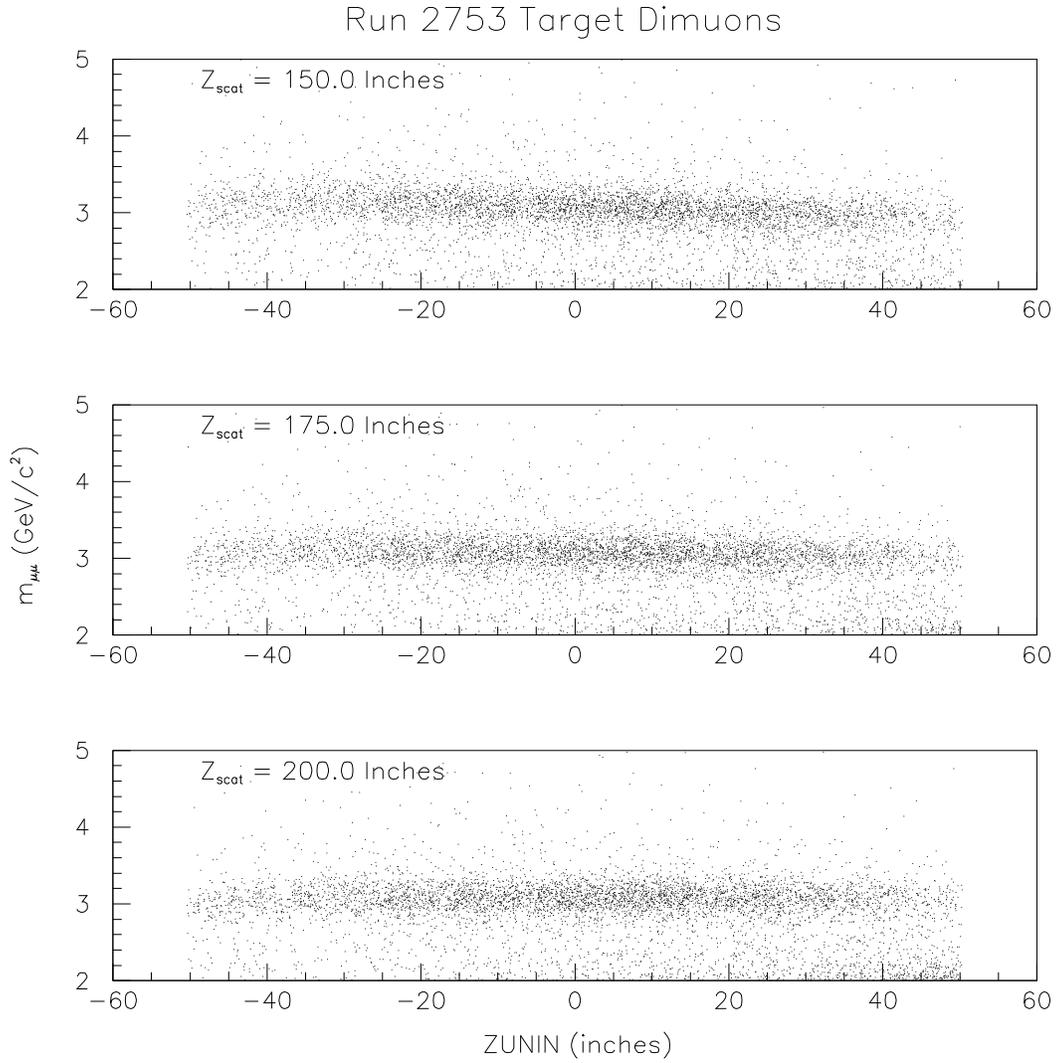}}
\caption[Plots of $m_{\mu\mu}$ vs. \textit{ZUNIN} For Run 2753 Target
Analysis.]{Reconstructed dimuon mass, $m_{\mu\mu}$
(\nscmass), versus the estimated $z$ vertex (inches) for the target
analysis of Run 2753. All plots use the same spectrometer
calibrations except the position for the scattering bend plane,
$Z_{scat}$, which was
set to $150.0$ inches for the top plot, $175.0$ inches for the
middle and $200.0$ inches on bottom. Plots were used to determine
the optimal position resulting in the minimum width of the mass
spectrum, which occured when there was no slope.}
\label{amvzunin}
\end{figure}
\end{center}

\begin{center}
\begin{figure}[!ph]
\resizebox{5.7in}{2.1in}
{\includegraphics[25,490][530,675]{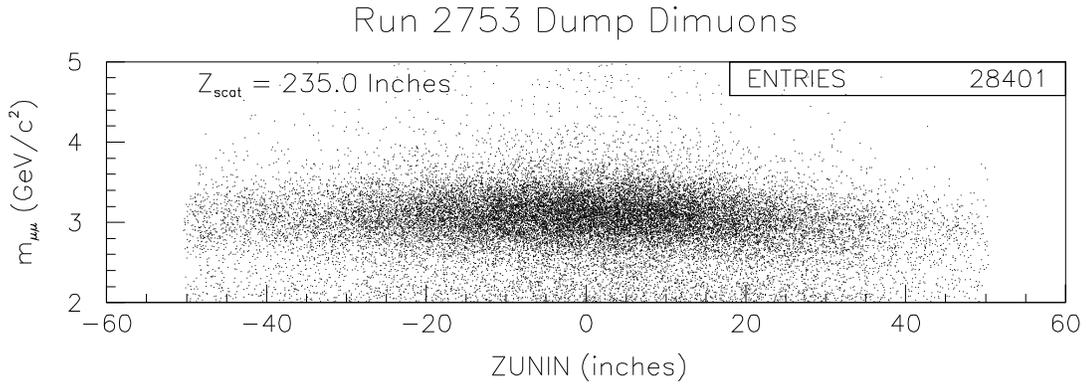}}
\caption[$m_{\mu\mu}$ vs. \textit{ZUNIN} Run 2753 Dump Analysis.]
{Reconstructed dimuon mass, $m_{\mu\mu}$ (\nscmass), versus
the estimated $z$ vertex (inches) for the dump analysis of Run 2753.
Similar plots and fitted mass spectrum plots to those in Figure
\ref{dmpwidths} were used to determine the optimal position of
$Z_{scat}$, resulting in the minimum width of the mass spectrum.}
\label{amvzunindmp}
\end{figure}
\end{center}

\begin{center}
\begin{figure}[!ph]
\resizebox{5.7in}{3.0in}
{\includegraphics[30,410][530,675]{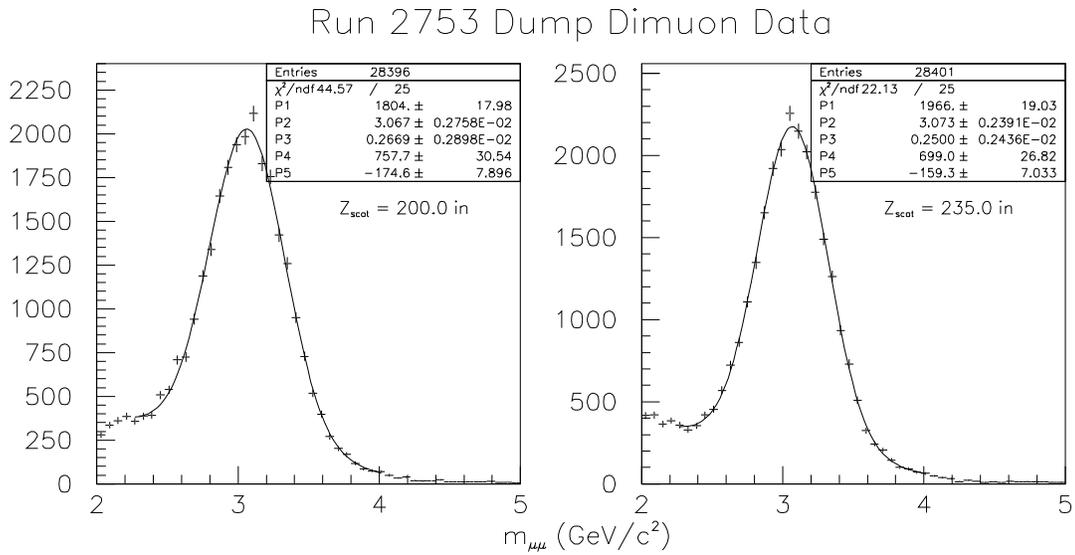}}
\caption[$m_{\mu\mu}$ For Run 2753 With Different $Z_{scat}$.]
{Reconstructed dimuon mass, $m_{\mu\mu}$ (\cmass) for the dump
analysis of Run 2753. Fits are to the $J/\Psi$ ($m_{J/\Psi} =
3.097$ \nscmass). Left is the reconstructed mass using
$Z_{scat}=200.0$ inches and right is the same data using
$Z_{scat}=235.0$ inches.}
\label{dmpwidths}
\end{figure}
\end{center}

The optimal $z$ position for the scattering plane, $Z_{scat}$, was
determined for target events by plotting the reconstructed dimuon
mass, $m_{\mu\mu}$, versus the estimated $z$ vertex, \textit{ZUNIN}.
A sample of the method used for target dimuon events from Run 2753
is shown in Figure \ref{amvzunin}. Top is $m_{\mu\mu}$ versus
\textit{ZUNIN} with $Z_{scat}=150.0$ inches, middle is the same data
with $Z_{scat}$ set to 175.0 inches and bottom is the result with
$Z_{scat}=200.0$, inches which was found to give the optimum value
for target events. Determination of the optimal value of $Z_{scat}$
was found by iteratively changing
the position, re-analyzing the dimuon data and fitting the
reconstructed mass peak around the $J/\Psi$. The optimal value was
found by picking the position giving the minimum width. The optimal
values were determined to be $Z_{scat}=200.0$
inches for target events and $Z_{scat}=235.0$ inches for dump
events. Figure \ref{amvzunindmp} shows a plot of $m_{\mu\mu}$ versus
\textit{ZUNIN} for dump events from Run 2753 with $Z_{scat}=235.0$
inches for comparison to the target analyses shown in Figure
\ref{amvzunin}. The width of the mass spectrum for dump events was
less sensitive to the location of $Z_{scat}$ than target events due
to the thickness  of the proton interaction region and the
limitation of having a fixed analysis plane. Figure
\ref{dmpwidths} shows two plots of the reconstructed mass spectrum
from run 2753 dump events. The left figure has $Z_{scat}=200.0$
inches and the right $Z_{scat}=235.0$ inches. Both plots use the
same calibrations otherwise.

\clearpage

\section{Single Muon Analysis}
\label{analysmu}

Once the spectrometer had been calibrated the analysis code was
initialized for single muon reduction. This entailed setting
several initial cuts to be made on all events as well as
initializing different output to the n-tuple files. The single muon
analysis was primarily interested in information regarding the
tracking and trace-back, and effort was made to develop an estimate
of the region along the $z$ axis where the parent hadron originated.

\subsection{Single Muon Changes}
\label{smuchanges}

The original code used pairs of tracks that had been traced back. One
criterion used to determine a single muon event was limiting the
number of candidate tracks to one. The code was modified to allow
single muon tracks to become valid events for the remaining sections
of code by creating a ghost muon having the same \xmom but reversing
the direction of \nsymom. All single muon events thus became opposite
sign dimuon events for analysis in the remaining sections of code,
with the exception that $ZUNIN = 0$. Since all dimuon information
was output for a  muon pair (usually opposite in charge), the single
muon code was changed to differentiate which track was the real
track and what charge that muon had. This was required since the
$y$ dependent ghost track information was reversed in sign and
placed into the output n-tuple as the opposite sign muon of that
pair, and the original code was developed for only one current
direction in SM12, creating an ambiguity in the analyzed events as
to which muon was really the \muplus and which was the \nsmuminus.
Since the original code reconstructed events using both muons, this
ambiguity had no effect on their results \cite{privater}.

\begin{center}
\begin{figure}[ht]
\resizebox{5.6in}{5.7in}
{\includegraphics[23,155][550,675]{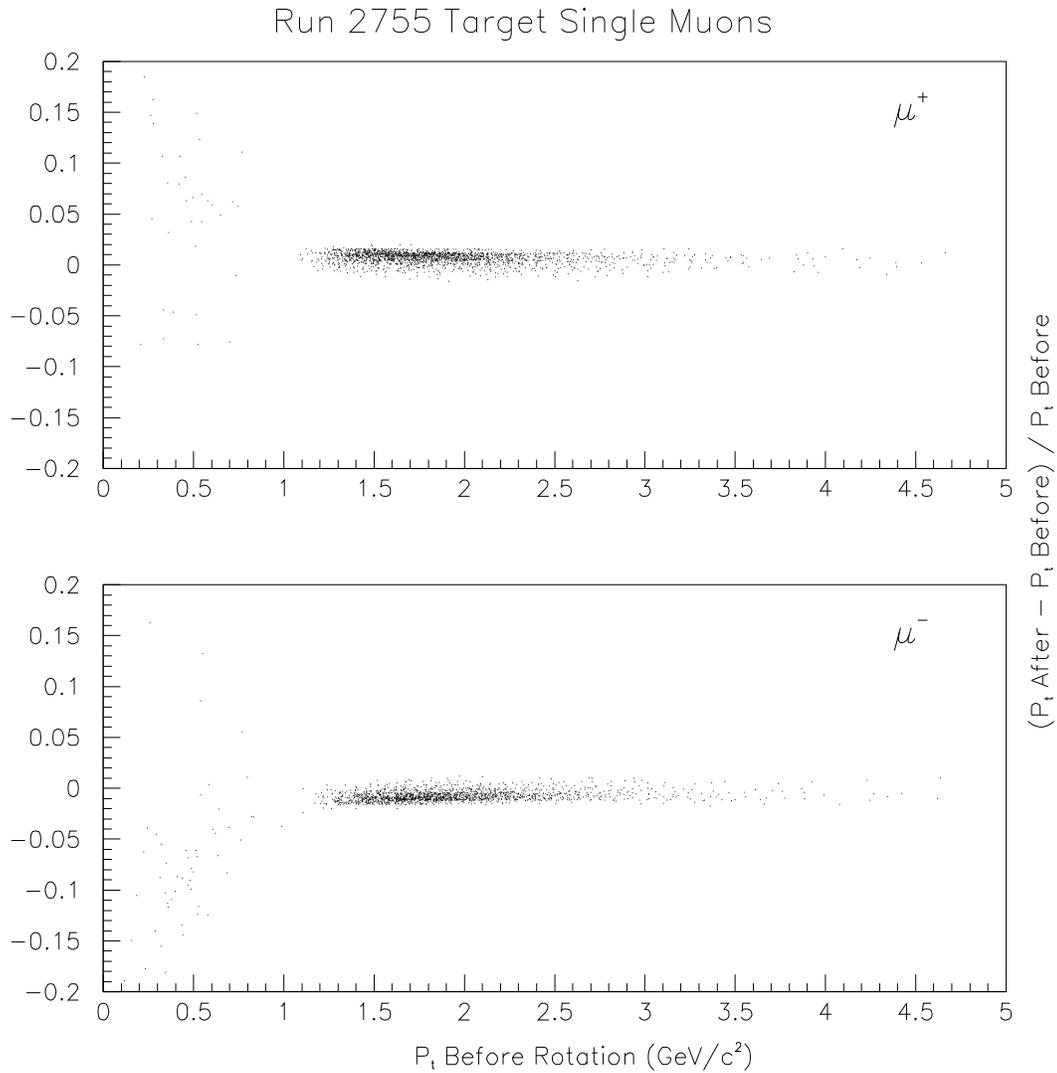}}
\caption[Effects Of Small Angle Correction.]{Effects of small angle
rotation correction applied to single muon events from the target
analysis of Run 2755. Events have the trigger, tracking, circle and
maximum $tan(\theta_{y})$ cuts applied. Top is \muplus and bottom
is \muminus.}
\label{rot-pt}
\end{figure}
\end{center}

A short section of code, after the
pair of muons had been fully traced back, allowed for rotation of
the $z$ axis to allow for beam angle corrections. The new momenta
were calculated from:
\begin{equation}
 | p_{x}^{\prime} |  = 
 cos(\theta_{x}^{\prime}) \;
 | \nsxmom |  + sin(\theta_{x}^{\prime}) \;
 | \nszmom |
\end{equation}
\begin{equation}
 | p_{y}^{\prime} |  =
 cos(\theta_{y}^{\prime}) \;
 | \nsymom |  + sin(\theta_{y}^{\prime}) \;
 | \nszmom |
\end{equation}
\begin{equation}
 | p_{z}^{\prime} |  =
 \sqrt{ | \nszmom |^{2} - | p_{x}^{\prime} |^{2} - 
 | p_{y}^{\prime} |^{2}}
\end{equation}
where primed implies after rotation. The angle corrections,
$\theta_{x}^{\prime}$ and $\theta_{y}^{\prime}$ were found using
a large number of dimuon events for each run as described in the
previous section. Figure \ref{rot-pt} shows the ratio
$(p_{t}^{\prime}-p_{t})/p_{t}$ versus $p_{t}$ for \muplus (top)
and \muminus (bottom) single muon events from a target analysis of
run 2755.

The original code was also altered to allow calculating the '$z$ of
closest approach' for three cases. The code was changed to calculate
the distance $x$, $y$ and $R=\sqrt{x^{2}+y^{2}}$ from the muon to
the $z$ axis at each section during the first trace-back. The code
then output the $z$ locations at which each of these became a
minimum. These minima gave a qualitative examination of the
ability of the tracing section of the code to determine the
origination of the muons in $z$, i.e. target or dump, for single
muon events.

\clearpage

\subsection{Single Muon Cuts}
\label{smucuts}

Initial analysis of the data for single muon events used a series
of cuts designed to insure that the event was the result of a
leptonic or semi-leptonic decay resulting in one muon, the decay
muon must have traversed the full length of the solid copper dump
and the event must have originated from the correct region along
$z$. Triggers and number of candidate tracks for the tracking
section of the code were used for the first part, and cuts
developed using both iterated as well as uniterated physics or
tracking variables were used for the others. Subsequent analysis
used cuts designed to further restrict these requirements as well
as ensure the validity of the tracking section of the analysis code.

\subsubsection{Trigger Cut}

A cut was placed after the trigger was read out and any event not
having a PhysB1 trigger set was cut. A further restriction on which
triggers were not allowed was placed so that any event having any of
the five dimuon triggers consisting of PhysA1, PhysA2, PhysA3,
PhysA4 and Diag3 was cut. This combination is referred to as the
trigger cut. This cut limited all events to those having the
required X134L/R coincidence and having no second track
setting a dimuon trigger.

\subsubsection{One Track Cut}

Once an event had been read in and passed the trigger cut, the code
began constructing all candidate tracks from hit clusters in
Stations 1, 2 and 3 for that event. Once all candidate tracks had
been found for that event the code cut any event having more than
one candidate track. A second cut was placed after the event was
passed from the retracing section of code to insure no event had
more than one retraced muon.

\subsubsection{Fixed Tracking and Retracing Cuts}

Any muon that tracked outside the apertures of any wire plane,
hodoscope layer or the volume in magnet SM3 was cut during the
tracking section of code. During the trace-back section of code
the analysis cut all events having $x$ and $y$ slopes,
\tantx and \nstanty, pointing outside the physical volume inside
magnet SM12. These physical apertures were checked continuously
during the trace-back section of code for each muon.

\subsubsection{Projection Cuts}

To insure that the single track could have actually set the
X134L/R coincidence trigger from the hit clusters used in the 
tracking section of code, the Data Summary Tapes were subsequently
analyzed with a cut imposed on the minimum projected distance in
$x$, referred to as $X_{S1}$, from the $z$ axis in the $x-z$ plane
at $z=770.72$ inches and the maximum distance in $x$, referred to
as $X_{S3}$, at $z=1822.0$ inches. The two $z$ positions
are the two hodoscope banks $X1$ and $X3$ respectively (see figure
\ref{xplane}, \pageref{xplane} for a visual explanation). Hodoscope
plane $X1$ set the minimum opening angle in $x$ that a muon must
have to set the X134L/R coincidence, and hodoscope plane $X3$ set
the maximum opening angle the muon could have to set the same
coincidence. This cut not only insured that the track would set
the trigger, but when combined with the circle cut described in a
later section, they also cut events that had scattered back into the
acceptance from the walls of the spectrometer magnet.

\begin{center}
\begin{figure}[ht]
\resizebox{5.7in}{6.4in}
{\includegraphics[45,140][530,680]{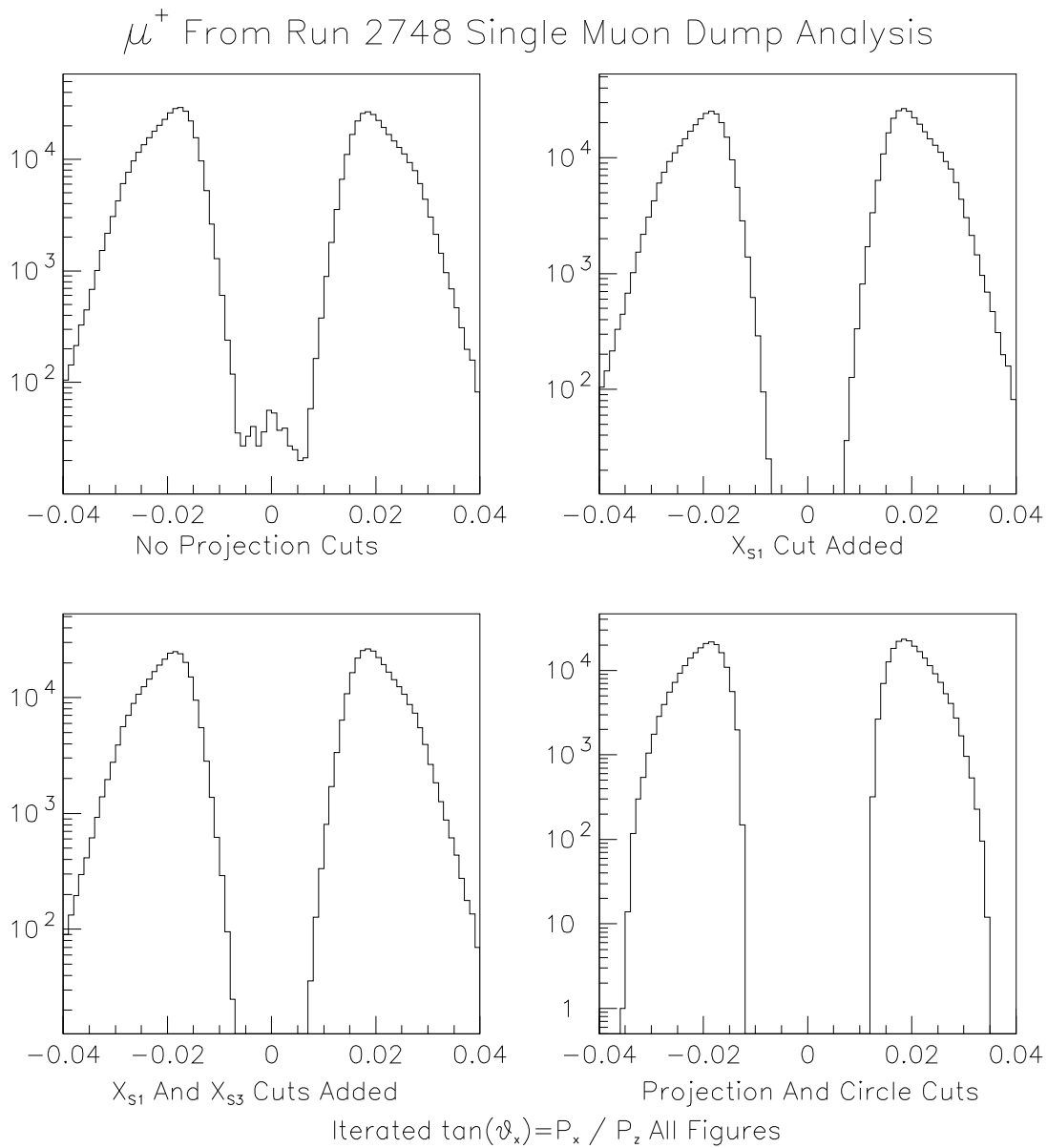}}
\caption[Effects Of Projection Cut.]
{Effects on the slope $tan(\theta_{x})$ with cuts applied on the
minimum distance for $X_{S1}$ (upper right), maximum distance for
$X_{S3}$ (lower left) and both with the circle cut added (lower
right). Upper left is all \muplus for single muon events from the
initial target analysis of Run 2748.}
\label{x-project-cuts}
\end{figure}
\end{center}

Figure \ref{x-project-cuts} shows the effect on
\tantx when these cuts were applied to \muplus from the
initial dump analysis of Run 2748. An event satisfying the X134L/R
trigger cut should, with no multiple scattering, have a minimum
$ | \mbox{\nstantx} | $ greater than 0.017, and a maximum of 0.028.
Data outside of these limits are events that, while traversing the
dump and absorber materials, either scattered into the maximum
acceptance angle $ ( | \mbox{\nstantx} |  \geq 0.028 ) $
before hitting the spectrometer or scattered into the minimum angle
$ ( | \mbox{\tantx} |  \leq 0.017 ) $. Similar effects were seen
from the target.

\afterpage{\clearpage}

\subsubsection{Circle Cut}

\begin{center}
\begin{figure}[th]
\resizebox{5.7in}{3.0in}
{\includegraphics[21,400][573,697]{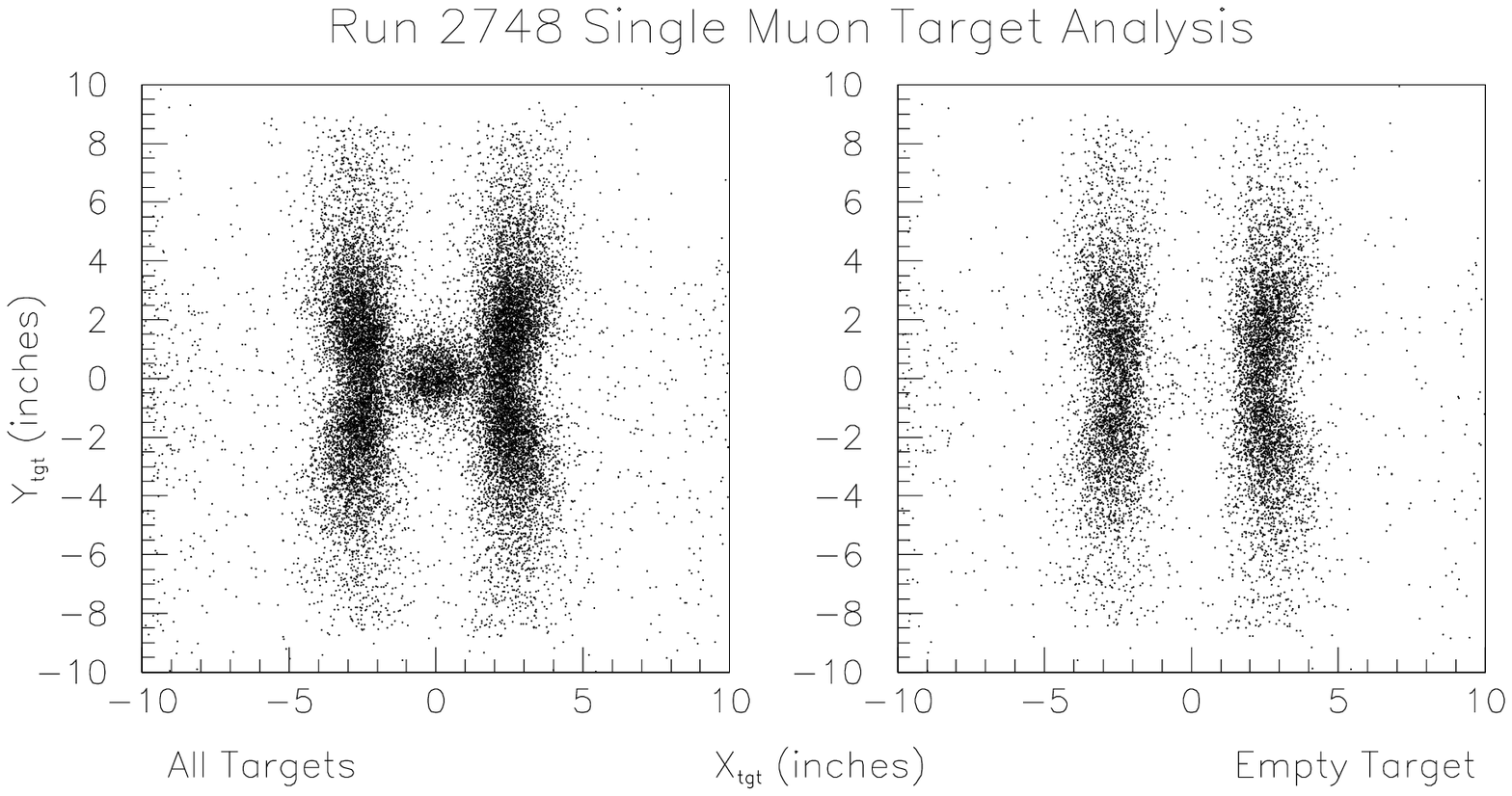}}
\caption[$Y_{tgt}$ vs $X_{tgt}$ For Target Analysis]
{$Y_{tgt}$ vs $X_{tgt}$ for data from Run 2748
analyzed at $Z_{tgt}=-24.0$ inches. Left plot shows data
taken with any target, right is the data with the target position
empty from the same run.}
\label{circle-data}
\end{figure}
\end{center}

A cut on the maximum allowable $ | X_{tgt} |$ and $ |Y_{tgt} |$ was
used to begin the separation of target and dump events in the
initial target and dump reductions. This initial cut was imposed
after inspection showed that events originating from $Z_{tgt}$
primarily retraced (in a target analysis) to a relatively small
distance from the $z$ axis when plotted as $Y_{tgt}$ versus
$X_{tgt}$ at $Z=Z_{tgt}$. Figure
\ref{circle-data} illustrates this effect on the data when the
analysis plane is set to the target location. The left plot is
events when any of the three targets were incident to the beam
during run 2748. The plot is limited to events where both
$X_{tgt}$ and $Y_{tgt}$ are less than 10 inches away from the $z$
axis. The data has the trigger and one track cuts applied. In
contrast, the plot on the right is the same data where the target
position was empty, which means that all of the events originated
from hadrons created in the dump.

The initial reduction cut all events having the magnitudes of
either $Y_{tgt}$ or $X_{tgt}$ greater than 2 inches. The final
reduction used a cut on the maximum distance defined as
$D=\sqrt{X_{tgt}^{2}+Y_{tgt}^{2}}$. Events having $D \geq 1.0$
inches were cut.

\afterpage{\clearpage}

\subsubsection{\tanty and Momentum Cuts}

To limit the open decay distance to one value the single muon
analysis restricted all muons to traverse the entire length of the
copper beam dump. There were two original methods under
consideration for this cut. The first was to cut any events that
exited the dump as they were traced back to $Z_{tgt}$. However, the
acceptance for dump events would have been considerably larger
since, for the most part, all events originating from hadrons
produced in the dump traversed the entire length left over. That was
not the case for target events.

The cut used consisted of determining the maximum and minimum
slopes in the $y-z$ plane combined with a minimum momentum muons
must have and still traverse the length of the dump. The same cuts
would then be applied for muons originating from either the targets
or the dump. Studies of target events using the traced $y$ position
of the muon at $z=86$ inches (referred to as $Y86$) were performed
using a variety of cuts in both
\tanty and momentum, \nspmag. It was decided the
cut should remove the fewest high transverse momentum events
possible and still meet the criteria of forcing all muons from
hadrons produced in the target to retrace the full length of the
dump. The cuts selected for data taken with the spectrometer
magnets having parallel fields were
$| \mbox{\nstanty} | \leq 0.030$ and $\nspmag \geq 55.0$ \nscmom.

\begin{center}
\begin{figure}[ht]
\resizebox{5.6in}{5.6in}
{\includegraphics[21,145][573,696]{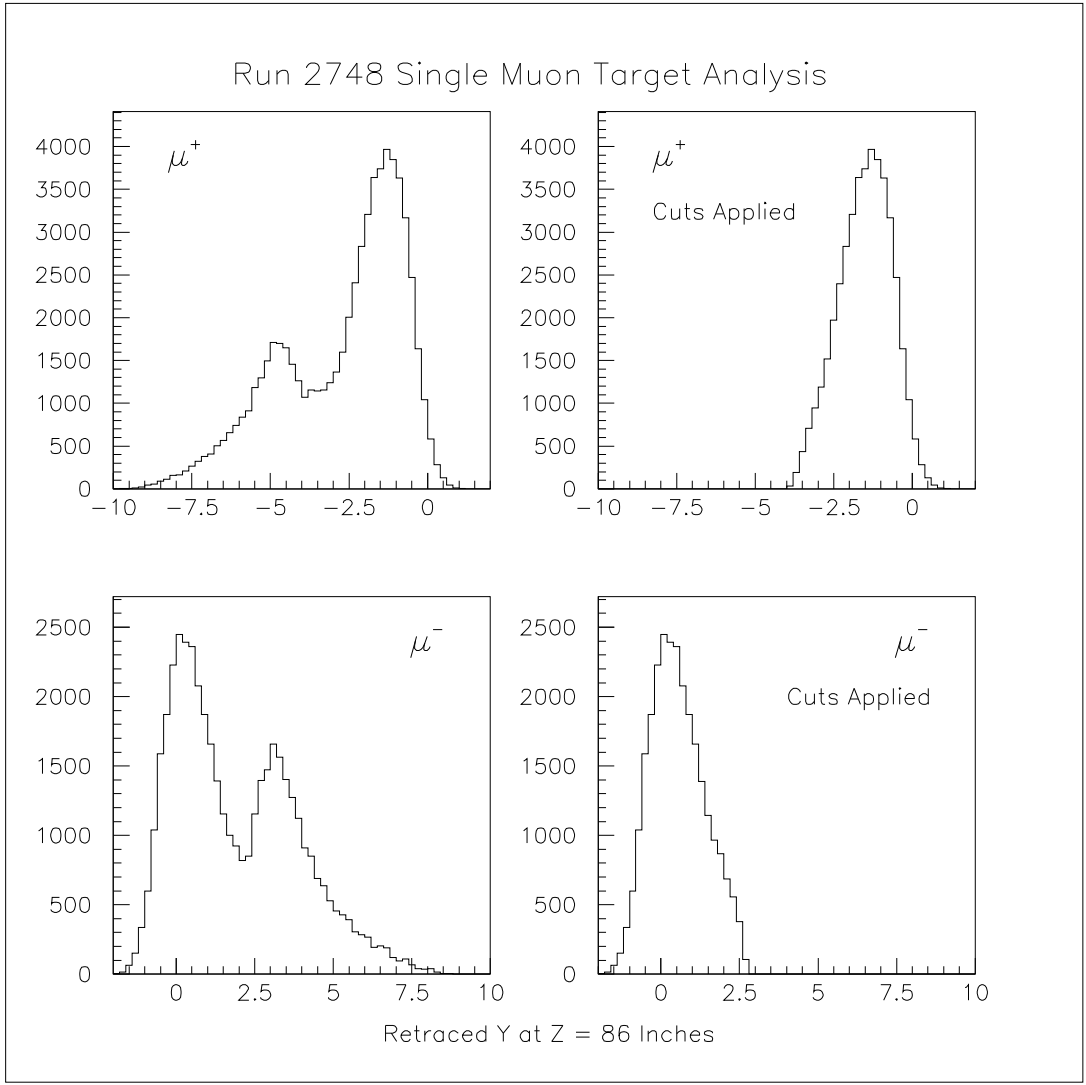}}
\caption[Effects Of The \tanty And Minimum Momentum
Cuts.]{The values of $Y86$ for \muplus (top figures) and
\muminus (bottom figures) from a target analysis of target events
in Run 2748. Left are events with the trigger, one track, aperture,
projection and circle cuts applied, right are the same events
with the \tanty and minimum \pmag cuts added. The two
figures on the left show the effects of the increased decay
length given light hadrons that may pass under or over the copper
beam dump. The decay length for events that must traverse the
entire length of the dump is 92 inches,
while for those passing above or below it is 259
inches. The decay muons from these hadrons are those
in the smaller peaks having larger mean distances from $Y86=0$.}
\label{tan-y-cut}
\end{figure}
\end{center}

Figure \ref{tan-y-cut} shows the effects of these cuts. The top two
plots show the \muplus from a target analysis of Run 2748 having
the trigger, tracking and circle cuts applied on the left, and the
additional \tanty and \pmag cuts applied on the
right. The bottom plots are the same data for the \nsmuminus.

The two figures on the left show the effects of the increased decay
length given light hadrons that may pass under or over the copper
beam dump. The decay length for events that must traverse the
entire length of the dump is 92 inches,
while for those passing above or below it is 259
inches. The decay muons from these hadrons are those
in the smaller peaks having larger mean distances from $Y86=0$.

Dimuon data had suggested that magnet SM12
had dropped approximately 0.40 inches in the $y$ direction. The
bottom of the dump would therefore be at $y=-3.9$ inches, and the
top at $y=3.1$ inches. The application of the \tanty and
\pmag cuts together force all decay muons from hadrons
originating from the target to trace between those values at
$Z=86$ inches.

\afterpage{\clearpage}

\subsection{Single Muon Cuts Summary}

The single muon analysis has initial cuts as well as some
intermediary and final cuts. These are:

\begin{itemize}
\item  Initial Cuts:

\begin{enumerate}
\item  PhysA1 trigger requirement.

\item  No dimuon trigger set.

\item  One candidate track.

\item  One traceable track.

\item  Spectrometer aperture cuts, generally;

\begin{description}
\item $ \nspmag  \geq 30.0 $ \nscmom.

\item $| \mbox{\tantx} | \leq 0.040$.

\item $| \mbox{\tanty} | \leq 0.050$.
\end{description}

\item  Loose radius or circle cut set to:

\begin{description}
\item $D=\sqrt{X_{tgt}^{2}+Y_{tgt}^{2}} \leq 2.0$ inches.
\end{description}
\end{enumerate}

\item  Intermediate Cuts:

\begin{enumerate}
\item  Loose projection cuts set to:

\begin{description}
\item $X_{S1} > 11.0$ inches.

\item $X_{S3} < 26.0$ inches.
\end{description}
\end{enumerate}

\item  Final Cuts:

\begin{enumerate}
\item  Tight radius or circle cut set to:

\begin{description}
\item $D=\sqrt{X_{tgt}^{2}+Y_{tgt}^{2}} \leq 1.0$ inches.
\end{description}

\item  Tight projection cuts set to:

\begin{description}
\item $X_{S1}>12.0$ inches.

\item $X_{S3}<25.5$ inches.
\end{description}

\item  Minimum momentum set to:

\begin{description}
\item $ \nspmag \geq 55.0$ \nscmom.
\end{description}

\item  Maximum $y$ slope set to:

\begin{description}
\item $ | \mbox{\tanty} | \leq 0.030$
\end{description}
\end{enumerate}
\end{itemize}

\section{Data Reductions}
\label{datareductions}

The analysis required four separate initial reductions for each
tape, one each for target and dump dimuon and one each for target
and dump single muon events. The results from each reduction were
stored as Data Summary Tapes, or DSTs. The basic procedure was to
first reduce the data for dimuon events from the targets, then
dimuon events from the dump. These DSTs were then used to
calibrate the spectrometer, after which the data tapes were reduced
to the two initial single muon DSTs.

\subsection{Final Analysis and Presentation of Results}

The final analysis of the data was performed using the CERNLIB
software Physics Analysis Workstation, or PAW. This software is
available from CERN for most platforms \cite{cernlib}. The final
output from the analysis code was in four forms, one a detailed
log file of the analyses performed and relevant information
to the run, a Data Summary Tape for successive analyses, an analysis
generated set of histograms for variables of interest under
several conditional criteria such as cuts and the charge of the muon
using the CERNLIB routine HBOOK, and an n-tuple file containing all
events that are of interest with the desired information for each
of those events as a database.

The information in each n-tuple file was searched for those events
of interest and placed into histograms. The histograms could be
defined for any binning width within reason, but the final
histograms for the transverse momentum had 40
bins from 0 to 10 \cmom resulting in a bin width of 0.25 \nscmom.
Reconstructed masses from dimuon events typically had bin widths of
0.60 \nscmass. The use of PAW and n-tuple files greatly
increased the ability to change search criteria without lengthy
re-running of the analysis code.

The single muon analysis increased the original number of variables
in the n-tuple file for each event to accommodate increased tracking
information as well as uniterated momenta and momenta before
rotation by the beam angle corrections.

\afterpage{\clearpage}

\chapter{Monte Carlo}
\label{mc}

Central to determining the open charm cross sections was the use of
the E866 Monte Carlo to produce expected muon decay spectra from the
hadrons that contribute. The hadrons that contribute significantly
to the single muon spectra are \nsbothpi, \nsbothk,
$ \left( \bothkzero \right) \rightarrow K_{L}$, \bothd and 
\nsbothdzero. The open charm cross section was limited to the
production of \bothd and \bothdzero since the spectrometer could not
differentiate between these open charm mesons
and any other open charm or heavier mesons such as the
$D^{\ast}$ or $D_{S}$ since those decay strongly to one of the four
open charm mesons used.

\begin{figure}[ph]
\resizebox{5.9in}{2.84in}
{\includegraphics[15,400][535,650]{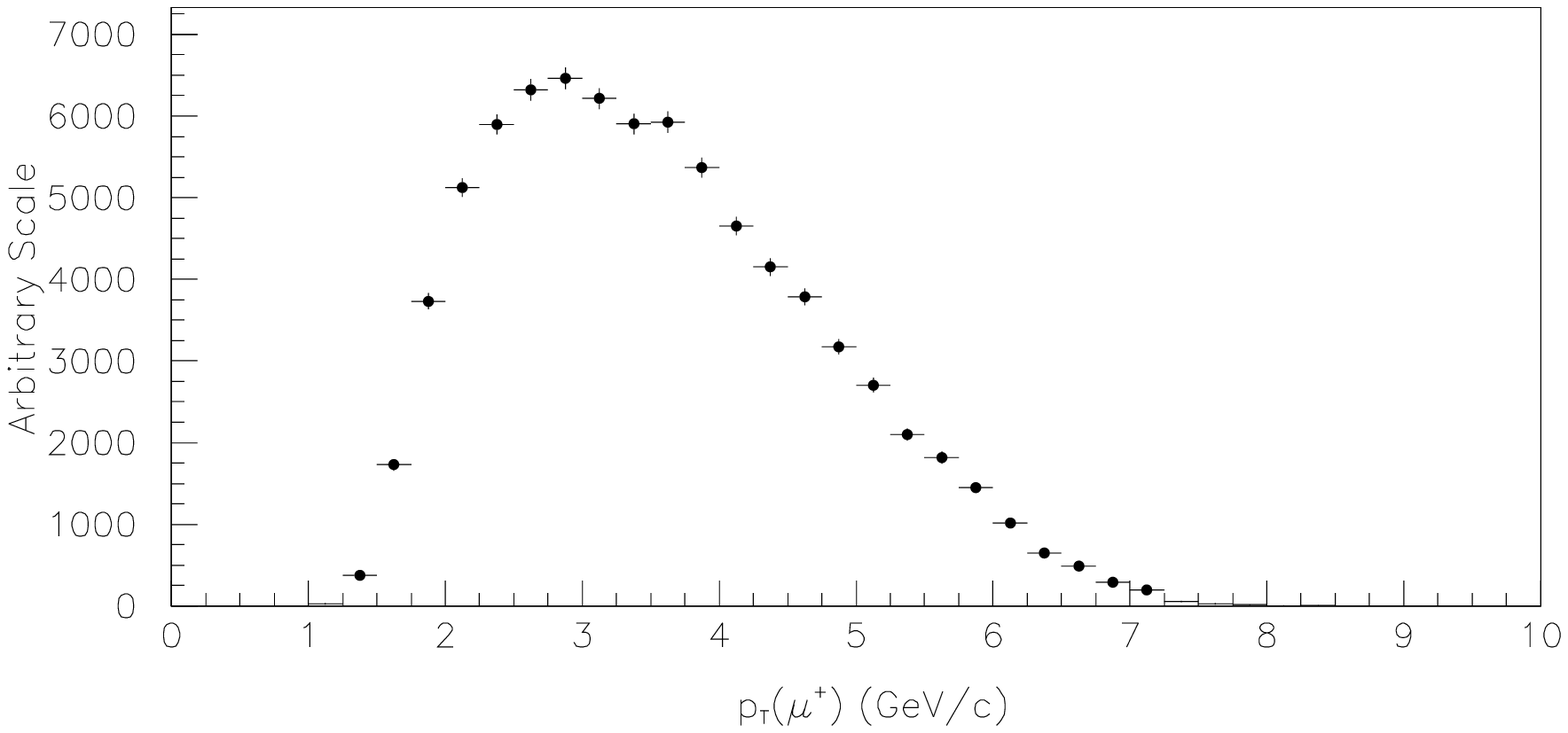}}
\caption[Transverse Momentum Shift]{The muon \pt distribution
from open charm hadrons thrown with random transverse momentum
between 8.00 and 8.25 GeV/c.}
\label{pt-shift}
\end{figure}

Muons resulting from the decay of mesons with large momentum usually
have smaller transverse momentum than the parent. This is due to
the fact that the momentum of the parent is 'shared' between the
decay products, and \pt is a function of the
momentum of the particle,
$\nspt=\nspmag \, \mbox{sin} \left( \theta \right) $. Figure
\ref{pt-shift} shows the resulting \mupt distribution from open charm
hadrons thrown with \hadpt between 8.00 and 8.25 \cmom.

For brevity, and to avoid confusion, the transverse momentum shift
will be referred to as the \pt shift, and the transverse momentum
of hadrons will be denoted \hadpt while the transverse momentum of
muon will be denoted \nsmupt. Also, since there are two charges of
muons, as well as hadron/anti-hadron, the usual reference will be
to hadrons and muons unless the distinction between
hadron/anti-hadron or \muplus and \muminus is required.

The \pt shift imposed the requirement that all hadrons needed to be
thrown to larger \hadpt than the maximum \mupt in the data.
Hadron production falls steeply as a function of \hadpt (see
Equation \ref{bscfunc} (page \pageref{bscfunc}) and Figures
\ref{pi-plus-minus} and \ref{k-plus-minus} on pages
\pageref{pi-plus-minus} and \pageref{k-plus-minus}.) The value of
the differential cross section, which falls exponentially or
nearly so as a function of \nspt, is a measure of the number of
hadrons that will be produced. Worse, the likelihood that a hadron
would decay also falls exponentially as a function of the momentum
of the hadron. The minimum momentum of the muons from the data
after final analysis was 55 \nscmom, and since the momentum is shared
between the decay products, the minimum momentum for hadrons would
necessarily have to be greater than the minimum for the muons. The
decision was made to throw the hadrons flat in transverse momentum,
center of mass rapidity $y$ $ \left(
y = \frac{1}{2} \, log \,
\left[ \frac{E/c + \nszmag}{E/c - \nszmag} \right] \right) $ and
$\phi$, and then use weights to shape the resulting spectra. Use
of the term flat means that a random value was thrown between a
lower and upper limit, so the distribution was 'flat' when it
was histogrammed.

The E866 Monte Carlo had two distinct parts in its original form,
an event generator section and a decay muon tracing section
for dimuon events. Usual practice was to use an outside source of
dimuon events generated in other custom event generators using
hadron production based on programs such as PYTHIA, where the
resulting dimuon kinematics were read directly into the tracing
section of the Monte Carlo. The muon tracing section of the Monte
Carlo was used to simulate the E866 spectrometer and produce
simulated tracking information from the four tracking stations as
well as trigger information from the hodoscopes. The output
trigger and track information was placed into an E866 data
format file for later analysis. Other useful information, such
as the weights used in this analysis, could be passed to the
analysis routine and read out for each event. The information could
then be used directly in the analysis and/or written to the
n-tuples if desired.

\begin{figure}[ph]
\resizebox{5.5in}{6.2in}
{\includegraphics[0,69][612,748]{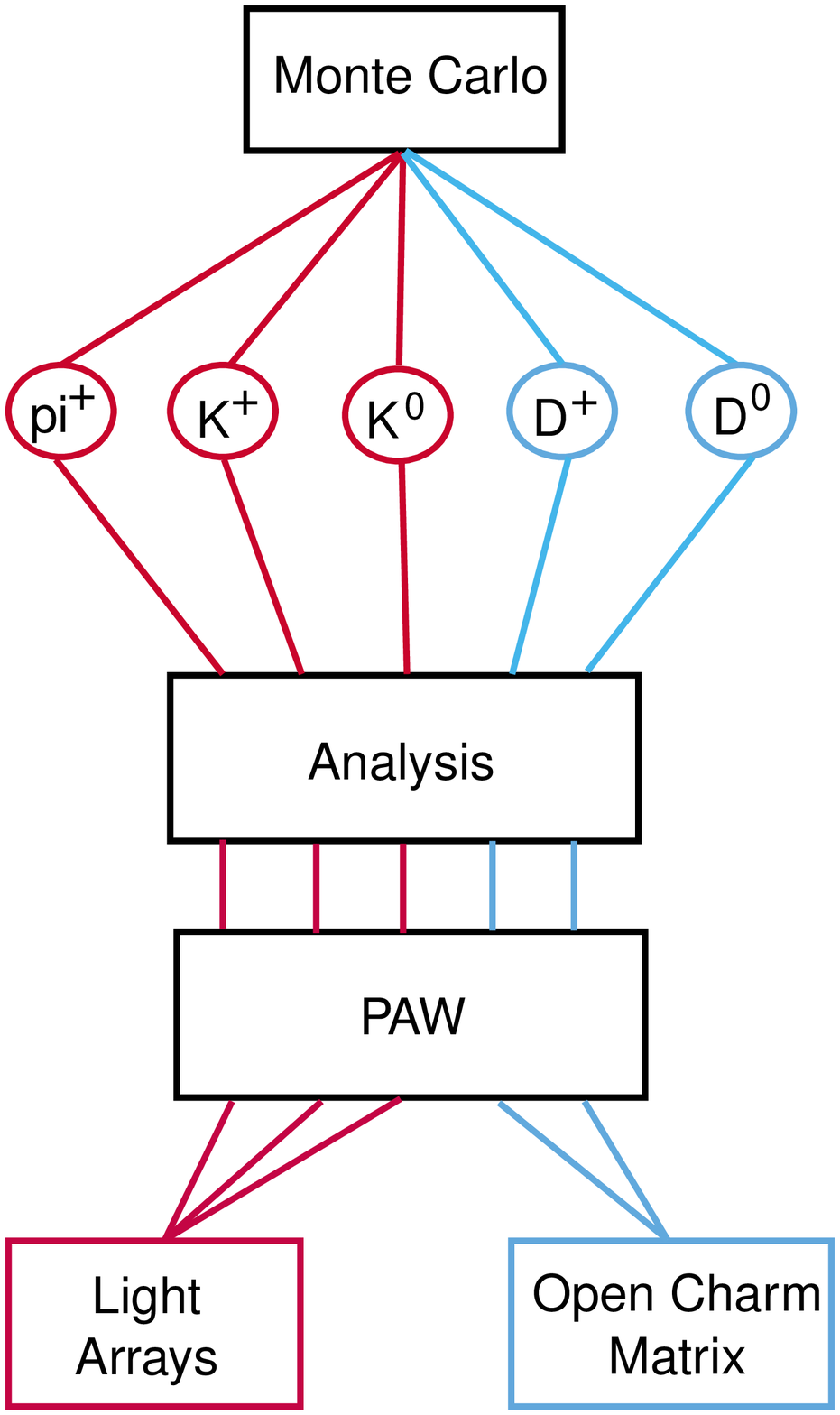}}
\caption[Monte Carlo Process]{Basic outline of the Monte Carlo
process to calculate the various contributions to the single
muon spectra.}
\label{monte-to-array}
\end{figure}

Several major modifications were made to the Monte Carlo to
simulate single muon decays from the hadrons of interest. The
event generator was modified to throw the desired hadrons using a
distribution flat in \nshadpt, center of mass rapidity, $y$, and
$\phi$. Once the hadron was thrown the event generator was required
to determine a decay point based on the decay properties of the
hadron and its momentum. Traversal of target or dump
materials and magnetic field effects of the spectrometer magnet
SM12 were simulated in a new section of the program for all
charged hadrons. Hadron decay was performed using a CERN routine
named GENBOD, and all decay products were transformed to
the lab frame using the CERN routine LORENB. Secondary hadron decays,
when required, were introduced by repeating the hadron tracing and
decay sections for each secondary hadron of interest resulting from
the primary hadron decay. These modifications allowed for large
numbers of events at high transverse momentum since all hadrons
that made it into the spectrometer were allowed to decay, and all
hadrons were thrown flat in \nshadpt. Figure \ref{monte-to-array}
shows a representation of the single muon Monte Carlo process.

Various cuts on primary hadrons, secondary hadrons and decay muons
were implemented throughout the event generator and muon tracing
sections of the Monte Carlo. The first set of cuts were loose
aperture and momentum cuts taken on the hadrons immediately after
they were thrown and histogrammed. This increased the efficiency of
the Monte Carlo since similar or harder cuts on the thrown
hadron were done later while the hadron was
being traced through the spectrometer magnet. All decay muons or
secondary hadrons having $\nspmag \le 50$ \cmom were cut
immediately after the decay and boost routines were done. All
hadrons that were allowed to contribute secondary muons that
resulted in more than one muon traversing the spectrometer were
cut. A cut was imposed on all muons that tracked through any of the
pulled x hodoscopes.

After a muon had passed all cuts the parent hadron and decay muon
information required for producing weighted spectra was written to
the MC data file.

\afterpage{\clearpage}

\section{Weighted Spectra}
\label{lightspectra}

Extraction of the open charm cross sections required a set of
\mupt spectra from all hadrons contributing to the data. These
spectra were required to be the muon spectra that would result
for a given number of protons interacting in the various materials
of the targets and copper dump. A set of weights were given each
muon as it was written to disk. These weights are shown in
Table \ref{weights-1}. The total weight of a muon is the product
of the individual weights (listed in Table \ref{weights-1})
appropriate for the event.

\afterpage{\clearpage}

\renewcommand{\arraystretch}{1.5}

\begin{table}[!ht]
\caption[Weights Applied To Events In The Monte Carlo]
{Weights calculated for muons in the Monte Carlo spectra. Hadrons
are divided into groups, light ($L$), open charm ($H$) and
both ($B$). $ E \, \frac{d^{3} \, \sigma}
{d \, p^{3}} \; (mb\,c^{3}\,GeV^{-2}) $ is the \ppp differential
cross section at the thrown \hadpt and c.m. $y$ of the hadron,
$A^{\alpha{\left(\nshadpt\right)}}$ is the nuclear dependency of
the hadron, $P_{D}$ is the probability that the hadron has decayed
in the spectrometer,$P_{NA}$ is the likelihood that a hadron
produced in the target did not interact in any remaining material in
the target and $br(mode)$ is the muonic branching fraction of the
decay mode.}
\label{weights-1}
\begin{center}
\begin{tabular}{|c|c|c|}

\hline

\multicolumn{2}{|c|}{Weight} &
 \multicolumn{1}{l|}{Hadron}\\
\hline
\hline

$W_{x}$ &
 $E \, \frac{d^{3} \, \sigma}{d \, p^{3}} \; (mb\,c^{3}\,GeV^{-2})$ &
  $L$\\
 
$W_{A}$ &
 $ A^{\alpha \left( \nshadpt \right)} $ &
  $L$\\

$W_{D}$ &
 $P_{D}$ &
  $B$\\

$W_{NA}$ &
 $P_{NA}$ &
  $B$\\

$W_{br}$ &
 $ br \left( mode \right) $ &
  $B$\\

\hline

\end{tabular}
\end{center}
\end{table}

\renewcommand{\arraystretch}{1.0}

\subsection{The Light Hadron \ppp Cross Section Weight}
\label{alperxsect}

All hadron distributions were parameterized
using \ppp differential cross sections, and the following
experimentally determined relationship was used to convert those
cross sections to \pA cross sections:
\begin{equation}
 E \frac{d^{3}\sigma^{pA}(h)}{dp^{3}} =
 A^{\alpha(h,\nshadpt)} \;
 E \,\frac{d^{3}\sigma^{\ppp}(h)}{dp^{3}} \qquad
 \mbox{\nsucrosswpar}
\label{hadxsects}
\end{equation} where $h$ is the hadron of interest and
$A^{\alpha(h,\nshadpt)}$ is referred to as the nuclear dependency
term.

A parameterization of the \ppp differential \bothpi and \bothk
cross sections as a function of transverse momentum, \nshadpt, and
c.m. rapidity $y$, developed by the British-Scandinavian
Collaboration (BSC) at CERN \cite{nucl-phys-b-100-237} was used to
calculate the cross section weight, $W_{x}$ \nsucrosswpar. The
parameterization of the differential cross sections developed by
the BSC is
\begin{equation}
E \, \frac{ d^{3} \, \sigma^{\nsppp} (h)}{dp^{3}}(\nshadpt,y) =
 A_{1} \, e^{ \left( -B \, \nshadpt \right) } \,
 e^{ \left( -Dy^{2} \right) } +
 A_{2} \frac{ \left( 1 - \frac{\nshadpt}{p^{cm}_{beam}}
\right) ^{m}} { \left( \nshadpt^{2} + M^{2} \right)^{n}}
\label{bscfunc}
\end{equation} where values for the
parameters $A_{1}$, $B$, $D$, $A_{2}$, $M$, $m$ and $n$
suggested for use by the BSC, their best central value and
undertainty are
listed in Table
\ref{xsectparams}. It should be noted that the exponent $n$ was
held fixed at 4 for all fits, and the collaboration suggested no
deviation for that parameter. The center-of-mass (c.m.)
momentum of the proton,
$p^{cm}_{beam}$, introduces the energy dependency of hadron
production and for this experiment was set to
$p^{cm}_{beam}=19.37$ \nscmom.

It was assumed that
$\sigma(\nskzero)+\sigma(\nsakzero)=
 \sigma(\nsplusk)+\sigma(\nsminusk)$
and that a similar production relationship existed
between the \kzero and \akzero as between the \plusk and \minusk
due to conservation of isospin. This was introduced into the
Monte Carlo by using $\sigma(\nskzero)=\sigma(\nsplusk)$ and
$\sigma(\nsakzero)=\sigma(\nsminusk)$. The Monte Carlo did not
throw \bothkzero but instead threw $K_{L}$ where the parent was
either a \kzero (50\%) or \akzero (50\%).

\renewcommand{\arraystretch}{1.5}

\begin{table}[ht]
\caption[Parameter Values Used In Calculating $W_{x}$.]
{Values of the parameters used in the BSC parameterization
of the \bothpi and \bothk differential cross sections. Fits were
to data with \pt between 0 and approximately 6 \nscmom. Taken from
 \cite{nucl-phys-b-100-237}.}
\label{xsectparams}
\begin{center}
\begin{tabular}[c]{|lcccccc|}
\hline

  &
 $A_{1}$ &
 $B$ &
 $D$ &
 $A_{2}$ &
 $M$ &
 $m$\\

\hline

\pluspi &
 $210 \pm 4$ &
 $7.58 \pm 0.11$ &
 $0.20 \pm 0.01$ &
 $10.7 \pm 0.7$ &
 $1.03 \pm 0.03$ &
 $10.9 \pm 0.4$\\

\minuspi &
 $205 \pm4$ &
 $7.44 \pm 0.09$ &
 $0.21 \pm 0.01$ &
 $12.8 \pm 0.9$ &
 $1.08 \pm 0.02$ &
 $13.1 \pm 0.4$\\

\plusk &
 $14.3 \pm 0.4$ &
 $6.78 \pm 0.21$ &
 $1.5 \pm 0.1$ &
 $8.0 \pm 1.1$ &
 $1.29 \pm 0.03$ &
 $12.1 \pm 0.8$\\

\minusk &
 $13.4 \pm 1.0$ &
 $6.51 \pm 0.23$ &
 $1.8 \pm 0.1$ &
 $9.8 \pm 1.7$ &
 $1.39 \pm 0.04$ &
 $17.4 \pm 1.0$\\

\hline
\end{tabular}
\end{center}
\end{table}

\renewcommand{\arraystretch}{1.0}

\begin{figure}[ht]
\resizebox{5.8in}{5.9in}
{\includegraphics[67,221][492,653]{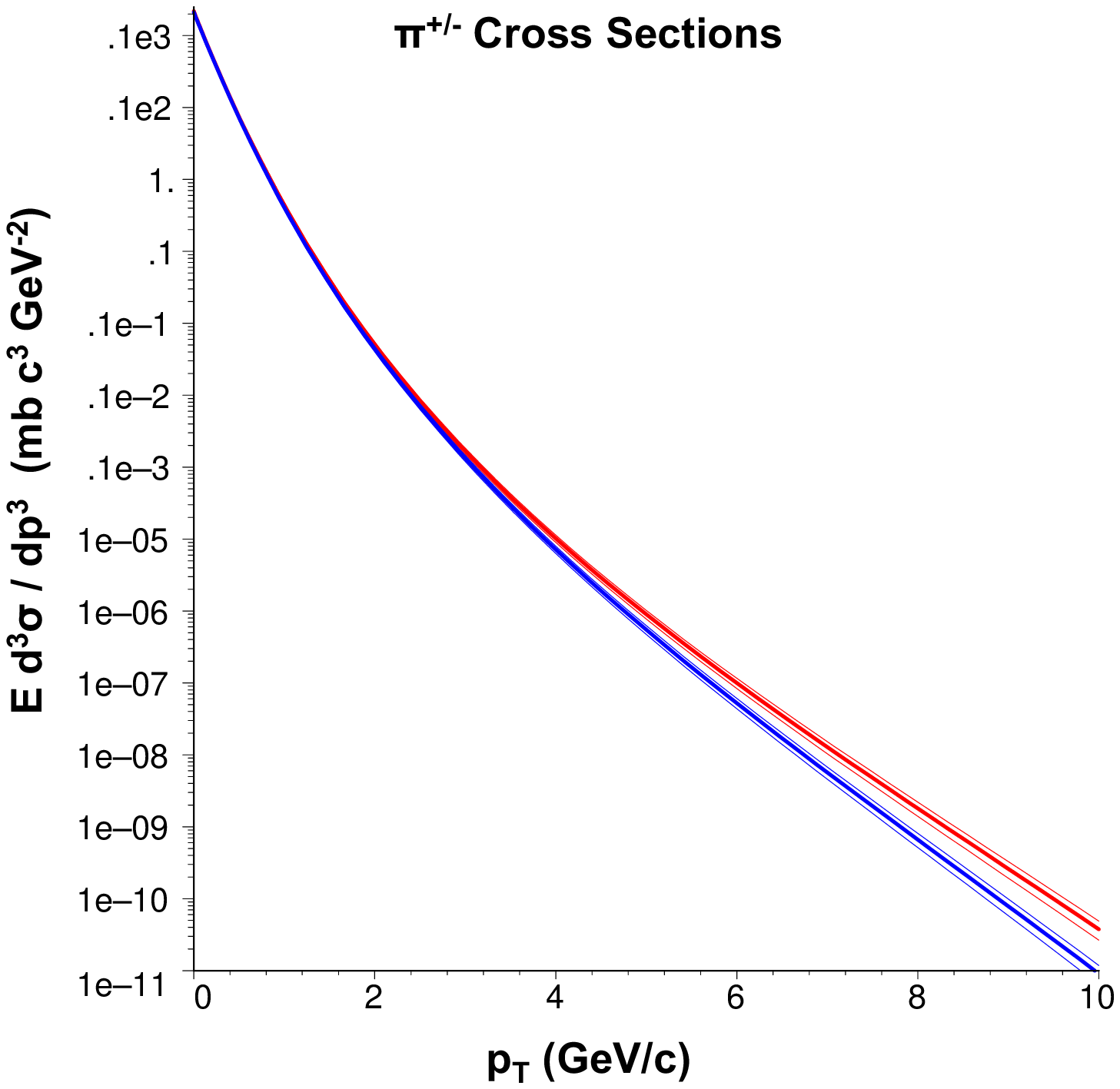}}
\caption[\bothpi Differential Cross Section.]{The \pluspi (red)
and \minuspi (blue) differential
cross section as used in the Monte Carlo. Distribution is taken at
c.m. $y=0$ and $p^{cm}_{beam}=19.37$ \nscmom. The value of the
differential cross section at the thrown \hadpt and $y$ was
calculated and passed to the output MC data file and later used as
a weight ($W_{x}$) for the event.}
\label{pi-plus-minus}
\end{figure}

\begin{figure}[ht]
\resizebox{5.8in}{5.9in}
{\includegraphics[63,219][488,651]{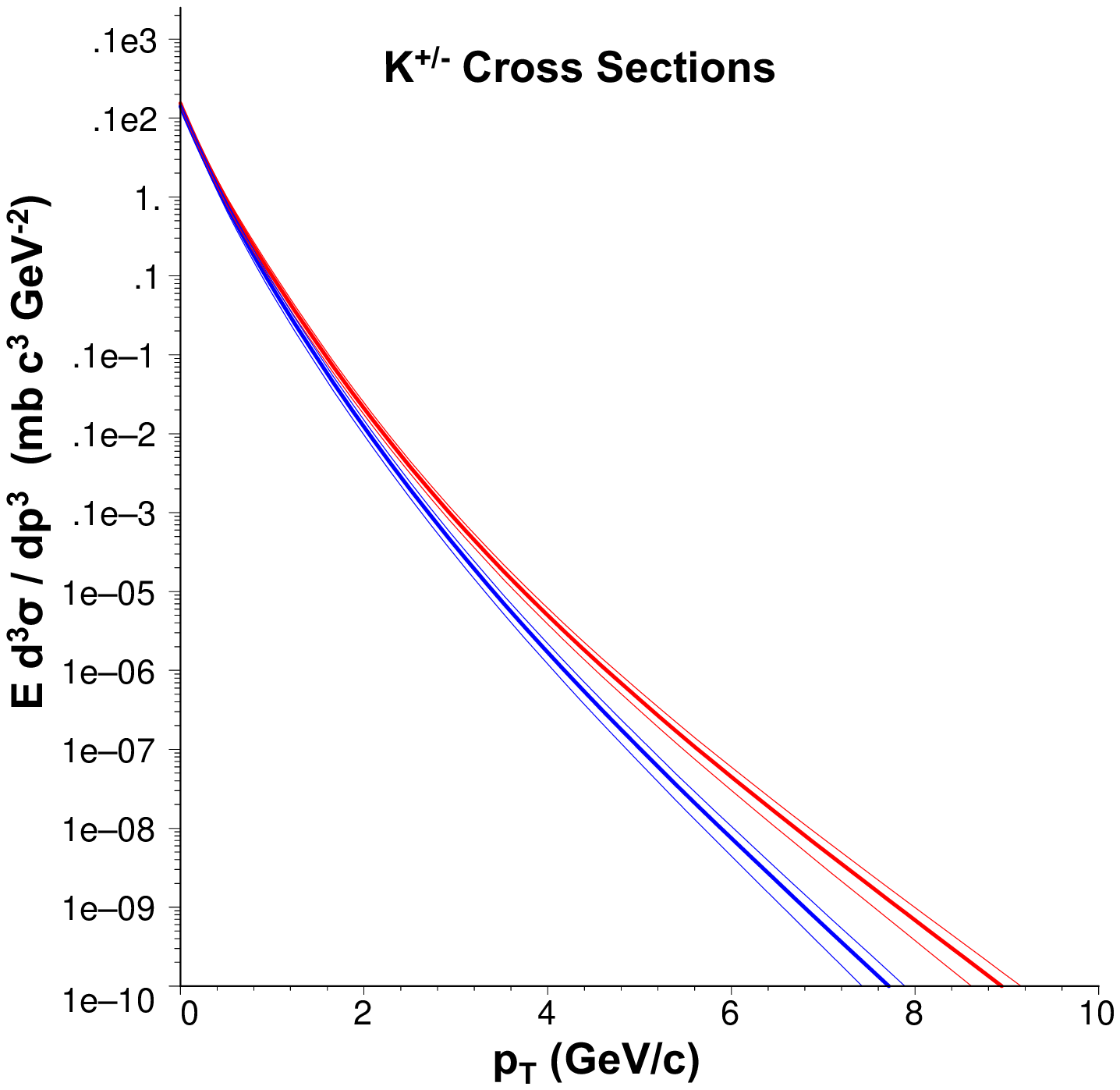}}
\caption[\bothk Differential Cross Section.]{The \plusk (red)
and \minusk (blue)
differential cross sections as used in the Monte Carlo. Distribution
is taken at c.m. $y=0$ and $p^{cm}_{beam}=19.37$ \nscmom. The
value of the
differential cross section for the thrown \hadpt and $y$
was calculated for each event and passed to the output MC
data file and later used as a weight ($W_{x}$) for the event.}
\label{k-plus-minus}
\end{figure}

Figures \ref{pi-plus-minus} and \ref{k-plus-minus} show the \bothpi
and \bothk cross sections for the parameterizations in
\cite{nucl-phys-b-100-237} at $y=0$ (c.m.) for the energy of this
experiment, $\sqrt{s}=38.8$ GeV.

\afterpage{\clearpage}

\subsection{Nuclear Dependency Term}
\label{atoalpha}

It is well known that the proton-nucleus (\nspA) cross section,
at least for light hadrons, does not scale to the proton-proton
(\nsppp) cross section by the number of nucleons, $A$, in the target
nucleus. The Chicago-Princeton Collaboration (C-P) studied the
production of light hadrons on various nuclear targets and developed
what has become known as the nuclear dependency term, relating
the \pA to \ppp cross sections by the relation \cite{prd-11-3105}:
$$ \crosspa = A^{\alpha(h,\nshadpt)} \, \crosspp $$ where $A$ is
the nuclear number (referred to as the atomic weight in
\cite{prd-11-3105} and $h$ is the species of hadron.

Light hadrons thrown in the Monte Carlo had the nuclear dependency
calculated as a weight, $W_{A}$, that was
passed to the Monte Carlo data files for use in weighting the
thrown spectra after analysis. Since the open charm cross sections
were determined as the \pcu and \pbe cross sections, no nuclear
dependency for the production of open charm was used.

\begin{figure}[th]
\resizebox{5.6in}{5.6in}
{\includegraphics[26,158][562,688]{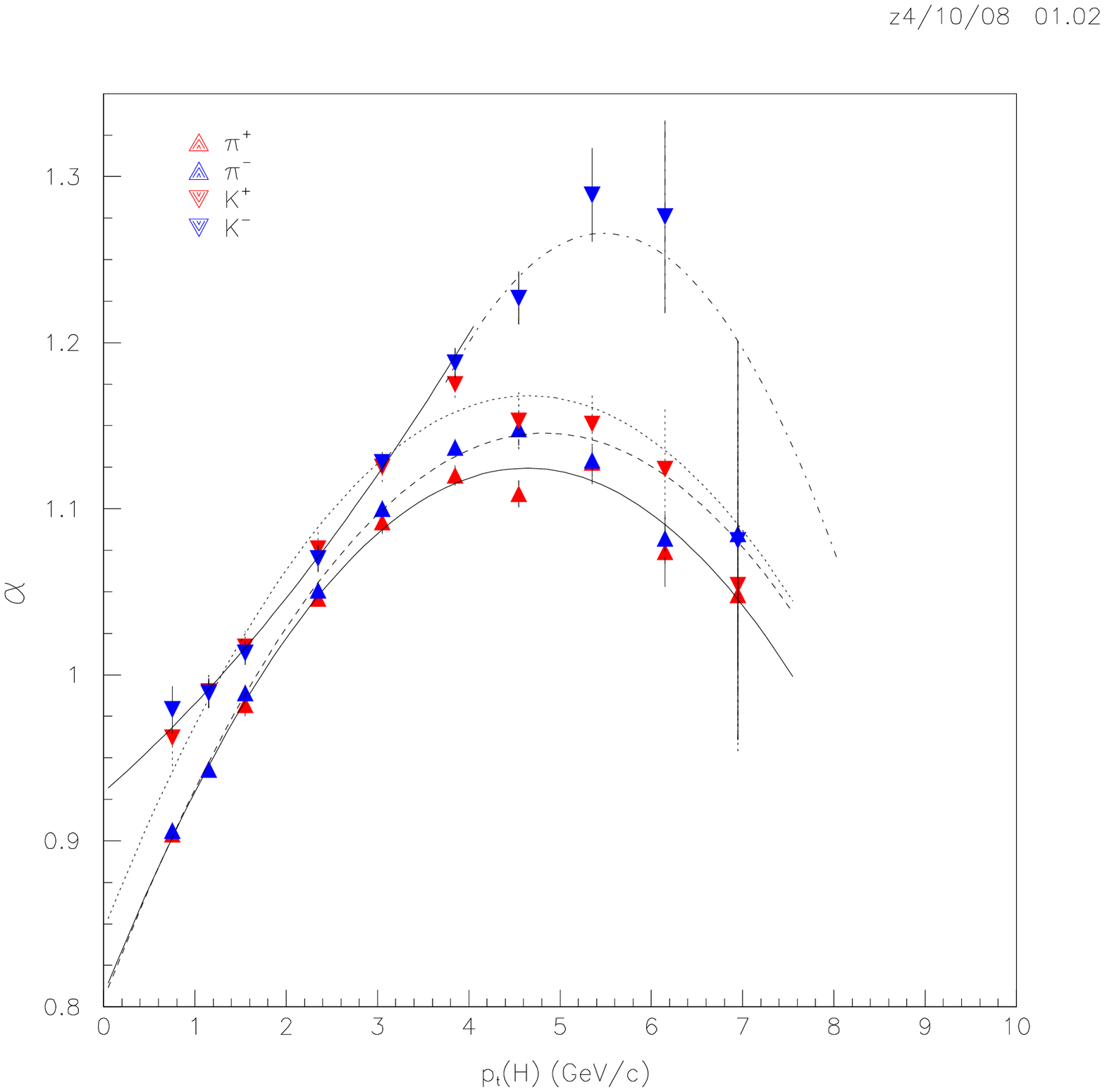}}
\caption[Fits To $A^{\alpha(h,\nshadpt)}$.]{Fits of
quadratic equations to the atomic-weight dependency for light
hadrons. Data points taken from \cite{prd-19-764}.}
\label{alphas}
\end{figure}

Data points from the Chicago-Princeton Collaboration (C-P) study
for the nuclear number (or equivalently the atomic-weight)
dependency for charged pion and kaon production from 400
\cmom \pA collisions \cite{prd-19-764} were fitted to simple
quadratic equations of the form
$p_{1} + p_{2} \, \nspt + p_{3} \, \nspt^{2}$ using the CERN
program MINUIT. The resulting parameterizations were used to
calculate the value of the dependency for each light hadron thrown.
Figure \ref{alphas}
shows the four fits to the C-P data, and Table \ref{alphavals}
gives the parameter values of the fits. The \minusk
parameterization was fitted to two regions in \nshadpt.

\renewcommand{\arraystretch}{1.5}

\begin{table}[ht]
\caption[Parameter Values Used To Calculate $W_{A}$.]
{Values for parameters of the equation
$\alpha(h,\nshadpt) = p_{1} + p_{2} \, \nspt + p_{3} \,
\nspt^{2}$. The \kzero used the same parameter values
as the \plusk and the \akzero used the same values as
the \nsminusk. The \minusk fit required using two regions; hadrons
thrown with transverse momentum less than 4.0 \nscmom,
and hadrons having transverse momentum greater than 4.0 \nscmom.}
\label{alphavals}
\begin{center}
\begin{tabular}[c]{|lccc|}
\hline

  &
 $p_{1}$ &
 $p_{2}$ &
 $p_{3}$\\

\hline

\nspluspi &
 0.8074 $\pm$ 0.0072 &
 0.1368 $\pm$ 0.0060 &
 -0.0148 $\pm$ 0.0010\\

\nsminuspi &
 0.8043 $\pm$ 0.0059 &
 0.1412 $\pm$ 0.0040 &
 -0.0146 $\pm$ 0.0005\\

\nsplusk &
 0.8464 $\pm$ 0.0154 &
 0.1380 $\pm$ 0.0113 &
 -0.01481 $\pm$ 0.0019\\

\nsminusk &
 0.9295 $\pm$ 0.0148 &
 0.0477 $\pm$ 0.0121 &
 0.0053 $\pm$ 0.0006\\

  &
 0.3649 $\pm$ 0.0654 &
 0.3289 $\pm$ 0.0266 &
 -0.0300 $\pm$ 0.0032\\

\hline
\end{tabular}
\end{center}
\end{table}

\renewcommand{\arraystretch}{1.0}

\afterpage{\clearpage}

\subsection{Probability To Decay}
\label{decayprob}

All hadrons thrown were given a weight, $W_{D}$, based on the
likelihood that the hadron would decay, $L_{D}$, in the
spectrometer. Since all hadrons must traverse some material, the
loss of hadrons due to catastrophic collisions was included in
the decay likelihood.

The probability for a hadron with momentum \pmag to decay in a
distance $z$ is given by
\begin{equation}
P_{D} = 1 - exp \, \left( -z \, \frac{m}
{p \, \tau} \right)
\label{pdeco}
\end{equation} where $z$ is the distance the hadron has travelled
(cm), $m$ is the mass of the
hadron (GeV/c$^{2}$), $p$ is the momentum (\nscmom) and $\tau$ is
the proper lifetime (sec). The
probability for a hadron to interact after traversing a distance $z$
in a material is calculated from
\begin{equation}
P_{I} = 1 - exp \, \left( -z \frac{\rho}{\lambda_{I}} \right)
\label{pint}
\end{equation} where
$\rho$ is the density of the material being traversed (\nsdensity)
and $\lambda_{I}$ (\nsilength) is the nuclear interaction length of
the material for the hadron, which is slightly dependent on the
hadron momentum. The two processes are independent of each other,
so the likelihood that a hadron has neither interacted nor decayed
in a length $z$ and will decay in the next differential thickness
$dz$ is given by
\begin{equation}
\frac{dP_{D}}{dz} = \frac{m}{p \, \tau} \,
 exp \left[ -z \left( \frac{m}{p \, \tau} +
 \frac{\rho}{\lambda_{I}} \right) \right]
\label{dpdec}
\end{equation} 

The fraction of hadrons that will decay while traversing a material
is found by integrating Equation \ref{dpdec} and normalizing the
result such that the likelihood for a hadron to either decay or
interact while traversing a target of infinite thickness is unity.
The fraction of hadrons that will decay while traversing a material
of thickness $z$ is then given by
\begin{equation}
F_{D} = \frac{m \, \lambda_{I}}
 {m \, \lambda_{I} + \rho \, p \, \tau} \,
 exp \left[ -z \left(
 \frac{m}{p \, \tau} + \frac{\rho}{\lambda_{I}} \right) \right]
\label{pdeca}
\end{equation}
while the fraction that will interact, $F_{I}$, is given by
\begin{equation}
F_{I} = \frac{\rho \, p \, \tau}
 {m \, \lambda_{I} + \rho \, p \, \tau} \,
 exp \left[ -z \left(
 \frac{m}{p \, \tau} + \frac{\rho}{\lambda_{I}} \right) \right]
\label{pinta}
\end{equation} The fraction of hadrons not interacting while
traversing a material, $F_{NI} = 1 - F_{I}$ was more useful in
calculating the likelihood for decay in the spectrometer.

The assumption was made that all hadrons traversing the dump would
either interact or decay, so the length of the dump was set to
infinity in Equation \ref{pdeca} for simplicity. This assumption
reduced Equation \ref{pdeca} to
$$F_{D} = \frac{m \, \lambda_{I}}
 {m \, \lambda_{I} + \rho \, p \, \tau}$$. It was also
assumed that all open charm hadrons thrown from the target would
decay if they didn't interact while traversing any remaining
material in the target.

The likelihood that the hadron thrown would decay in the spectrometer
was then determined as follows:
\begin{itemize}

\item If a light hadron was thrown in the target, the likelihood for
  decay was calculated using
$$ W_{D} = F_{NI} \, P_{D} + \left( 1 - F_{NI} \, P_{D} \right) \,
F_{D} $$

\item If an open charm hadron was thrown in the target, then
$$ W_{D} = F_{NI}$$

\item All hadrons thrown in the dump used
$$ W_{D} = F_{D} $$

\end{itemize}

In all cases, the value of $F_{NI}$ used the difference between
the thickness of the target and the $z$ location where the hadron
was produced (Section \ref{distribs} below), and the distance
used in calculating $P_{D}$ was set to 92 inches (233.68 cm), the
distance between the targets and the beam dump. 

The nuclear interaction length, $\lambda_{I}$, for light hadrons
was calculated using the results published by A. S. Carrol,
\textit{et al.} \cite{phys-lett-80b-319}. The code initialized
three interaction lengths associated with three momentum ranges of
the hadron, $0 < \pt \leq 130$, $130 < \pt \leq 240$ and
$\pt > 240$ \nscmom. The nuclear interaction length, calculated
using the absorption cross sections determined by A. S. Carroll,
\textit{et al.}, are given in Table \ref{nucil}. The nuclear
interaction lengths for open charm was calculated by decreasing
the absorption cross section for strange particles by
10 percent.

\renewcommand{\arraystretch}{1.25}

\begin{table}[ht]
\caption[$\sigma_{a}$ And $\lambda_{I}$]{Absorption cross section,
$\sigma_{a}$, and nuclear interaction length, $\lambda_{I}$, for
hadrons thrown in the Monte Carlo. All hadron momenta are in
\nscmom. Nuclear interaction length for neutral kaons and
\bothdzero are the same as their charged counterparts. The density
for beryllium used was 1.848 \nsdensity, and for copper
8.96 \density. The nuclear interaction lengths for protons was taken
directly from the tables provided by the Particle Data Group.}
\label{nucil}
\begin{center}
\begin{tabular}[c]{|lc|cc|cc|}
\hline

$H$ &
 \pt &
 \multicolumn{2}{|c|}{Copper} &
 \multicolumn{2}{c|}{Beryllium}\\

  & Range &
 $\sigma_{a}$ (mb) &
 $\lambda_{I}$ \ilength &
 $\sigma_{a}$ (mb) &
 $\lambda_{I}$ \ilength\\

\hline

\pluspi &
 $0\,-\,130$ &
 $611.6 \pm 34.8$ &
 $172.5 \pm 9.8$ &
 $137.5 \pm 6.2$ &
 $108.8 \pm 4.9$\\

  & $130 \, - \, 240$ &
 $616.6 \pm 35.2$ &
 $171.0 \pm 9.8$ &
 $139.3 \pm 6.2$ &
 $107.4 \pm 4.8$\\

  & $> \, 240$ &
 $608.6 \pm 37.0$ &
 $173.4 \pm 10.5$ &
 $140.4 \pm 6.6$ &
 $106.6 \pm 5.0$\\

\hline

\minuspi &
  &
 $611.8 \pm 34.7$ &
 $172.5 \pm 9.8$ &
 $138.1 \pm 6.1$ &
 $108.4 \pm 4.8$\\

  &
  &
 $611.4 \pm 34.7$ &
 $172.6 \pm 9.8$ &
 $138.8 \pm 6.2$ &
 $107.8 \pm 4.8$\\

  &
  &
 $610.6 \pm 36.6$ &
 $172.8 \pm 10.4$ &
 $139.7 \pm 6.4$ &
 $107.1 \pm 4.9$\\

\hline

\plusk &
  & $543.5 \pm 31.1$ &
 $194.1 \pm 11.1$ &
 $115.9 \pm 5.2$ &
 $129.1 \pm 5.8$\\

  &
  &
 $562.7 \pm 32.3$ &
 $187.5 \pm 10.8$ &
 $122.2 \pm 5.5$ &
 $122.5 \pm 5.6$\\

  &
  &
 $555.8 \pm 36.8$ &
 $189.9 \pm 12.6$ &
 $122.3 \pm 6.2$ &
 $122.3 \pm 6.2$\\

\hline

\minusk &
  &
 $572.7 \pm 33.0$ &
 $184.2 \pm 10.6$ &
 $125.1 \pm 5.7$ &
 $119.7 \pm 5.5$\\

  &
  &
 $571.3 \pm 32.8$ &
 $184.7 \pm 10.6$ &
 $125.2 \pm 5.7$ &
 $119.5 \pm 5.4$\\

  &
  &
 $570.0 \pm 35.6$ &
 $185.1 \pm 11.6$ &
 $126.4 \pm 6.2$ &
 $118.4 \pm 5.8$\\

\hline

\plusd &
  &
  &
 $215.7 \pm12.3$ &
  &
 $143.4 \pm6.5$\\

  &
  &
  &
 $208.4 \pm12.0$ &
  &
 $136.1 \pm6.2$\\

  &
  &
  &
 $211.0 \pm14.0$ &
  &
 $135.9 \pm6.9$\\

\hline

\minusd &
  &
  &
 $204.7 \pm11.8$ &
  &
 $133.0 \pm6.1$\\

  &
  &
  &
 $205.2 \pm11.8$ &
  &
 $132.8 \pm6.0$\\

  &
  &
  &
 $205.7 \pm12.9$ &
  &
 $131.5 \pm6.4$\\

\hline

$p$ &
  &
  &
 $134.9$ &
  &
 $75.2$\\

\hline
\end{tabular}
\end{center}
\end{table}

\renewcommand{\arraystretch}{1.0}

\afterpage{\clearpage}

\subsection{Proton Interaction And Hadron Decay Distributions}
\label{distribs}

Two routines were used to model the distribution of hadron
interactions in the materials; one using a two dimensional Gaussian
distribution to model the axial dispersion of the proton beam,
and one using a distribution based on the likelihood that a
hadron had not interacted in a randomly chosen distance. As before,
the distance along the $z$ axis was used as the total distance
ignoring any small $x$ or $y$ components. The beam distribution
was provided in the original Monte Carlo. All lengths and distances
throughout this section are in cm.

All hadrons thrown assumed that a proton-nucleon interaction had
occured, and that all hadrons would decay (the decay weight,
$W_{D}$, described in \ref{decayprob} above provided the proper
scaling for loss due to interactions). The Monte Carlo used an
iterative process where a trial distance in the material at which
the interaction or decay would occur $(\Delta_{z}$, and then tested
the trial distance against the interaction or decay distribution. If
the trial distance fell within the distribution, the routine then
determined the point $(\Delta_{x},\Delta_{y},\Delta_{z})$ at which
the decay or interaction would have occured based on either the
distributed beam diffusion and beam correction angles, or the
momentum of the hadron \nspmom. If the trial distance failed to fall
within the distribution, the routine initiated a new trial
$\Delta_{z}$ until a distance was accepted.

Specifically, to determine the point along the $z$ axis at which
the proton had interacted for hadrons thrown from the target, the
routine threw a random number (all random numbers were between
0 and 1) and multiplied the number to the thickness of the target,
providing the trial interaction distance, $\Delta_{z}$. The
likelihood that the proton had survived to the trial distance was
calculated using:
$$ L_{NI} = exp \, \left( - \Delta_{z} \, \frac{\rho_{A}}
{\lambda_{I}(p,A)} \right) $$ where $\rho_{A}$ (\nsdensity) is the
density of the target material $A$ and
$\lambda_{I}(p,A)$ (\nsilength) is the nuclear interaction length
for 800 \cmom protons in the material. $L_{NI}$ was referred to as
the test function. A second random number,
$rn$ was thrown and compared to the value $L_{NI}$. If
$rn \le L_{NI}$ the trial distance $\Delta_{z}$ was accepted and
the routine then determined the interaction point
$(X_{I},Y_{I},Z_{I})$ where $X_{I}=\delta_{x} + XSLP \, \Delta_{z}$.
$\delta_{x}$ was the distance in $x$ away from the $z$ axis
determined by the beam dispersion routine, and XSLP is the beam
angle in the $x-z$ plane defined
in \ref{calibrations}. If $rn > L_{NI}$, the process was repeated
until the test distance $\Delta_{x}$ was accepted. $Y_{I}$ was
calculated by substituting $y$ for $x$, and $Z_{I} = -60.96 +
\Delta_{z}$.

For hadrons thrown in the dump, the process to determine where the
proton interacted was the same, except the trial distance was
limited to the distance at which 95 percent of the proton
interactions would have occured, and magnetic field effects
were included. The maximum distance
$\Delta^{max}_{z}$ was calculated using
$$ \Delta^{max}_{z} = - \, ln(0.05) \, \frac{\lambda_{I}(p,Cu)}
{\rho_{Cu}} $$ The trial distance was then calculated using
$\Delta_{z} = rn \, \Delta^{max}_{z}$. Once a $z$ decay distance
had been found, the routine projected the proton through the
spectrometer to the point $Z_{I}= 203.2 + \Delta_{z}$. The proton
was given initial $\delta_{x}$ and $\delta_{y}$ components as well
as the correct beam angle before the projection was performed, so
the interaction point, $(X_{I},Y_{I},Z_{I})$, was the values of
the position of the proton at $Z_{I}$ after the projection.

The distribution of hadron decays for hadrons produced in the dump
were handled in the much the same manner. The test function used
was
$$ L_{NI} = exp \left[ -z \left( \frac{m}{p \, \tau} +
 \frac{\rho_{Cu}}{\lambda_{I}(h,Cu)} \right) \right] $$ where $m$
is the mass (\nscmass), $p$ the magnitude of the momentum
\nspmom (\nscmom),
$\tau$ the proper lifetime (s) and $\lambda_{I}(h,Cu)$ the nuclear
interaction length (\nsilength) of the hadron $h$. The maximum
allowable distance, $\Delta^{max}_{z}$ was determined by setting
$L_{NI} = 0.05$ and solving for $z$.

\begin{figure}[!ph]
\resizebox{5.6in}{2.7in}
{\includegraphics[80,290][505,485]{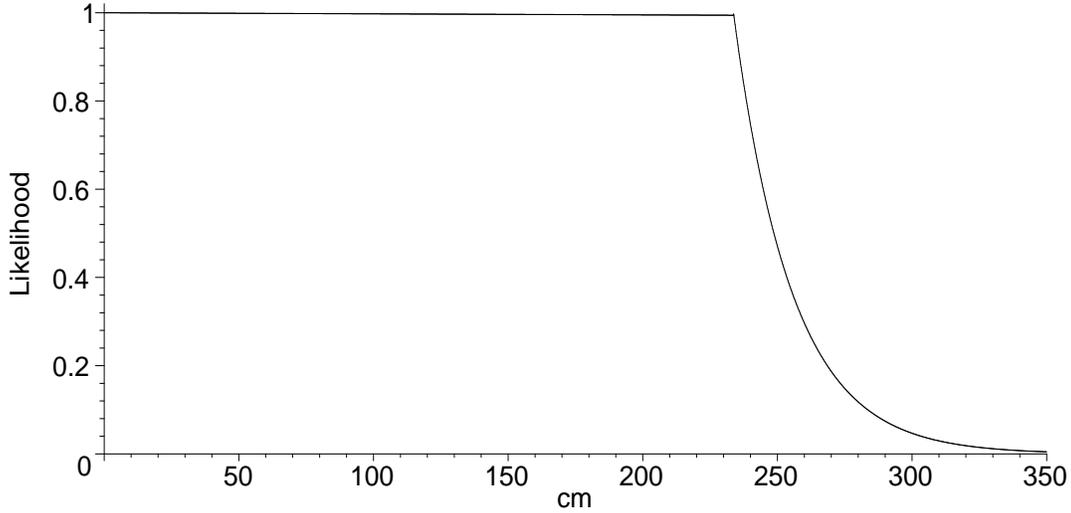}}
\caption[Decay Distribution Test Functions.]{The two test functions
used to determine the decay distribution for 55.0 \cmom charged
kaons thrown from a target. The value of the test function at
the randomly chosen decay distance was compared to a random
number between 0 and 1. If the random number was smaller than or
equal to the value of the test function, the decay distance was
accepted, if not a new decay distance was calculated and the
process repeated until a distance met the required test. Distance is
the distance downstream of the target ($Z_{tgt}=-60.96$ cm)}
\label{tdecdist}
\end{figure}

Two test functions were required for light hadrons thrown from the
dump, one for the distribution of decays in the open decay region
between the target and the dump, and one for the distribution of
decays in the dump if the hadron survived the open decay length. The
test function used for $\Delta_{z} \le 233.68$ (cm) was:
$$ L^{odl}_{I} = exp \, \left( - z \, \frac{m}
{p \, \tau} \right) $$ where $odl$ signifies the open decay length
(233.68 cm). The test function for $ 233.68 < \Delta_{z} \le
\Delta^{max}_{z}$ (cm) was given as
$$ P^{odl}_{S} \, exp \left[ -z \left( \frac{m}{p \, \tau} +
 \frac{\rho_{Cu}}{\lambda_{I}(h,Cu)} \right) \right] $$ where
$P^{odl}_{S} = 1 - exp \, \left[ \left( -233.68 \, m \right)
\, / \, \left( p \, \tau \right) \right]$ (the probability the
hadron did not decay in the open decay region) and
$\Delta^{max}_{z}$ was calculated by setting $L_{NI} = 0.05$ in
the second test function and solving for $z$ and adding 233.68 (cm).
Figure \ref{tdecdist} shows a plot of the two test functions
used to determine the decay distribution of charged kaons having
$\nspmag = 55.0$ \nscmom.

\afterpage{\clearpage}

\subsection{Decay Modes And Branching Fractions}
\label{modes}

The Monte Carlo used a set of decay mode routines to determine the
mode and branching fraction of the mode chosen. Since the
Monte Carlo spectra were weighted spectra, only semi-muonic
decay modes were allowed for open charm hadrons. Table
\ref{decmodes} shows the decay modes allowed for each hadron species
thrown in the Monte Carlo. The weight given each decay was the
branching fraction listed. Strange hadrons were allowed to decay
to charged pions, which were subsequently passed through the
decay routine as secondary decays. However, if more than one muon
was accepted the event was cut due to the requirement in the
analysis section where only one valid track was allowed for each
event.

\renewcommand{\arraystretch}{1.5}

\begin{table}[!ht]
\caption[Decay Modes]{The decay modes allowed in the Monte Carlo.
The weight $W_{br}$ given any muon that was accepted was the
branching fraction as shown. Only the hadron decay modes are shown,
the anti-hadron modes are the charge conjugate of the mode listed.
The $K_{L}$ hadrons were assumed to be from \kzero 50 percent of
the time, and \akzero 50 percent of the time. Decay modes
of the anti-hadrons are the charge conjugate of those shown. }
\label{decmodes}
\begin{center}
\begin{tabular}{|ccc|}

\hline

  &
 $ \qquad \mbox{Decay} \qquad $ &
 $ \qquad \mbox{Branching} \qquad $\\

\raisebox{2.0ex}[0pt]{$ \qquad \mbox{Hadron} \qquad $} &
 Mode &
 Fraction\\

\hline

\nspluspi &
 $ \nsmuplus \, \nu_{\mu} $ &
 1.0000\\

\hline

\nsplusk &
 $ \nsmuplus \, \nu_{\mu} $ &
 0.6351\\

  &
 $ \pi^{0} \, \nsmuplus \, \nu_{\mu} $ &
 0.0318\\

  &
 $ \nspluspi \, \pi^{0} $ &
 0.2116\\

  &
 $ \nspluspi \, \nspluspi \, \nsminuspi $ &
 0.0559\\

  &
 $ \nspluspi \, \pi^{0} \, \pi^{0} $ &
 0.0173\\

\hline

$ K_{L} $ &
 $ \bothpi \mu^{\mp} \nu $ &
 0.2717\\

  &
 $ \nspluspi \, \nsminuspi \, \pi^{0} $ &
 0.1256\\

  &
 $ \nsbothpi \, e^{\mp} \, \nu_{e} $ &
 0.3878\\

\hline

\nsplusd &
 $ \nsmuplus \, \nu_{\mu} \, \nsakzero $ &
 0.0650\\

  &
 $ \nsmuplus\,\nu_{\mu}\, \overline{K}^{*} ( 892 )^{0} $ &
 0.0440\\

  &
 $ \nsmuplus \, \nu_{\mu} \, \phi $ &
 0.0370\\

  &
 $ \nsmuplus \, \nu_{\mu} \, \overline{K}_{1} ( 1270)^{0} $ &
 0.0350\\

\hline

\dzero &
 $ \nsmuplus \, \nu_{\mu} \, \nsminusk $ &
 0.0343\\

  &
 $ \nsmuplus \, \nu_{\mu} \, \overline{K}^{*} ( 892 )^{-} $ &
 0.0214\\

  &
 $ \nsmuplus \, \nu_{\mu} \, {K}^{-} \, \pi^{0} $ &
 0.0031\\

\hline

\end{tabular}
\end{center}
\end{table}

\renewcommand{\arraystretch}{1.0}

\afterpage{\clearpage}

\chapter{Method Used To Extract The Open Charm Cross Sections}
\label{fitting}

This analysis used the single muon data and Monte Carlo
muon spectra to extract four inclusive open charm differential
cross sections, \nsdcucross,
\nsadcucross, \dbecross and \adbecross by fitting the Monte Carlo
spectra to the data. The Monte Carlo spectra were referred to as
total Monte Carlo spectra. The open charm contribution to a total
Monte Carlo spectrum could be varied by using
open charm differential cross sections parameterized by functions
of the hadrons \nspt. The
open charm cross sections were
fit to the data using a least-squares minimization routine where
each data point contributed an individual \chisquare calculated
as the square of the difference between the data and the total
Monte Carlo spectrum for that data point, divided by the sum of the
squares of the errors:
\begin{equation}
\nschij = \frac{\left( N^{\mu}_{j} - W^{MC}_{j} \right)
^{2}} {\varepsilon^{2}_{j} + \epsilon^{2}_{j}}
\label{chiindv}
\end{equation} where $N^{\mu}_{j}$ is the number of muons from the
data in a histogram bin $j$ and $W^{MC}_{j}$ is the number of muons
determined from weighted spectra in bin $j$ of a histogram of the
total Monte Carlo spectrum corresponding to the data. The errors
$\varepsilon$ and $\epsilon$ are the errors for the data and
composite Monte Carlo, respectively.

Use of single muons prevented the analysis from determining which
species of hadron decayed to a given muon. In the case of open charm
production, hadrons containing a charm quark and a light anti-quark
\cqbar are designated \nshadron, and those having an anti-charm
quark and a light quark \qcbar are designated by \nsahadron. The
inclusive \hadron cross sections were extracted using \muplus
spectra, and the inclusive \ahadron cross sections were
extracted using \muminus spectra.

Previous experiments have fitted various functions to their results
based on
theoretical predictions of the shape of the spectra as a function
of \hadpt.\footnote{References
\cite{hep-ph-9711337} - \cite{nucl-ex-0410038} are a few
examples.} From the literature, the functions used to parameterize
the open charm cross sections for this analysis were:

\begin{equation}
\funcexp\label{expfunc}
\end{equation}

\begin{equation}
\functhree\label{3varfunc}
\end{equation}

\begin{equation}
\funcfour\label{4varfunc}
\end{equation}

\vspace{0.2in}

Another exponential form, \nsfuncexptwo, used in \cite{prl-72-2542}
and \cite{prl-77-2392},
was also attempted but resulted in very high total \chisquare
indicating it should not be used to describe the open charm
cross sections for
this data. Function \ref{expfunc} is referred to as the exponential
form and the functions \ref{3varfunc} and \ref{4varfunc} are
referred to as the 3 and 4 parameter forms, respectively. There were
4 fits performed using \functhree as the open charm
cross sections, one where all parameters were free parameters, and
one each where $n$ was held fixed at the integer values 4, 5 and
6. These were chosen from theoretical predictions as well as fits
used in previous experiments \cite{hep-ph-9711337},
\cite{prl-77-2392}, \cite{nucl-phys-b-100-237},
\cite{hep-ph-9701256}, \cite{nucl-phys-b-438-261},
\cite{prl-31-1153}, \cite{prd-10-2973}, \cite{prd-12-3469},
\cite{prl-33-719}. Fits to \funcfour were done for the same
fixed values of $n$ only. All fits using the 3 and 4 parameter
functions used $m_{c}=1.5$ \nscmass. The 4 parameter function
includes an energy dependency term,
$\frac{\nshadpt}{p_{beam}}$ where $p_{beam}$ is the center of mass
momentum of the proton beam. For this experiment,
$p_{beam}=19.36$ \nscmom.

Each target provided four independent but related muon
spectra; \muplus and \muminus from interactions in the
target, and \muplus and \muminus from interactions in the dump. All
four spectra were related by the number of protons incident on the
target. Each fit required four total Monte Carlo spectra to be
compared against the four data spectra simultaneously.

A total Monte Carlo spectrum is the number of muons resulting from
$2 \times 10^{7}$ proton interactions in either copper or beryllium.
More precisely, the total Monte Carlo copper target \muplus
spectrum is the number of \muplus resulting from the interactions
of $2 \times 10^{7}$ protons in the copper target. The spectrum was
determined from adding the number of \muplus from light hadron
production, called the light contribution, plus the number of
\muplus resulting from production of open charm, called the
open charm contribution. The light contribution is
the sum of the muons from the production of \nspluspi, \plusk and
$\nskzero/\nsakzero \rightarrow K_{L} \rightarrow \nsmuplus$
copper target Monte Carlo spectra. The light contributions were
normalized since all the light hadrons were thrown using fully
normalized differential cross sections. The open charm
contribution was calculated using a set of values for the free
parameters of the function being fitted.

The four data spectra
were related by the total number of incident protons, $N_{p}$. The
Monte Carlo generated all spectra using $2 \times 10^{7}$ incident
protons, so a scaling factor $N$ was introduced to scale the total
MC spectra to the proper number of incident protons. The
spectrometer lacked a beam intensity monitor which was sensitive to
the requested
number of protons per spill, so the parameter $N$ (referred to
as the scale factor $N$) was introduced as an extra free parameter
for all fits. The relationship between $N$ and $N_{p}$ is given in
section \ref{tgtscale}. The requirement that the four data and four
total MC spectra be used in calculating \nschisquare, and the
use of normalized light contributions resulted in normalized
open charm cross sections. Figure \ref{monte-to-array} (page
\pageref{monte-to-array}) shows a
representation of the process used to calculate the light
MC contributions and the open charm transformation arrays used
to extract the open charm cross sections. The open charm arrays
and their use are described in Chapter \ref{transmats}. For
reference, Figure \ref{data-to-array}
(Chapter \ref{analysis} page \pageref{data-to-array}) shows the
method used to analyze the single muon data and place the resulting
muon spectra into arrays used in the least-squares minimization
routines.

\begin{center}
\begin{figure}[ph]
\resizebox{5.7in}{5.7in}
{\includegraphics[27,158][562,688]{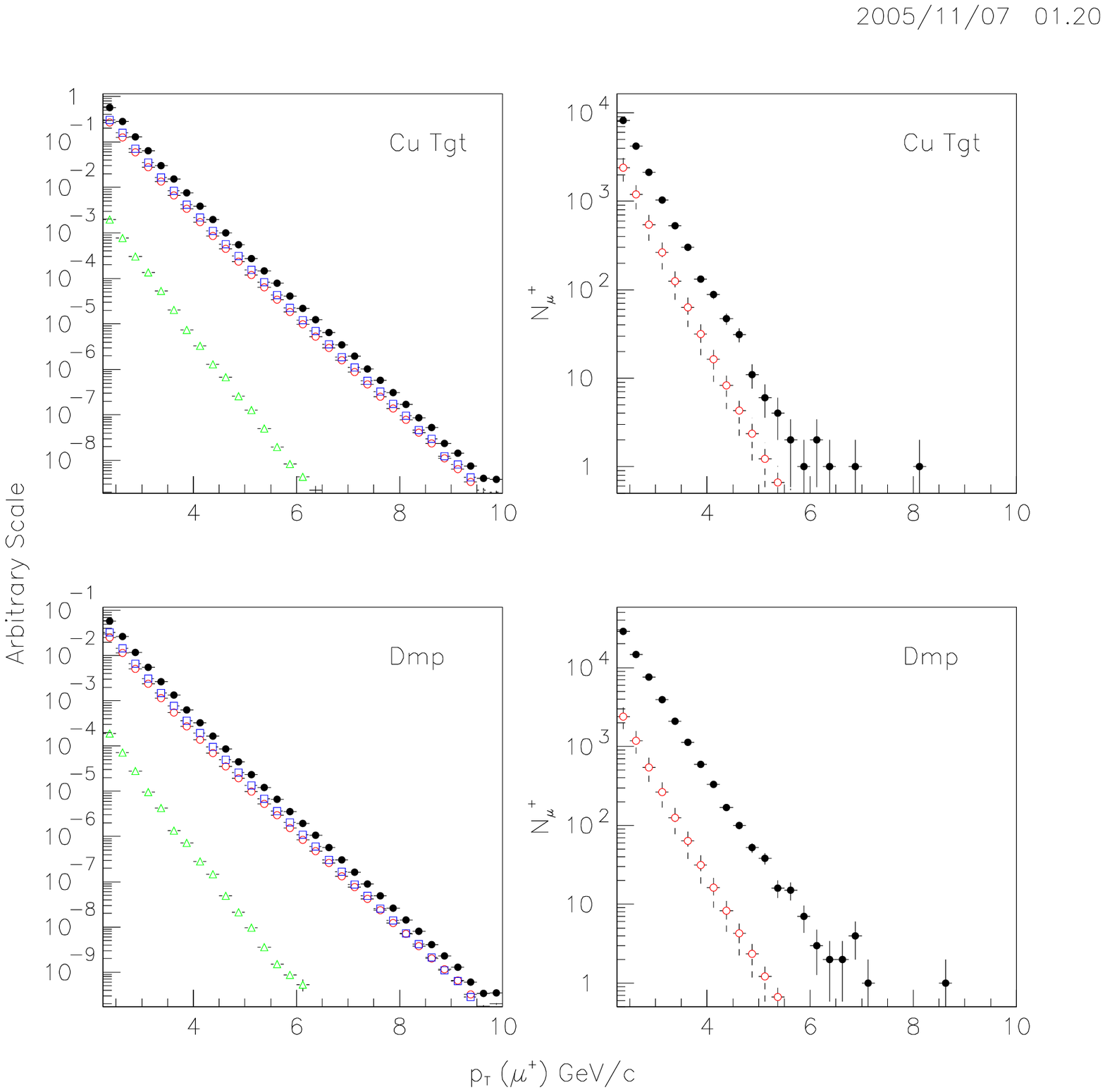}}
\caption[MC \nspcu Light Spectra]{Left Figures: light Monte
Carlo \muplus contribution (black filled circles) from the copper
target (top) and beam dump (bottom), composed of the sum of 
\muplus spectra from $\pi^{+}$ (red open circles), $K^{\pm}$
(blue open squares) and $K^{0} \, / \, \overline{K^{0}}$ (green
open triangles). Right Figures: the same light Monte Carlo
contribution after fitting the 3 variable function with $N$
a free parameter (red open
circles) and the \muplus data (black closed circles). Errors are
statistical only for the figures on the left. For figures on the
right, the errors for the data
are statistical only, the scaled light MC contributions have all
systematic errors except normalization added in quadrature to the
statistical errors.}
\label{light-plus-cu}
\end{figure}
\end{center}

\begin{center}
\begin{figure}[ph]
\resizebox{5.7in}{5.7in}
{\includegraphics[27,158][562,688]{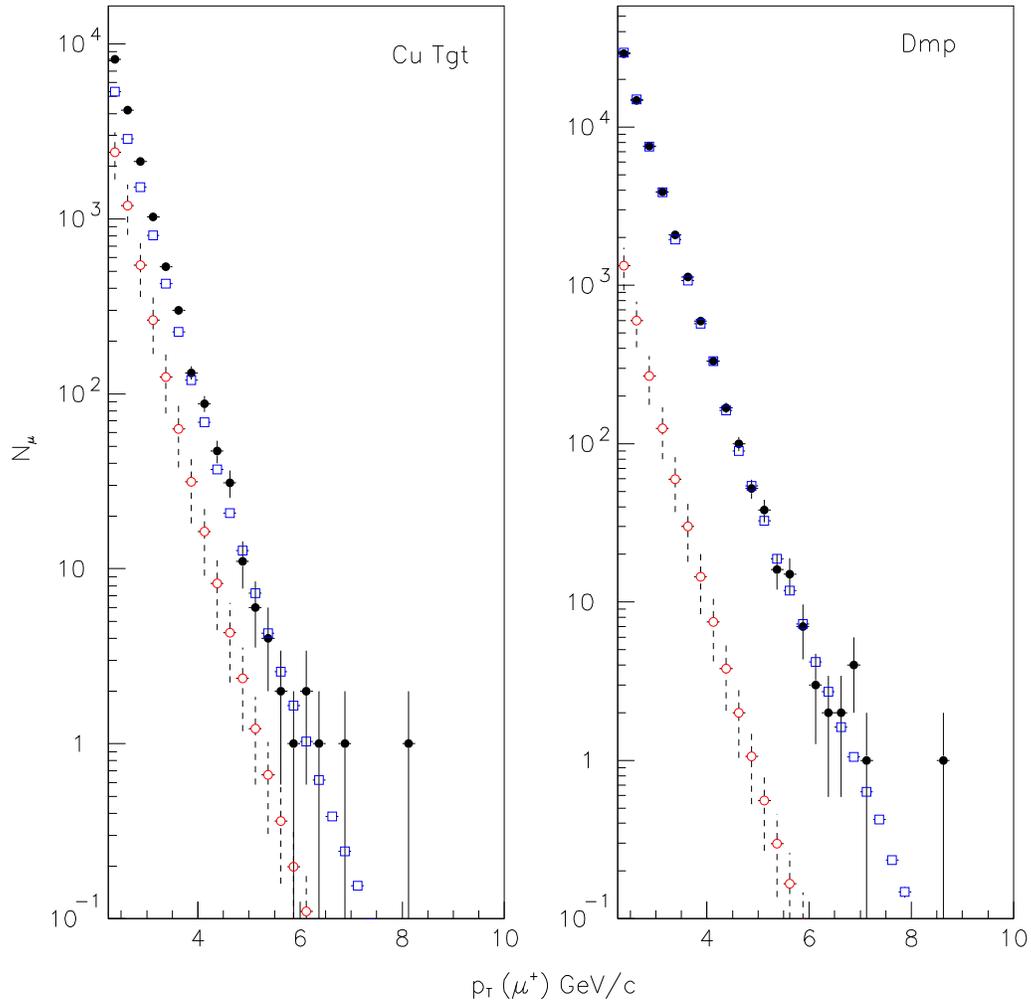}}
\caption[Open Charm And Total MC Spectra From \nspcu]{Total
open charm Monte Carlo \muplus contribution (blue open squares),
total light MC \muplus spectra from Figure
\ref{light-plus-cu} (red open circles) and the the data (black
filled circles) from the copper target (left) and the dump (right).
Data is the \muplus spectra from the copper target, and the
Monte Carlo spectra are from fitting the 3 variable function with
$n$ a free parameter. Data errors are statistical only. MC spectra
have all systematic errors except normalization added in quadrature
to the statistical error.}
\label{heavy-plus-cu}
\end{figure}
\end{center}

Figure \ref{light-plus-cu} shows
the \muplus spectra for the individual and the light Monte Carlo
contribution used for all fits for the copper target (top left)
and dump (lower left). The upper and lower right histograms show
the light Monte Carlo contribution scaled to the data after the
3 parameter function with $n$ a free parameter had been fitted.
Figure \ref{heavy-plus-cu} shows the \muplus data
(black circles), total Monte Carlo \muplus spectra (blue open
squares) and the open charm Monte Carlo \muplus contribution (red
open circles) for the open charm cross section determined from the
fit in Figure \ref{light-plus-cu}. Left is the spectra for the
copper target and right is the dump.

\afterpage{\clearpage}

\section{Minimization}
\label{minimization}

The three parameterizations of the open charm differential cross
sections used in this analysis were:
$$\funcexp$$
$$\functhree$$
$$\funcfour$$

\vspace{0.2in}

The analysis histogrammed the data into 1-dimensional histograms
with 40 bins from 0 to 10 \cmom in \nspt. The contents of the bins
in the
histograms were placed into arrays for use in minimization. The
light Monte Carlo spectra were placed into similar arrays, and the
contents of \mupt versus \hadpt histograms were placed into
40 $\times$ 40 2-dimensional arrays, referred to as hadron to
muon \pt transformation arrays.

A change in any free parameter resulted in four
total MC spectra to compare against the data. A \chisquare, called
an individual \chisquare, \nschij, was calculated for
all data points in the four data spectra using 
Equation \ref{chiindv} (page \pageref{chiindv}). The total
\chisquare, \nstotchi, was the sum of the individual \chisquare: 
\begin{equation}
\nstotchi=\sum_{j=1}^{k}\nschij
\end{equation} where $k$ is the total number of elements in the
four data arrays
containing three or more muons.\footnote{The number of events in a
bin needed to make a data point statistically significant varies
between authors. This analysis chose the number of muons necessary
to be 3.}

The spectrometer had the middle half of all x measuring hodoscope
layers pulled, creating two inner acceptance edges (see Figure
\ref{xplane} page \pageref{xplane}). These acceptance edges
combined with the single bend approximation used to correct for
multiple scattering resulted in large uncertainties in muon
\pt below 2.25 \nscmom. Attempts to correct this uncertainty failed
because no decay vertex was available. As a consequence, the
fits were limited to $\nsmupt \ge 2.25$ \nscmom.

The minimization routines used a grid search method to insure that
local minima were avoided. The routines fitted each cross section
individually in a rotating sequence. Normal
initialization began with fitting the hadron cross section, 
\nsdcross, while the anti-hadron cross section, \nsadcross,
remained constant using a set of initial parameter values.
Once a set number of passes through the grid search had been
performed on the hadron cross section, the routine re-initialized
to begin fitting the anti-hadron cross section holding the
hadron cross section constant at the values of the parameters that
returned the smallest total \chisquare (referred to as \nsminchi)
during its fit. The scale factor $N$ was allowed
to float freely during fits to either cross section,
with the requirement that all four total Monte Carlo spectra be
scaled by the same $N$ simultaneously at each point in the routine
where \totchi was calculated.

Because the target and dump were the same material in the copper
target data, fits to the \pcu data were performed
first. Once the \pcu data had been fit to the various functions,
and the \minchi values
of the parameters of the \pcu cross sections had been
determined, fits of the \pbe cross sections were performed to the
two \pbe target spectra. To accomplish this,
the  dump data were compared to scaled total Monte Carlo
spectra calculated using a previous \pcu result. The open charm
contributions from the dump for fits to the \pbe data
were calculated using the 4 parameter function with $n=6$ and
the \minchi values of the parameters found from the fit to the
\pcu data. The scaling factor $N$ remained a free parameter.

\section{Scaling Monte Carlo Spectra To The Target Material And
Thickness}
\label{tgtscale}

The Monte Carlo spectra needed to be scaled by the integrated
luminosity \L, before they could be compared against the data,
where
$$ \mbox{\L}_{tgt} = \frac{N_{p} \, N_{A}}{A} \, \lambda_{I} \,
 \left[ 1 - exp \, \left( -l \, \rho \, / \, \lambda_{I} \right)
\right] \qquad \left( \mbox{cm}^{-2} \right) $$ for the target
spectra, and
$$ \mbox{\L}_{dmp} = \frac{N_{p} \, N_{A}}{A} \, \lambda_{I} \,
 exp \, \left( -l \, \rho \, / \, \lambda_{I} \right)
 \qquad \left( \mbox{cm}^{-2} \right) $$ for the dump spectra
where $N_{p}$ is the total number of protons incident on the target,
$N_A$ (mole$^{-1}$) is the Avogadro constant, $\lambda_{I}$
(\nsilength) is the
nuclear interaction length for protons in the material,
$A$ (gm mole$^{-1}$) is the atomic weight of the material, $\rho$
(\nsdensity) is the density of
the material, and $l$ (cm) is the thickness of the
target. The difference between the two
is simply stating that all protons incident on the dump interacted
in the dump, and the number of protons incident on the dump was
the number incident on the target times the probability that the
proton did not interact while traversing the target material. The
target scaling includes loss of protons due to interactions in the
target material.

It is important to note that $N_{p}$ is commmon to both. Since the
experiment lacked a beam intensity monitor, the total
number of protons incident on the target was unknown. To overcome
this, the experiment split the total number of incident
protons from the remaining terms of the two integrated luminosities
such that $\mbox{\L} = N_{p} \, L$ where $L$ is
$$ L_{tgt} = \frac{N_{A}}{A} \, \lambda_{I} \,
 \left[ 1 - exp \, \left( -l \, \rho \, / \, \lambda_{I} \right)
\right] \qquad \left( \mbox{cm}^{-2} \right) $$ for the target
spectra, and
$$ L_{dmp} = \frac{N_{A}}{A} \, \lambda_{I} \,
 exp \, \left( -l \, \rho \, / \, \lambda_{I} \right)
 \qquad \left( \mbox{cm}^{-2} \right) $$ for the dump spectra.

The minimization routine scaled the Monte Carlo spectra by the
appropriate value of $L$ to match the data being fitted at
initialization. To estimate the number of incident protons $N_{p}$,
the total nuclear cross section weight thrown for \pcu \pluspi
(either the target or dump)
is divided by the total \pcu \pluspi cross section.
The minimization routines input $N$ as a free
parameter, and using the conversion from the thrown weight,
$N_{p} \sim 2 1.42 \times 10^11 \, N$. The \minchi
value for the parameter $N$ thus determined the
integrated beam flux corrected for dead time.

\section{Calculating The Open Charm Contributions}
\label{transmats}

The open charm contribution to a total MC spectrum was calculated
using an array whose elements,
$\mathbf{W}^{D \rightarrow \mu}_{ij}$, were the weight of the
muons having transverse momentum between 0.250$j$ and 0.250($j+1$)
\nscmom, from open charm hadrons thrown with transverse momentum
between 0.250$i$ and 0.250($i+1$) \nscmom. The array used for
calculating the \muplus spectrum from \plusd and \dzero produced in
the copper target was composed of the elements
$$ \mathbf{W}^{D \rightarrow \nsmuplus}_{ij}(T,Cu) =
\mathbf{W}^{\nsplusd \rightarrow \nsmuplus}_{ij}(T,Cu) +
\mathbf{W}^{\nsdzero \rightarrow \nsmuplus}_{ij}(T,Cu) $$

\begin{center}
\begin{figure}[pht]
\resizebox{4.4in}{3.1in}
{\includegraphics[-168,166][563,688]{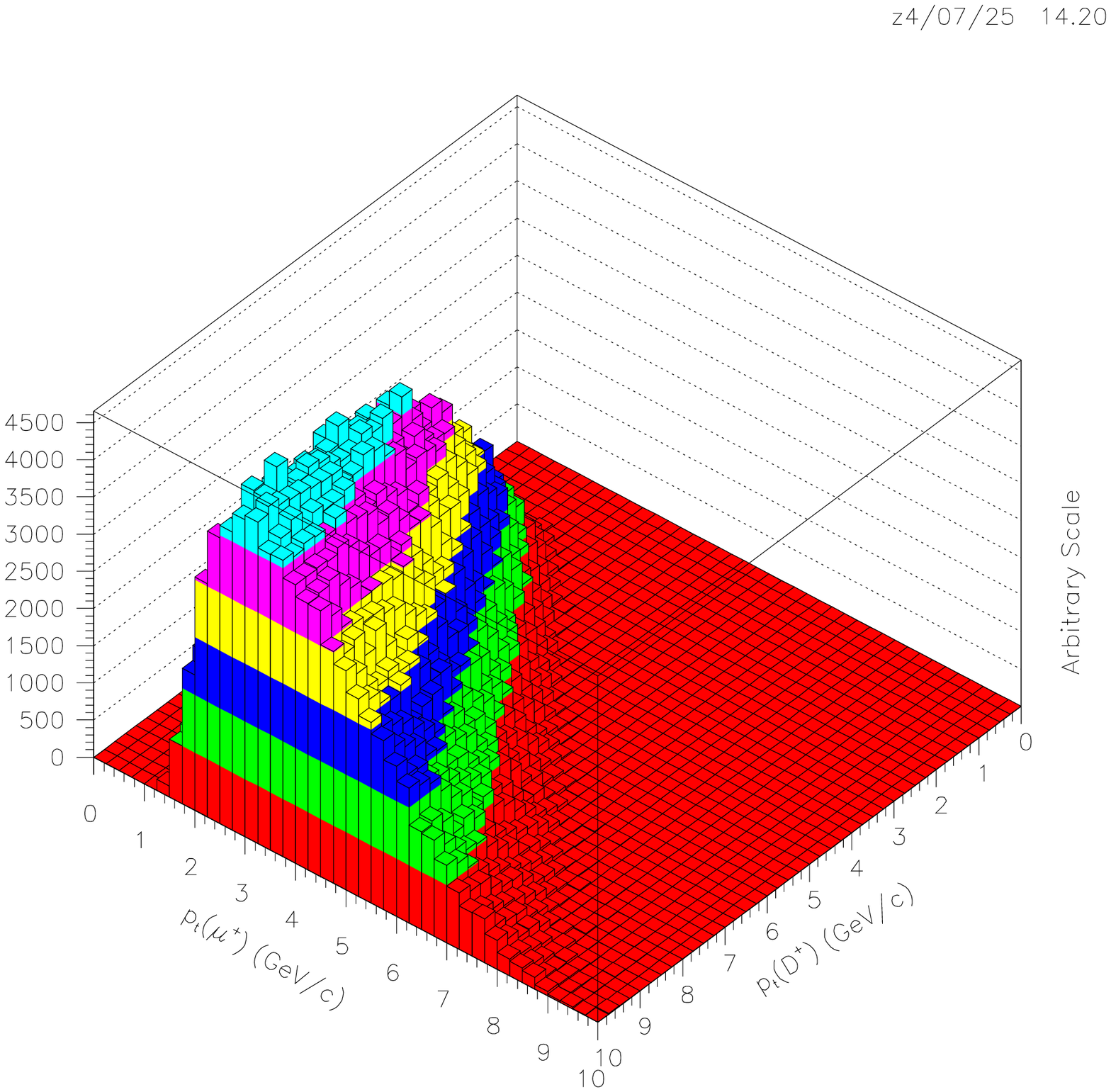}}
\caption[\mupt Versus \nspt Histogram w/o Cross Section]{The 
\mupt versus $ \nspt(D) $ histogram of Monte Carlo
$ D \rightarrow \muplus $ used in the calculation of the muon
contribution from D production from the copper target in
the fitting routine. Weight does not include any cross section
weight.}
\label{htomudplustgtnoxsect}
\end{figure}
\end{center}

The Monte Carlo was used to develop a set of 6 arrays used
to calculate the open charm contributions, one each \muplus and
\muminus from the copper target, the beryllium target and the
copper beam dump. The dump required only one set of arrays
because the difference between the arrays from the dump with
either target presented to the beam
would be the target thickness normalization
presented in Chapter \ref{tgtscale}. The elements of the arrays
were the values of the weight in the bins of
2-dimensional \mupt versus \hadpt histograms of the individual
open charm hadrons thrown in the Monte Carlo. Figure
\ref{htomudplustgtnoxsect} shows the 2-dimension \mupt
versus \hadpt histogram resulting from 20
million \plusd and \dzero thrown from the copper target. Table
\ref{mainmats} gives the 12 individual
arrays and the six transformation arrays used for calculating the
open charm contributions.

\renewcommand{\arraystretch}{1.5}

\begin{table}[ph]
\caption[Transformation Arrays]{The six transformation arrays
used in the minimization routines to determine the open charm
contributions from the copper target ($Cu$), beryllium
target ($Be$) and dump ($dmp$). The contribution
is determined from the transformation array which is the sum of the
weights of the two individual arrays listed, after an arbitrary
open charm cross section is applied. A fit required four
contributions to be determined for each change in any of the free
parameters.}
\label{mainmats}
\begin{center}
\begin{tabular}{|c@{\extracolsep{0.5in}}cc|}

\hline

\multicolumn{1}{|c}{Spectrum} &
 \multicolumn{1}{c}{Transformation} &
 \multicolumn{1}{c|}{Individual}\\

\multicolumn{1}{|c}{Desired} &
 \multicolumn{1}{c}{Array} &
 \multicolumn{1}{c|}{Array}\\

\hline

\multicolumn{1}{|c}{} &
 \multicolumn{1}{c}{} &
 \multicolumn{1}{c|}
  {$ \mathbf{W}^{ \nsplusd \rightarrow \nsmuplus }(Cu) $}\\

\multicolumn{1}{|c}
 {\raisebox{2.0ex}[0pt]{$ \nsmuplus(Cu) $}} &
 \multicolumn{1}{c}
  {\raisebox{2.0ex}[0pt]
   {$ \mathbf{W}^{ D \rightarrow \nsmuplus }(Cu) $}} &
  \multicolumn{1}{c|}
   {$ \mathbf{W}^{ \nsdzero \rightarrow \nsmuplus }(Cu)$ }\\

\hline

\multicolumn{1}{|c}{} &
 \multicolumn{1}{c}{} &
 \multicolumn{1}{c|}
  {$ \mathbf{W}^{ \nsminusd \rightarrow \nsmuminus }(Cu) $}\\

\multicolumn{1}{|c}
 {\raisebox{2.0ex}[0pt]{$ \nsmuminus(Cu) $}} &
 \multicolumn{1}{c}
  {\raisebox{2.0ex}[0pt]
   {$ \mathbf{W}^{ \overline{D} \rightarrow \nsmuminus }(Cu) $}} &
  \multicolumn{1}{c|}
   {$ \mathbf{W}^{ \nsadzero \rightarrow \nsmuminus }(Cu)$ }\\

\hline

\multicolumn{1}{|c}{} &
 \multicolumn{1}{c}{} &
 \multicolumn{1}{c|}
  {$ \mathbf{W}^{ \nsplusd \rightarrow \nsmuplus }(Be) $}\\

\multicolumn{1}{|c}
 {\raisebox{2.0ex}[0pt]{$ \nsmuplus(Be) $}} &
 \multicolumn{1}{c}
  {\raisebox{2.0ex}[0pt]
   {$ \mathbf{W}^{ D \rightarrow \nsmuplus }(Be) $}} &
  \multicolumn{1}{c|}
   {$ \mathbf{W}^{ \nsdzero \rightarrow \nsmuplus }(Be)$ }\\

\hline

\multicolumn{1}{|c}{} &
 \multicolumn{1}{c}{} &
 \multicolumn{1}{c|}
  {$ \mathbf{W}^{ \nsminusd \rightarrow \nsmuminus }(Be) $}\\

\multicolumn{1}{|c}
 {\raisebox{2.0ex}[0pt]{$ \nsmuminus(Be) $}} &
 \multicolumn{1}{c}
  {\raisebox{2.0ex}[0pt]
   {$ \mathbf{W}^{ \overline{D} \rightarrow \nsmuminus }(Be) $}} &
  \multicolumn{1}{c|}
   {$ \mathbf{W}^{ \nsadzero \rightarrow \nsmuminus }(Be)$ }\\

\hline

\multicolumn{1}{|c}{} &
 \multicolumn{1}{c}{} &
 \multicolumn{1}{c|}
  {$ \mathbf{W}^{ \nsplusd \rightarrow \nsmuplus }(dmp) $}\\

\multicolumn{1}{|c}
 {\raisebox{2.0ex}[0pt]{$ \nsmuplus(dmp) $}} &
 \multicolumn{1}{c}
  {\raisebox{2.0ex}[0pt]
   {$ \mathbf{W}^{ D \rightarrow \nsmuplus }(dmp) $}} &
  \multicolumn{1}{c|}
   {$ \mathbf{W}^{ \nsdzero \rightarrow \nsmuplus }(dmp)$ }\\

\hline

\multicolumn{1}{|c}{} &
 \multicolumn{1}{c}{} &
 \multicolumn{1}{c|}
  {$ \mathbf{W}^{ \nsminusd \rightarrow \nsmuminus }(dmp) $}\\

\multicolumn{1}{|c}
 {\raisebox{2.0ex}[0pt]{$ \nsmuminus(dmp) $}} &
 \multicolumn{1}{c}
  {\raisebox{2.0ex}[0pt]
   {$ \mathbf{W}^{ \overline{D} \rightarrow \nsmuminus }(dmp) $}} &
  \multicolumn{1}{c|}
   {$ \mathbf{W}^{ \nsadzero \rightarrow \nsmuminus }(dmp)$ }\\

\hline

\hline

\end{tabular}
\end{center}
\end{table}

\renewcommand{\arraystretch}{1.0}

The value of an open charm (or anti-charm) cross section,
$ W^{D}_{x,i} $, was calculated at the center of each bin
using $p_{T,i} = 0.250 \left(
i - 1 \right) + 0.125 \quad (\mbox{\nscmom}) $ for $i$ from 1 to
40. The muon spectrum expected from throwing the open charm hadrons
with the cross section being tested was then determined by
$$ W^{\mu}_{j} = \sum_{i=1}^{40}
\left[ W^{D}_{x,i} \, \mathbf{W}^{D \rightarrow \mu}_{i \, j}
\right] $$ 

\begin{center}
\begin{figure}[ht]
\resizebox{4.4in}{3.1in}
{\includegraphics[-156,166][563,688]{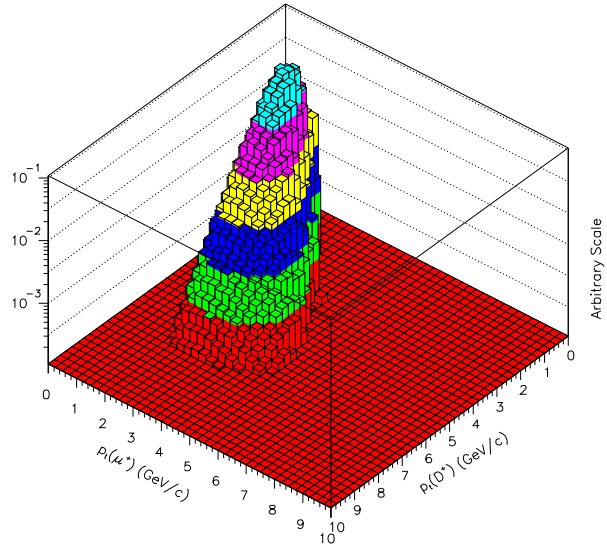}}
\caption[\mupt Versus \nspt Distribution With Cross Section
Applied]{Same hadron to muon \pt distribution as in Figure
\ref{htomudplustgtnoxsect} except trial input cross section weights
$W^{\sigma(D)}_{i}$ have
been calculated and applied to all columns $i$ in 
\hadpt during a minimization. The muon distribution $W^{\mu}_{j} $
expected from the input cross section is the projection onto the
\mupt axis.}
\label{htomudplustgtwxsect}
\end{figure}
\end{center}

Figure \ref{htomudplustgtwxsect} shows the result of applying a
cross section weight for \plusd to the histogram shown in Figure
\ref{htomudplustgtnoxsect}.

\begin{center}
\begin{figure}[ht]
\resizebox{5.6in}{5.6in}
{\includegraphics[27,158][562,670]{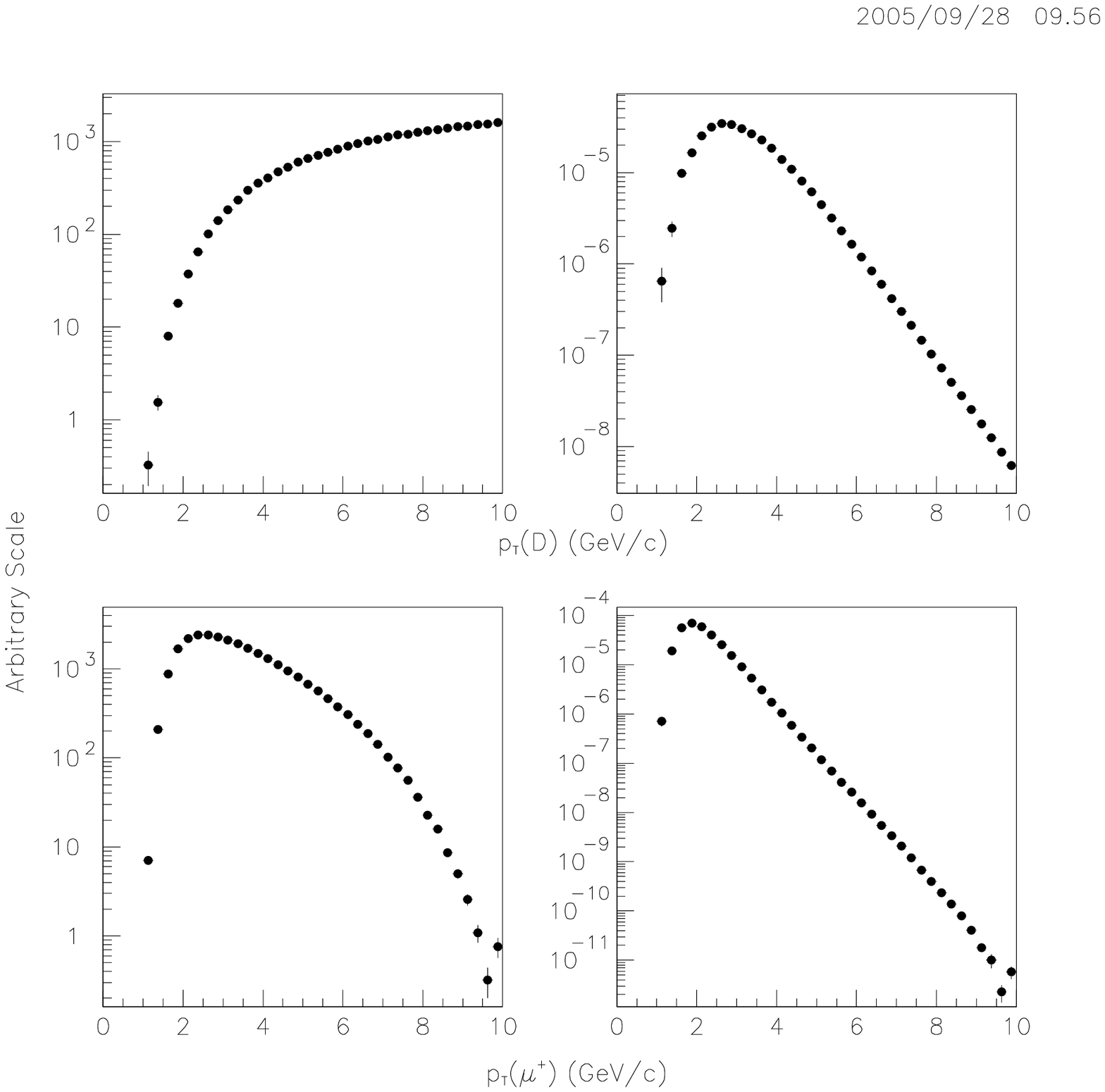}}
\caption[Hadron And Muon Spectra]{The $D$ spectrum of Figures
\ref{htomudplustgtnoxsect} (top left) and \ref{htomudplustgtwxsect}
(top right) and the \muplus spectra corresponding to the same
Figures on bottom. All spectra are the projection of the
2-dimensional \mupt versus \hadpt histograms onto their respective
axes. Errors may be smaller than the symbols used, and include
all systematic errors associated with the decay and branching
fraction parameterizations used to calculate those weights as
described in \ref{chijerrors}.}
\label{htomuspectra}
\end{figure}
\end{center}

Figure \ref{htomuspectra} shows the projections of the histograms
onto their respective axes to show the effect of applying a cross
section (in this case an arbitrary $D$ cross section found during a
minimization).

\afterpage{\clearpage}

\section{Errors Used For Calculating \nschisquare}
\label{chijerrors}

Only the statistical error was used for the data,
$ \varepsilon _{j} = \sqrt{s_{j}} $,
where $s_{j}$ is the total number of muons in element $j$.
Errors for Monte Carlo spectra included the systematic and
statistical errors of the parameterizations used to calculate the
primary weights $W_{x}$, $W_{A}$, $W_{D}$, $W_{NA}$ and $W_{br}$ for
light hadrons and $W_{D}$, $W_{NA}$ and $W_{br}$ for the open charm
hadrons. See table \ref{weights-1}, page \pageref{weights-1}.

The total weight given an event in the Monte Carlo was defined as
$$W_{k} = W_{x,k} \, W_{A} \, W_{D} \, W_{NA} \, W_{br}$$ for light
hadrons and
$$W_{k} = W_{D} \, W_{NA} \, W_{br}$$ for open charm hadrons.
Hadrons thrown from the dump had $W_{NA} = 1$. The
statistical error of $s$ weighted events in a bin $j$ is
determined from \cite{lyons}:
$$ \epsilon_{j} = \sqrt{ \sum^{s}_{k=1}
\left( W^{2}_{k} \right)} $$.

All statistical or systematic errors of the parameterizations used
in calculating the primary weights, $\sigma_{x}$, $\sigma_{A}$,
$\sigma_{D}$, $\sigma_{NA}$ were also calculated. As an example,
the systematic  error, $\sigma_{x}$, of the differential cross
section weight, $W_{x}$, of a light hadron was calculated as
two parts, the errors of the parameters used to calculate the cross
section, shown in Table \ref{xsectparams} (page
\pageref{xsectparams}), plus
the normalization uncertainty, both of which were taken from the
reference. The error due to the uncertainties in the values of the
parameters was calculated
by use of the error propagation equation, excluding covariant terms 
$$ \sigma _{x,p}^{2} \simeq \sigma _{A_{1}}^{2} \left(
\frac{\partial x}{\partial A_{1}} \right) ^{2} +
\sigma _{B}^{2}\left( \frac{\partial x}{\partial B} \right) ^{2}
+ \cdots +\sigma _{m}^{2} \left(
\frac{\partial x}{\partial m} \right) ^{2} $$ where $x,p$ is used
to show that the error is the error of the
parameterization used for the light hadron differential
cross section, $A_{1}$, $B$ and $m$ are three of those parameters
and $ \sigma _{A_{1}}$ is the error of the parameter $A_{1}$
taken from the reference. The second error associated with the
light hadron differential cross section weight was the uncertainty
in the absolute normalization, given by the reference as a relative
error, $R_{x}$. The error due to the normalization uncertainty was
calculated by $\sigma_{x,n} = W_{x} \, R_{x}$. The error associated
with the light hadron cross section weight, $\sigma_{x}$ was then
calculated as
$\sigma^{2}_{x} = \sigma^{2}_{x,p} + \sigma^{2}_{x,n}$. Additional
errors for all other
weights were calculated using the error propagation equation and the
errors of variables from the references. The error on $W_{br}$,
$\sigma_{br}$ (strange and open charm hadrons only)
was 0.2 percent (strange hadrons) and 11.5 percent (open charm
hadrons) of the value of the branching fraction weight.

Each muon from the Monte Carlo contributed the square of its total
weight as a statistical error, plus the square of the error of the
weights for use in calculating the $\epsilon^{2}_{j}$
$$ \epsilon^{2}_{j} = \sum^{s}_{k=1} \left[ W^{2}_{k} +
\sigma^{2}_{k} \right] $$ where $W_{k}$ is the total weight of muon
$k$ and $\sigma^{2}_{k}$ is the square of the errors of the weights
$$ \sigma^{2}_{k} = W^{2}_{k} \left[
\frac{\sigma^{2}_{x,k}}{W^{2}_{x,k}} +
\frac{\sigma^{2}_{A,k}}{W^{2}_{A,k}} +
\frac{\sigma^{2}_{D,k}}{W^{2}_{D,k}} +
\frac{\sigma^{2}_{NA.k}}{W^{2}_{NA,k}} +
\frac{\sigma^{2}_{br,k}}{W^{2}_{br,k}} \right]$$

The squared error, $\epsilon^{2}_{j}$ for $s$ muons in a bin $j$
is then
\begin{equation}
\epsilon^{2}_{j} = \sum^{s}_{k=1}
\left[
 W^{2}_{k} \,
\left(
1 +
\left\{
\frac{\sigma^{2}_{x,k}}{W^{2}_{x,k}} +
\frac{\sigma^{2}_{A,k}}{W^{2}_{A,k}} +
\frac{\sigma^{2}_{D,k}}{W^{2}_{D,k}} +
\frac{\sigma^{2}_{NA,k}}{W^{2}_{NA,k}} +
\frac{\sigma^{2}_{br,k}}{W^{2}_{br,k}}
\right\}
\right)
\right]
\label{totalmcerror}
\end{equation}

\section{Parameter Errors}
\label{errorcode}

Error routines were developed based on MINUIT, the method used for
error analysis of function minimizations used by CERN \cite{minuit}.
The errors of an $n$ dimensional function, including all
correlations, can be determined using contours of equal likelihood.
By choosing the appropriate amount to add to the value of \nsminchi,
referred to as \textit{UP}, the errors on the \minchi values of the
parameters can be determined by minimization to $n-1$ dimensions of
the original function, where one parameter is held fixed. In
general, the method involves increasing or decreasing the value of
the fixed parameter and then minimizing the remaining parameters
until the smallest value of the total \chisquare returned (referred
to as  \testchi here) remains smaller than or equal to \nsmaxchi,
where $ \nsmaxchi = \left( \nsminchi + \mbox{\textit{UP}} \right) $.
The maximum and minimum values the free parameters had to take in
order that \testchi remain smaller than or equal to \maxchi during
the entire process were returned as their errors.

The routines systematically increased/decreased the fixed parameters
value by a pre-set amount until the returned value of \testchi
increased to more than \nsmaxchi. Once this condition was met, the
routine decreased the size of increment that the fixed parameter
would increase/decrease and began minimization of the remaining free
parameters where the value for the fixed parameter was the previous
maximum or minimum value plus or minus the new increment size. This
procedure was repeated until the increment size became smaller than
$10^{-5}$. This experiment used values for \textit{UP} for a 70\%
confidence level for both the $nv=3$ and $nv=4$ functions as given
in Table \ref{upvals}.

\renewcommand{\arraystretch}{1.5}

\begin{table}[ht]
\caption[Values Of UP]{Values of \textit{UP} for given confidence
level and number of free parameters $n$. Taken from \cite{minuit}.}
\label{upvals}
\begin{center}
\begin{tabular}{|cc|c|c|c|c|c|}
\hline

 \multicolumn{2}{|c|}{} &
   \multicolumn{5}{c|}{Confidence level (probability contents
    desired}\\
 
 \multicolumn{2}{|c|}{Number of} &
  \multicolumn{5}{c|}{ inside hypercontour of
   $\nschisquare = \minchi + \mbox{\textit{UP}}$) }\\
 
\cline{3-7}

 \multicolumn{2}{|c|}{Parameters} &
  \makebox[50pt][c]{50\%} &
  \makebox[50pt][c]{70\%} &
  \makebox[50pt][c]{90\%} &
  \makebox[50pt][c]{95\%} &
  \makebox[50pt][c]{99\%}\\

\hline
\hline

 \makebox[25pt][c]{} &
  \makebox[25pt][c]{1} &
  0.46 &
  1.07 &
  2.70 &
  3.84 &
  6.63\\

  &
  2 &
  1.39 &
  2.41 &
  4.61 &
  5.99 &
  9.21\\
 
  &
  3 &
  2.37 &
  3.67 &
  6.25 &
  7.82 &
  11.36\\
 
  &
  4 &
  3.36 &
  4.88 &
  7.78 &
  9.49 &
  13.28\\
 
  &
  5 &
  4.35 &
  6.06 &
  9.24 &
  11.07 &
  15.09\\
 
  &
  6 &
  5.35 &
  7.23 &
  10.65 &
  12.59 &
  16.81\\ 

  &
  7 &
  6.35 &
  8.38 &
  12.02 &
  14.07 &
  18.49\\ 

  &
  8 &
  7.34 &
  9.52 &
  13.36 &
  15.51 &
  20.09\\ 

  &
  9 &
  8.34 &
  10.66 &
  14.68 &
  16.92 &
  21.67\\ 

  &
  10 &
  9.34 &
  11.78 &
  15.99 &
  18.31 &
  23.21\\ 

  &
  11 &
  10.34 &
  12.88 &
  17.29 &
  19.68 &
  24.71\\

\hline

\end{tabular}
\end{center}
\end{table}

\renewcommand{\arraystretch}{1.0}

\clearpage

\section{Other Errors}

This analysis did not require correcting the data for losses from
acceptances, and the large error introduced by the addition of the
two inside acceptance edges was minimized by cutting the first
5 data points from all spectra. Corrections for data lost due
to data acquisition limitations, referred to as signal busy losses,
are contained in the parameter $N$. Monte Carlo studies determined
the uncertainty in the transverse momentum of the muons was less
than 5 percent for the entire region above 2.0 \nscmom. This
uncertainty has the effect of shifting the \pt spectrum to slightly
higher values due to the fact that more events will 'feed down' than
will 'feed up' because the spectrum is (approximately) exponentially
decreasing as a function of \nspt. The same effect is present in
all Monte Carlo spectra as well, so the relative effect between the
two is, for the most part, cancelled.

Loss of data due to inefficiencies in the hodoscopes was modeled in
the Monte Carlo based on studies by the previous dimuon analysis.
For that data all hodoscopes had efficiencies of over 96 percent,
and for this analysis only the x measuring hodoscopes were needed.
Similarly, loss from bad wires and noise in the wire planes was
modeled in the Monte Carlo spectra based on studies from the same
analysis. As a check, the data were plotted using the variable
\nstantx. This variable is very sensitive to the opening angle the
muon trajectory had in the $x-z$ plane, and any significant loss of
events from a defective hodoscope or phototube should be clearly
discernable. No such defects in the spectra were found. (See Figure
\ref{x-project-cuts} page \pageref{x-project-cuts}.)

Light hadrons have a small uncertainty at moderate to high transverse
momentum associated with the parameterization used. The fits
determined by the BSC are valid up to approximately 6 \nscmom in
\nspt \cite{nucl-phys-b-100-237}. The light hadron distributions
above 6 \cmom have an additional uncertainty estimated to be less
than 10 percent. This uncertainty can be neglected since the light
hadron contributions are less than 1 percent beyond 6 \cmom
(see figure \ref{cu-pi-k}).

Production of hadrons containing open bottom were ignored when
calculating the total Monte Carlo spectra. Semi-leptonic decays
of open bottom hadrons introduces a small contamination since
the decay may give rise to both a \muplus and a \muminus (the
semi-leptonic decay $B^{+} \rightarrow \nsmuplus \nu_{\mu} \adzero$
where the \adzero then decays semi-leptonically to a \muminus is
one example). The branching fractions of these modes are small,
typically less than 3 percent, so contamination from
production of open bottom is negligible.

Table \ref{oerrs} lists the estmated uncertainty of the fits due
to the analyzed muon transverse momentum, hodoscope inefficiencies,
light hadron normalization, choice of the transverse momentum of the
open charm hadrons at the mid-point value of the bin and
contamination of the spectra through semi-leptonic decay of
open bottom production.

\renewcommand{\arraystretch}{1.5}

\begin{table}[ht]
\caption[Other Errors]{Estimated errors on final cross sections
from various sources
not included in the errors of the parameters. The Muon and average
hadron \pt are \pt dependent, and the error shown is the maximum
for these errors at any \nspt.}
\label{oerrs}
\begin{center}
\begin{tabular}{|lc|}
\hline

\multicolumn{1}{|c}{} &
 \multicolumn{1}{c|}{Estimated}\\

\multicolumn{1}{|c}{\raisebox{2.0ex}[0pt]{Source}} &
 \multicolumn{1}{c|}{Uncertainty (\%)}\\

Muon \nspt &
 7\\

Average Hadron \nspt &
 5\\

Hodoscopes &
 3\\

Light Hadron Normalization &
 $<$ 1\\

Contamination From Bottom &
 $<$ 1\\

\cline{1-2}

Total &
 9.2\\

\hline

\end{tabular}
\end{center}
\end{table}

\renewcommand{\arraystretch}{1.0}

\clearpage

\chapter{Results}
\label{results}

This analysis used single muon data to extract the open charm
differential cross sections as a function of the hadron transverse
momentum. Eight variations of three functions were fit to the data;
an exponential, a three parameter polynomial with $n$ a free
parameter, and both the three and four parameter polynomials with
$n$ fixed at the integer values 4, 5 and 6 (see Chapter
\ref{fitting}, page \pageref{fitting}).

\begin{center}
\begin{figure}[!ht]
\resizebox{5.8in}{5.8in}
{\includegraphics[27,158][528,645]{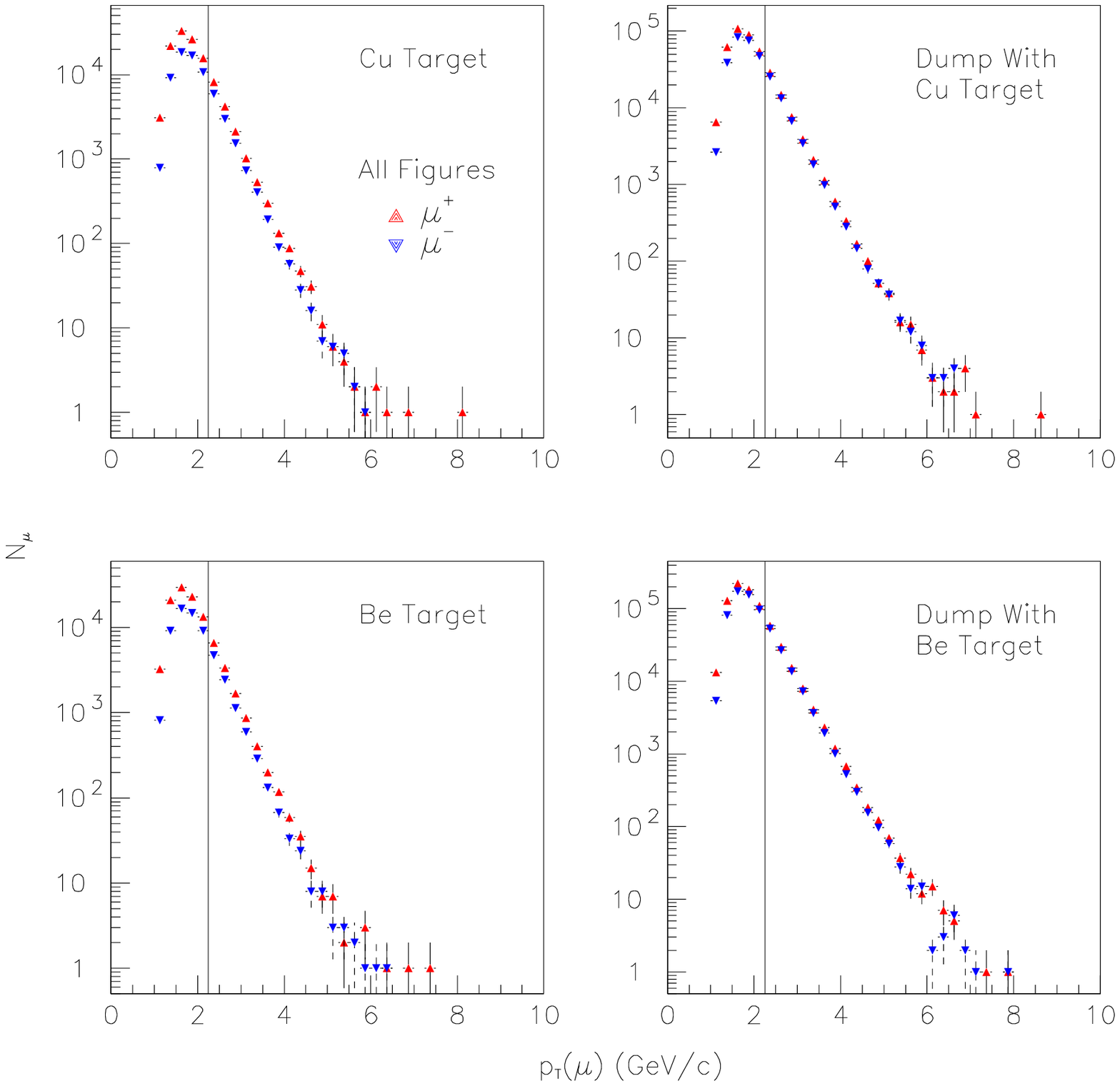}}
\caption[The E866 Data Spectra]
{The single muon E866 data spectra. Top two figures are the \muplus
(red triangles) and \muminus (blue upside down triangles) for the
copper target data set from the target (left) and dump (right).
Bottom row is the spectra from the beryllium target data set.
Vertical lines show lower limit
($\nsmupt \ge 2.25 \nscmom$) of the muon transverse-momentum
used during fitting. Both dump figures are the events from
the copper dump when the designated target (Cu or Be) was presented
to the beam. Errors are statistical only.}
\label{dataplot}
\end{figure}
\end{center}

Table \ref{npoints} gives the total number of muons in each data
spectrum, the total number of muons in the data and the number of
data points that each spectrum provided. Figure \ref{dataplot} shows
the \mupt distributions
from the copper target (top) and beryllium target
(bottom). The vertical line represents the minimum transverse
momentum used (2.25 \nscmom) during fitting.

\renewcommand{\arraystretch}{1.5}

\begin{table}[th]
\caption[The Single Muon Data]{The number of muons, $N_{\mu}$, and
the number of data points, $N_{DP}$, after final cuts
were applied for all spectra used in this analysis. T indicates
events from the targets, and D indicates events from the dump.}
\label{npoints}
\begin{center}
\begin{tabular}
[c]{|lcrc|}\hline

\multicolumn{1}{|c}{} &
 Spectrum &
 \multicolumn{1}{c}{$N_{\mu}$} &
 $N_{DP}$\\

\hline
\hline

  &
 $T^{+}$ &
 16657 &
 14\\

  &
 $T{-}$ &
 11954 &
 14\\

  &
 $D^{+}$ &
 59576 &
 20\\

 \raisebox{4.0ex}[0pt]{Copper} &
 $D{-}$ &
 53322 &
 18\\

\cline{3-4}

\multicolumn{2}{|c}{Total} &
 141689 &
 66\\

\hline
\hline

  &
 $T^{+}$ &
 13368 &
 16\\

  &
 $T{-}$ &
 9482 &
 14\\

  &
 $D^{+}$ &
 120262 &
 19\\

 \raisebox{4.0ex}[0pt]{Beryllium} &
 $D{-}$ &
 109017 &
 19\\

\cline{3-4}

\multicolumn{2}{|c}{Total} &
 252129 &
 68\\

\hline

\end{tabular}
\end{center}
\end{table}

\renewcommand{\arraystretch}{1.0}

\afterpage{\clearpage}

\section{Fits To The Data}

The values for the parameters for the best fits to the data using
the
exponential function are presented in Table \ref{exp-table}, for
the 3 parameter polynomial function in Table \ref{f3-table} and
the four parameter polynomial in Tables \ref{f4-cu-table} (copper)
and \ref{f4-be-table} (beryllium). The total \nschisquare,
\nsminchi, and reduced \nschisquare, \nspdfchi, are also shown for
the fit for each parameterization. From the tables, the author
concludes:

\begin{itemize}

\item The 3 parameter function with either $n$ a free parameter or
 $n$ fixed at $n=6$ as well as the 4 parameter function with $n$
 fixed at $n=6$ provided the best fits to the data.

\item Though no errors were calculated for floating values
 of $n$, the weighted average value of $n$ for all fits with $n$ a
 free parameter was 6.18 $\pm$ 0.19.

\item The exponential function provided a remarkably good fit for
 a large range in \hadpt, $2.0 \lesssim \pt \lesssim 7.0$ \nscmom.

\item The 4 parameter function provides no additional information
 beyond that determined from fitting the 3 parameter function.

\end{itemize}

All fits of functions having 4 free parameters, the 3 parameter
floating $n$ and all 4 parameter functions, had large correlations
between the parameters. No errors were
calculated for fits using the 3 parameter function with $n$ a free
parameter. Because of the large correlations, two sets of errors
were presented for fits using the fixed $n$
4 parameter function; one where the parameter $\alpha$ was
held constant at its \minchi value (referred to as the fixed
$\alpha$ errors), and one set where the parameter $m$ was held
fixed at its \minchi value (fixed $m$ errors).

\begin{center}
\begin{figure}[!ht]
\resizebox{5.8in}{6.0in}
{\includegraphics[27,158][522,668]{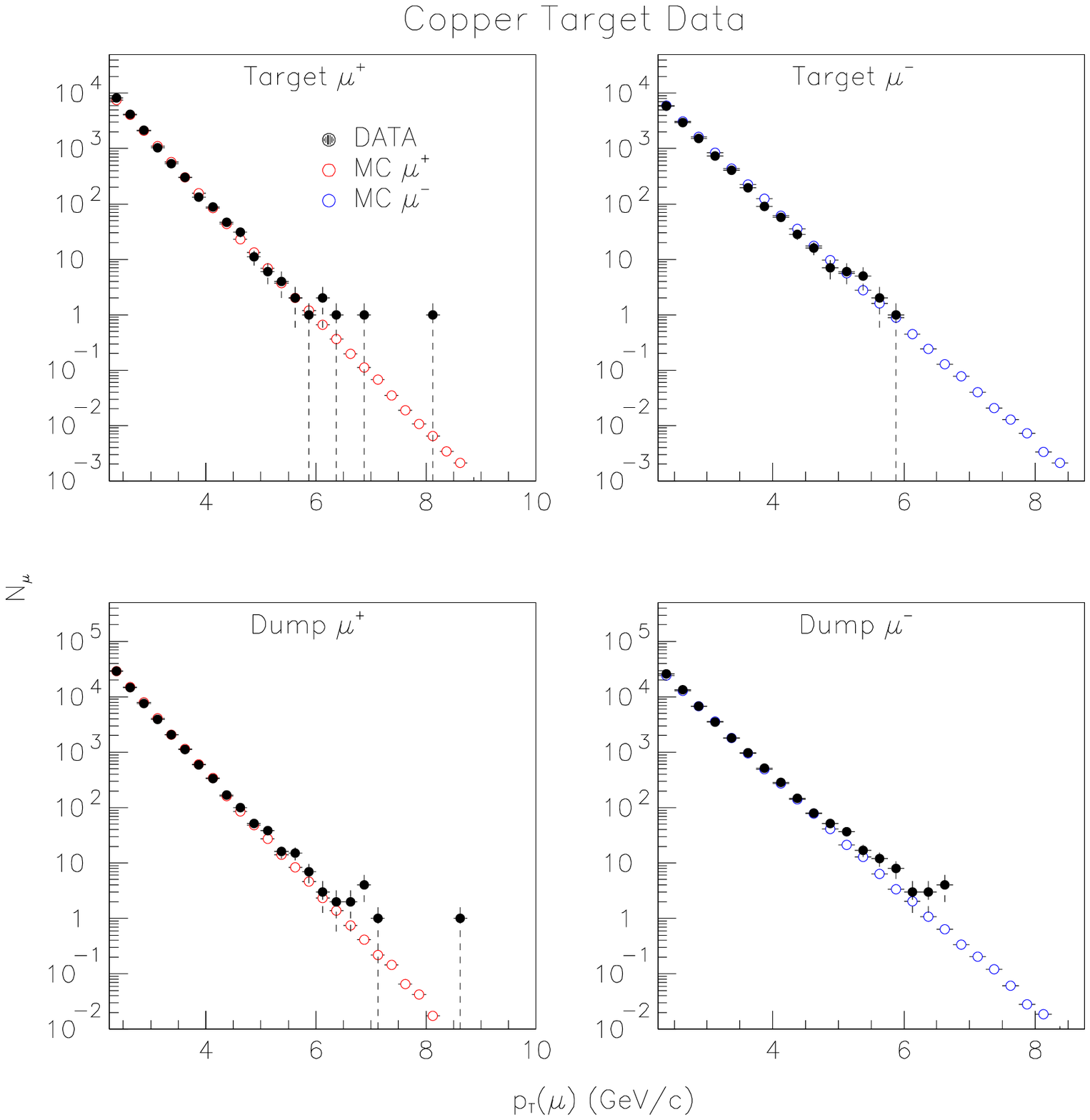}}
\caption[Cu. Data And MC From \nsfuncexp]
{The copper target data (black closed circles) and total
Monte Carlo spectra at \minchi (red open circles) from the fit
where the $D/\overline{D}$ cross sections were of the form
\nsfuncexp. Histograms are \muplus (left) and \muminus (right)
for the copper target (top) and copper dump (bottom). Errors for
the data are statistical only. Errors for the Monte Carlo include
all errors used for calculating \nschisquare (see text). Error
bars may be smaller than symbols used.}
\label{cuexp}
\end{figure}
\end{center}

\begin{center}
\begin{figure}[!ht]
\resizebox{5.8in}{6.0in}
{\includegraphics[27,158][522,668]{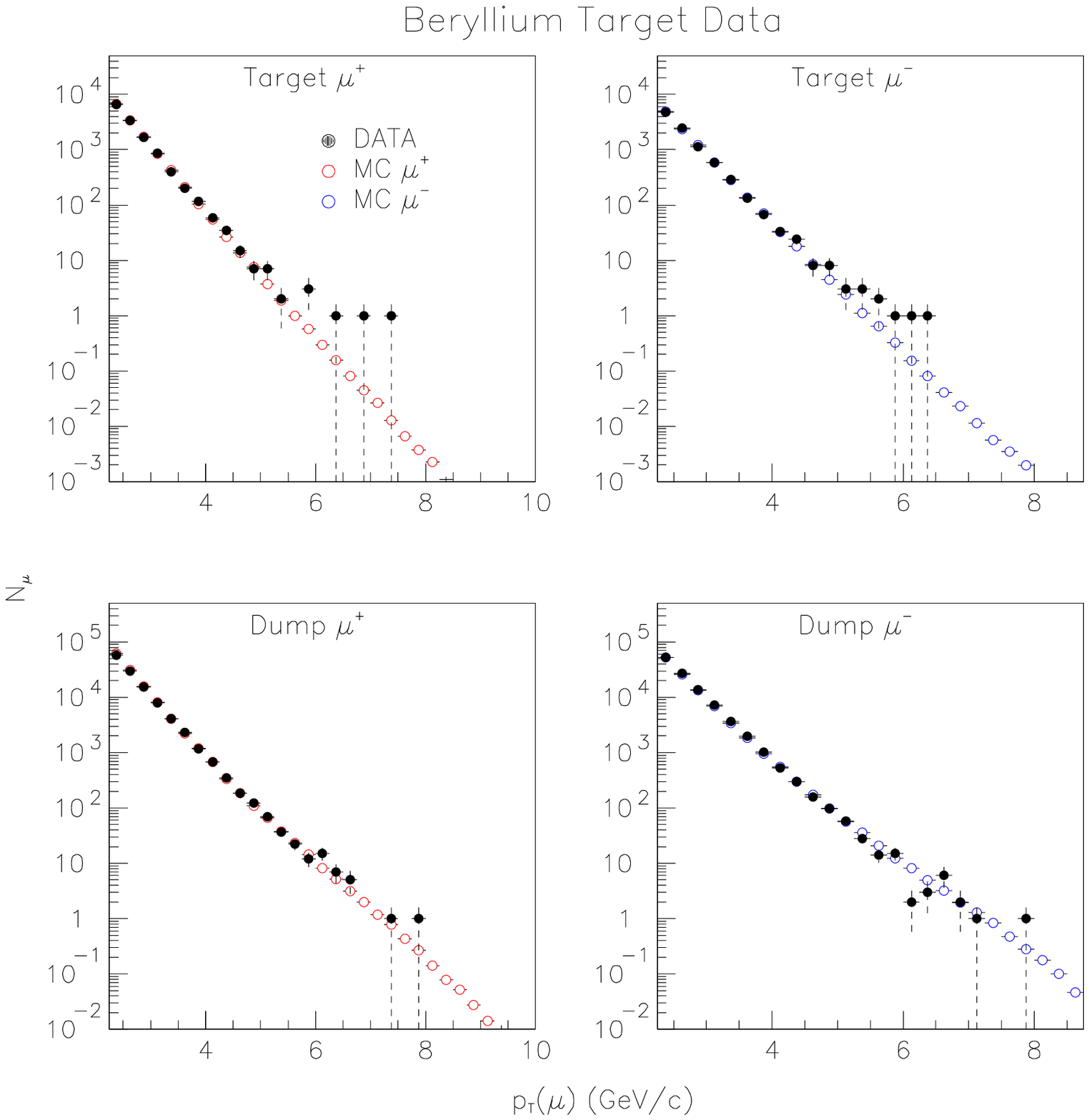}}
\caption[Be. Data And MC From \nsfuncexp]
{The beryllium target data (black closed circles) and total
Monte Carlo spectra at \minchi (red open circles) from the fit
where the $D/\overline{D}$ cross sections were of the form
\nsfuncexp. Histograms are \muplus (left) and \muminus (right)
for the beryllium target (top) and copper dump (bottom). Errors for
the data are statistical only. Errors for the Monte Carlo include
all errors used for calculating \nschisquare (see text). Error
bars may be smaller than symbols used.}
\label{beexp}
\end{figure}
\end{center}

\begin{center}
\begin{figure}[!ht]
\resizebox{5.8in}{6.0in}
{\includegraphics[27,158][522,668]{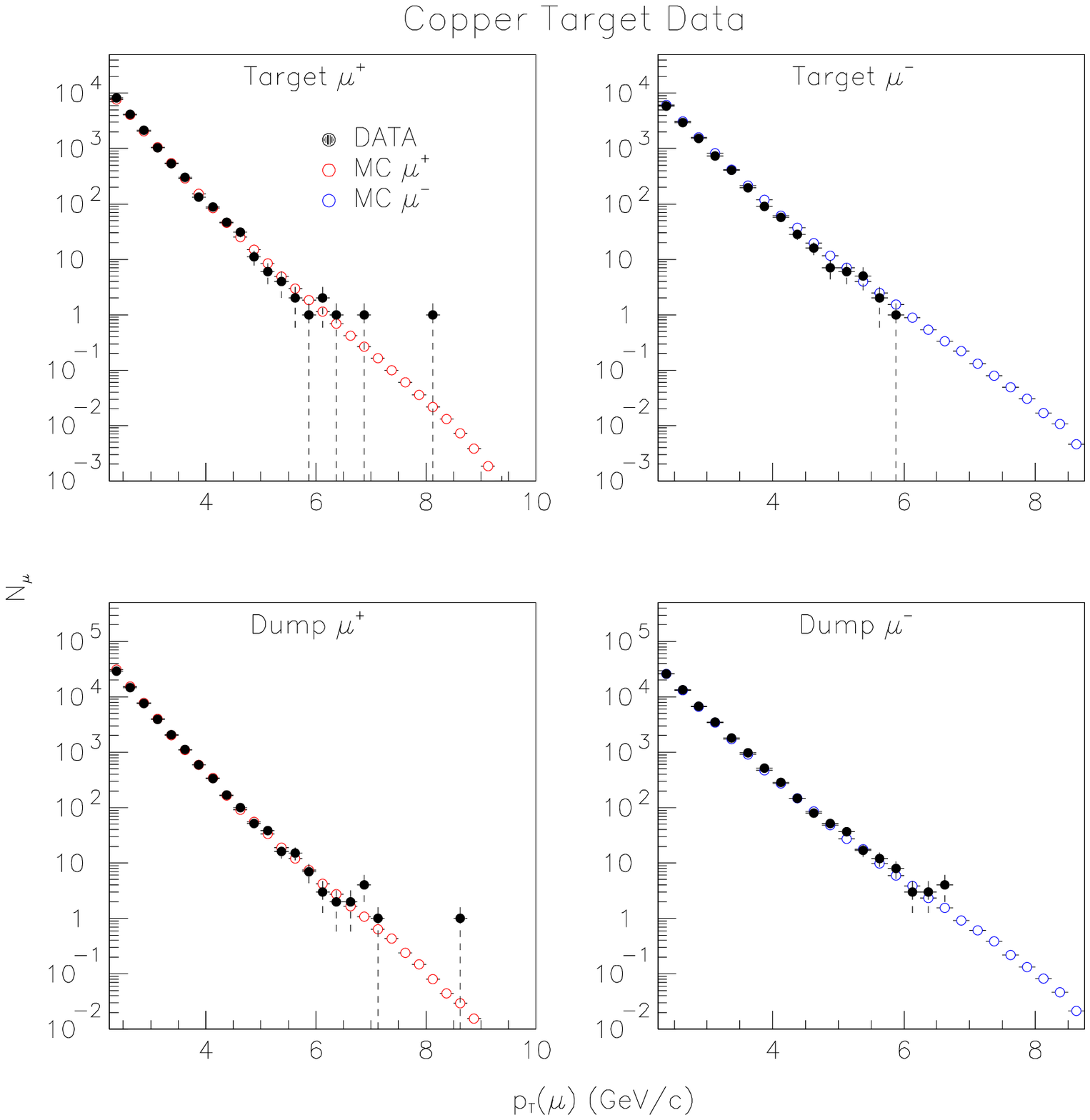}}
\caption[Cu. Data And MC From \nsfuncthree]
{The copper target data (black closed circles) and total
Monte Carlo spectra at \minchi (red open circles) from the fit
where the $D/\overline{D}$ cross sections were of the form
\nsfuncthree. The exponent $n$ was a free parameter for this fit.
Histograms are \muplus (left) and \muminus (right)
for the copper target (top) and copper dump (bottom). Errors for
the data are statistical only. Errors for the Monte Carlo include
all errors used for calculating \nschisquare (see text). Error
bars may be smaller than symbols used.}
\label{cuf3f}
\end{figure}
\end{center}

\begin{center}
\begin{figure}[!ht]
\resizebox{5.8in}{6.0in}
{\includegraphics[27,158][522,668]{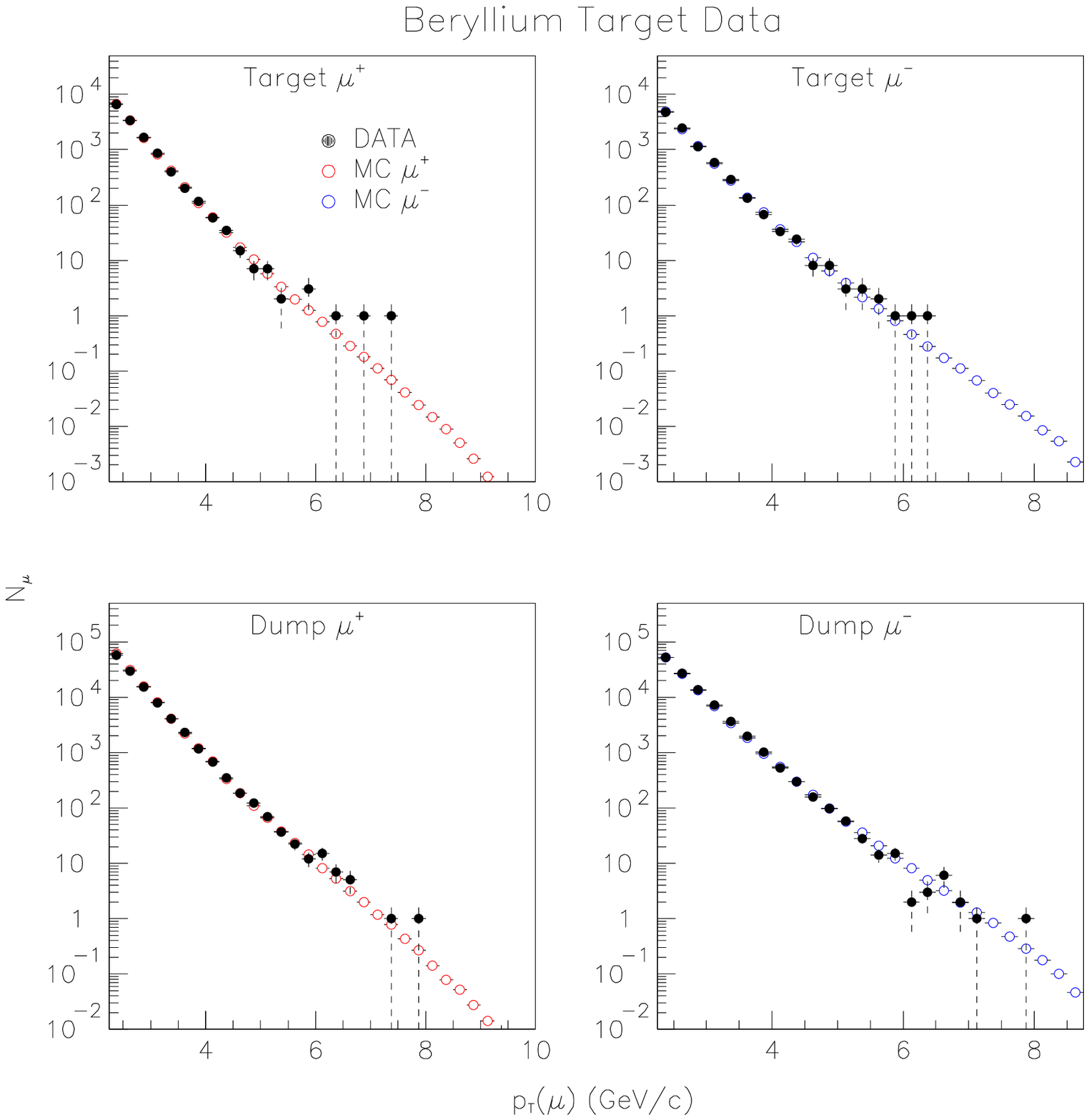}}
\caption[Be. Data And MC From \nsfuncthree]
{The beryllium target data (black closed circles) and total
Monte Carlo spectra at \minchi (red open circles) from the fit
where the $D/\overline{D}$ cross sections were of the form
\nsfuncthree. The exponent $n$ was a free parameter for this fit.
Histograms are \muplus (left) and \muminus (right)
for the beryllium target (top) and copper dump (bottom). Errors for
the data are statistical only. Errors for the Monte Carlo include
all errors used for calculating \nschisquare (see text). Error
bars may be smaller than symbols used.}
\label{bef3f}
\end{figure}
\end{center}

\begin{center}
\begin{figure}[!ht]
\resizebox{5.8in}{6.0in}
{\includegraphics[27,158][522,668]{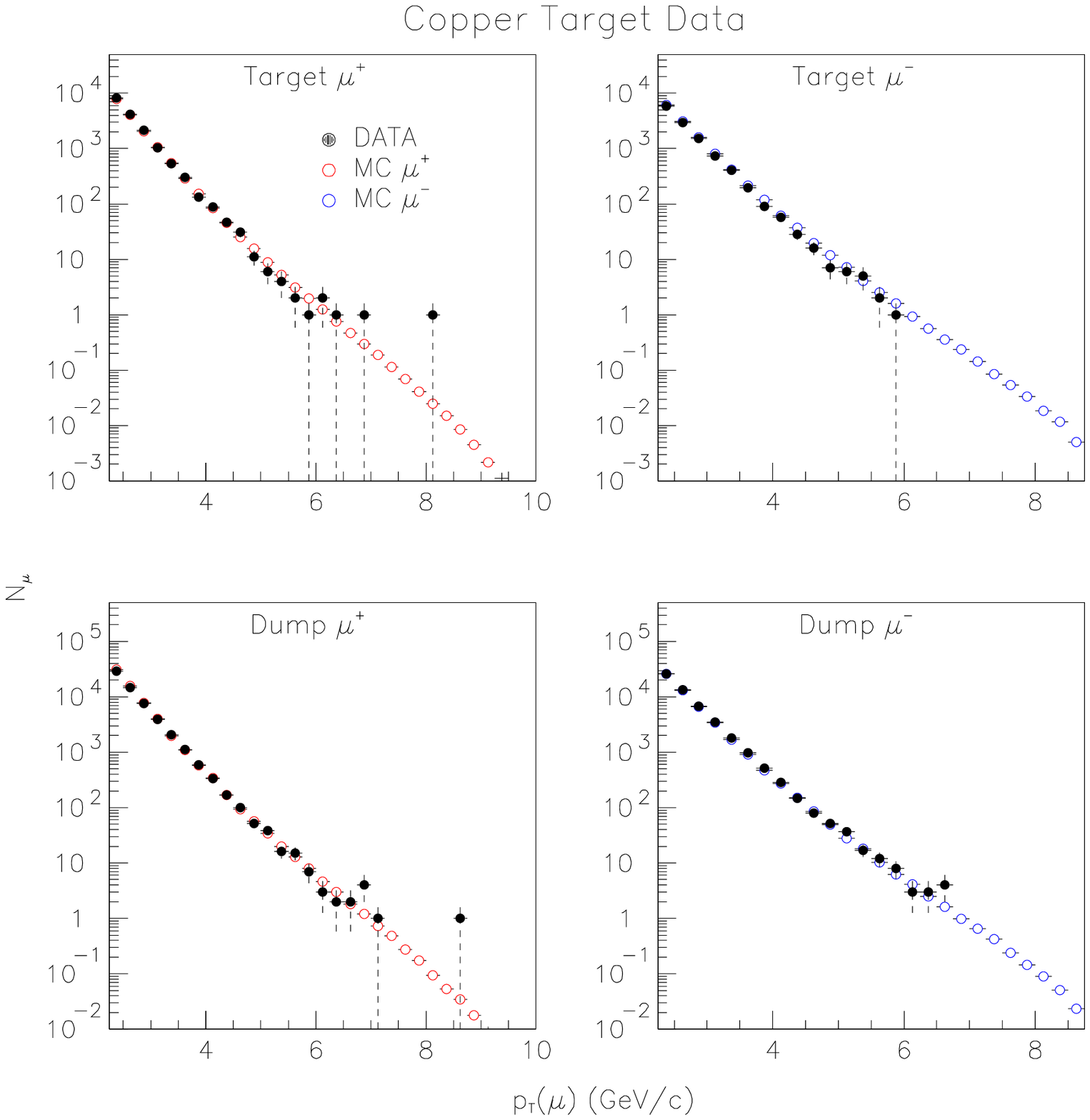}}
\caption[Cu. Data And MC From \nsfuncthree]
{The copper target data (black closed circles) and total
Monte Carlo spectra at \minchi (red open circles) from the fit
where the $D/\overline{D}$ cross sections were of the form
\nsfuncthree. The exponent $n$ was fixed at $n=6$ for this fit.
Histograms are \muplus (left) and \muminus (right)
for the copper target (top) and copper dump (bottom). Errors for
the data are statistical only. Errors for the Monte Carlo include
all errors used for calculating \nschisquare (see text). Error
bars may be smaller than symbols used.}
\label{cuf36}
\end{figure}
\end{center}

\begin{center}
\begin{figure}[!ht]
\resizebox{5.8in}{6.0in}
{\includegraphics[27,158][522,668]{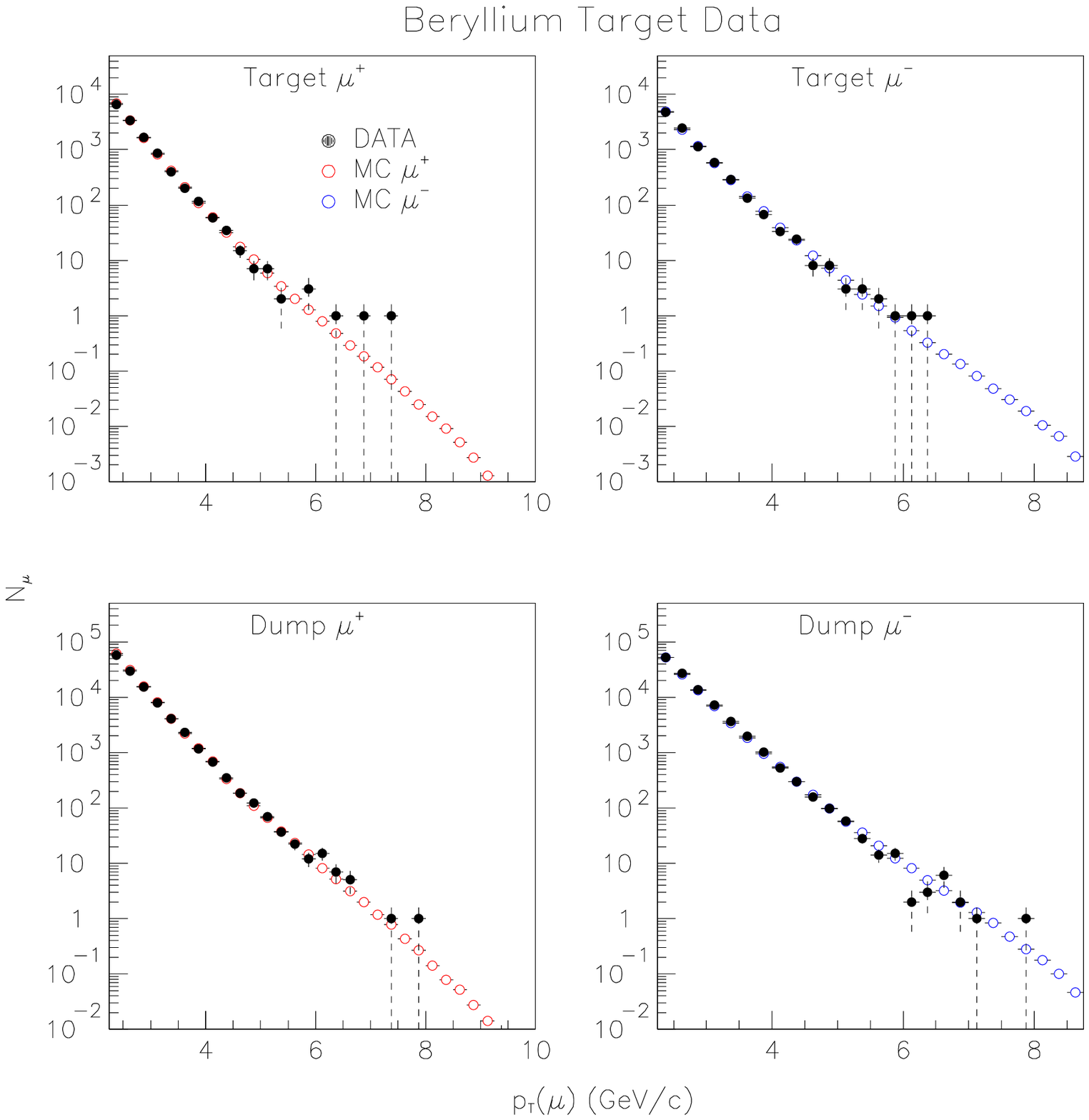}}
\caption[Be. Data And MC From \nsfuncthree]
{The beryllium target data (black closed circles) and total
Monte Carlo spectra at \minchi (red open circles) from the fit
where the $D/\overline{D}$ cross sections were of the form
\nsfuncthree. The exponent $n$ was fixed at $n=6$ for this fit.
Histograms are \muplus (left) and \muminus (right)
for the beryllium target (top) and copper dump (bottom). Errors for
the data are statistical only. Errors for the Monte Carlo include
all errors used for calculating \nschisquare (see text). Error
bars may be smaller than symbols used.}
\label{bef36}
\end{figure}
\end{center}

\begin{center}
\begin{figure}[!th]
\resizebox{5.8in}{6.0in}
{\includegraphics[27,158][522,668]{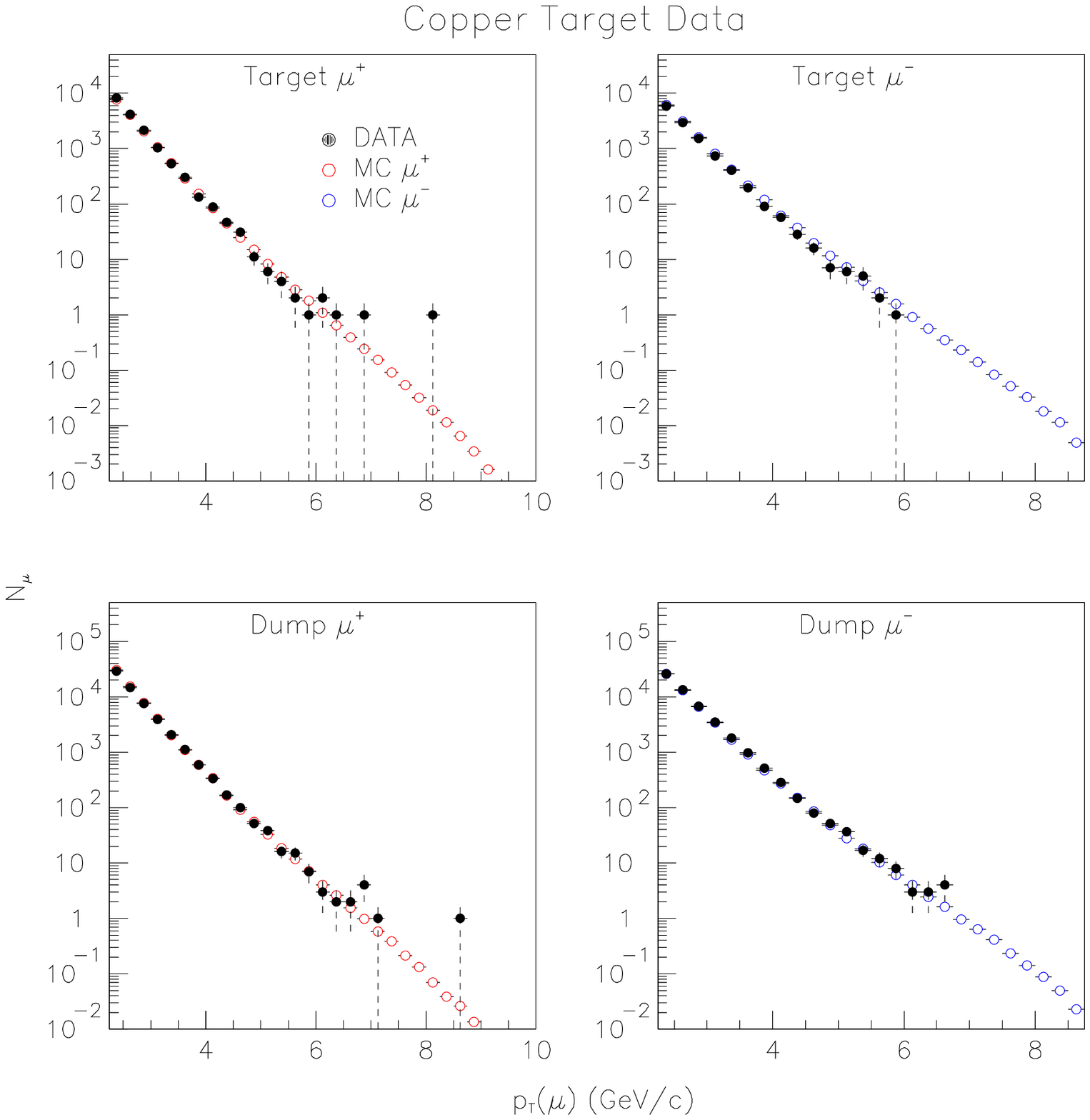}}
\caption[Cu. Data And MC From \nsfuncfour]
{The copper target data (black closed circles) and total
Monte Carlo spectra at \minchi (red open circles) from the fit
where the $D/\overline{D}$ cross sections were of the form
\nsfuncfour. The exponent $n$ was fixed at $n=6$ for this fit.
Histograms are \muplus (left) and \muminus (right)
for the copper target (top) and copper dump (bottom). Errors for
the data are statistical only. Errors for the Monte Carlo include
all errors used for calculating \nschisquare (see text). Error
bars may be smaller than symbols used.}
\label{cuf46}
\end{figure}
\end{center}

\begin{center}
\begin{figure}[!th]
\resizebox{5.8in}{6.0in}
{\includegraphics[27,158][522,668]{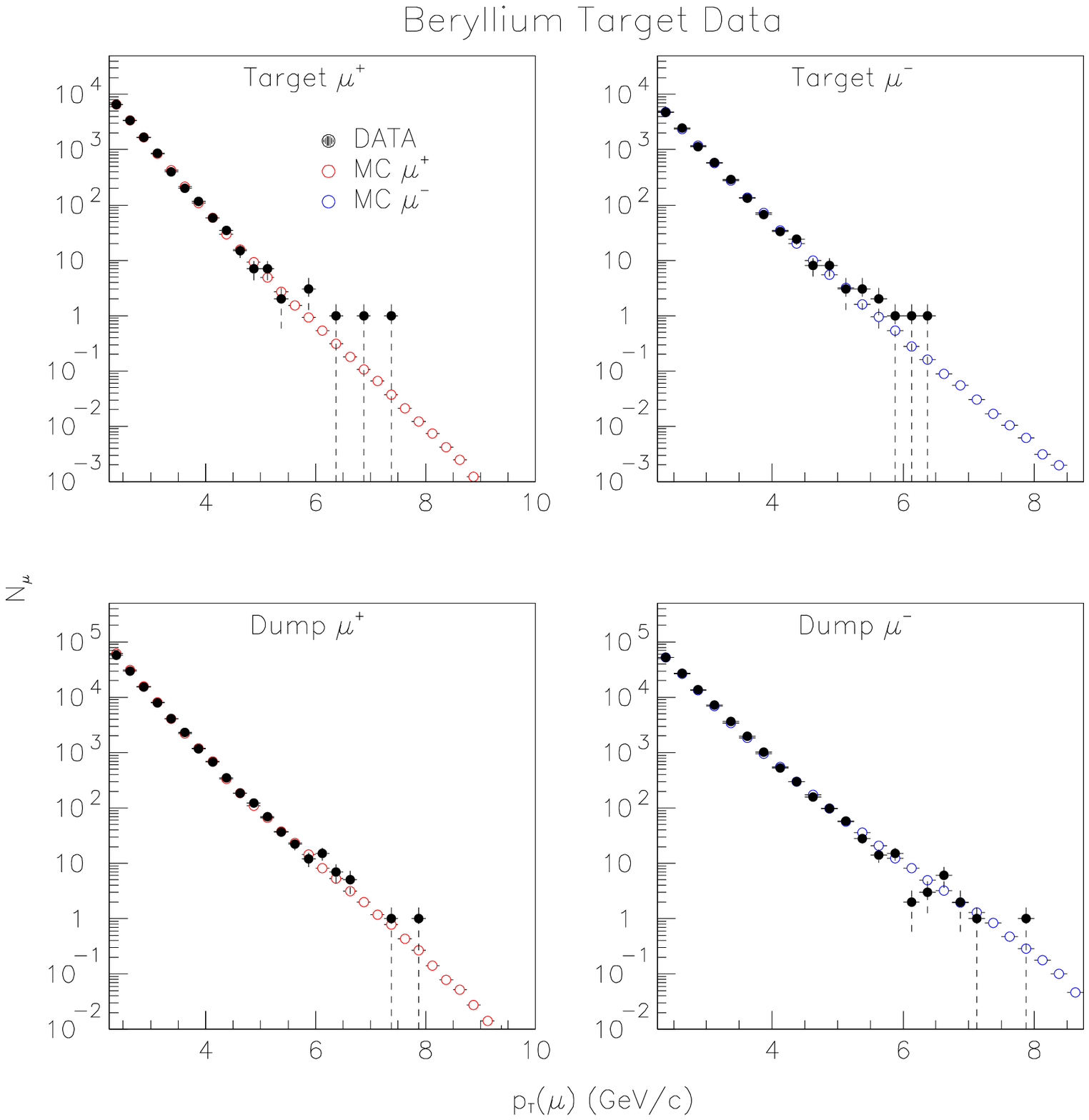}}
\caption[Be. Data And MC From \nsfuncfour]
{The beryllium target data (black closed circles) and total
Monte Carlo spectra at \minchi (red open circles) from the fit
where the $D/\overline{D}$ cross sections were of the form
\nsfuncfour. The exponent $n$ was fixed at $n=6$ for this fit.
Histograms are \muplus (left) and \muminus (right)
for the beryllium target (top) and copper dump (bottom). Errors for
the data are statistical only. Errors for the Monte Carlo include
all errors used for calculating \nschisquare (see text). Error
bars may be smaller than symbols used.}
\label{bef46}
\end{figure}
\end{center}

\begin{center}
\begin{figure}[!ht]
\resizebox{5.8in}{6.0in}
{\includegraphics[27,158][522,668]{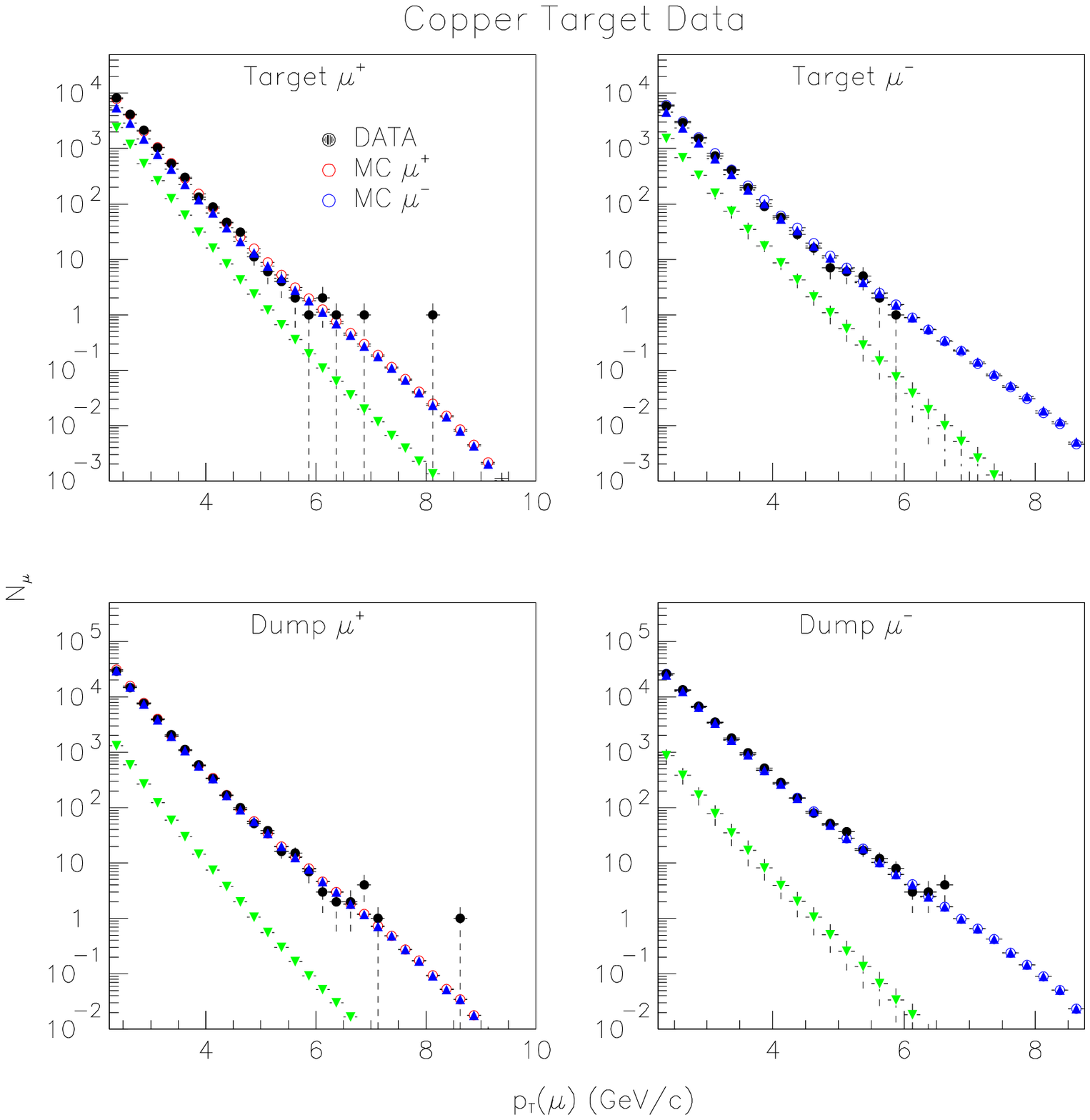}}
\caption[Cu. Data And MC Contributions
From \nsfuncthree]
{The copper target data (black closed circles) and total
Monte Carlo spectra at \minchi (red open circles) from the fit
where the $D/\overline{D}$ cross sections were of the form
\nsfuncthree. Blue triangles are the open charm contribution and
the green upside-down triagles are the contribution from light
hadrons. The exponent $n$ was fixed at $n=6$ for this fit.
Histograms are \muplus (left) and \muminus (right)
for the copper target (top) and copper dump (bottom). Errors for
the data are statistical only. Errors for the Monte Carlo include
all errors used for calculating \nschisquare (see text).
Contributions contain all systematic errors used in
calculation of the weights added in quadrature to the statistical
error, but do not include errors of the fit. Error
bars may be smaller than symbols used.}
\label{cu-w-light}
\end{figure}
\end{center}

Figures \ref{cuexp} through \ref{bef46} show the total Monte Carlo
spectra plotted against the data for four selected fits: the
exponential function, the 3 parameter polynomial function with $n$
free and $n$ fixed to $n=6$, and the 4 parameter polynomial
function with $n$ fixed to $n=6$. All Monte Carlo spectra are those
at \minchi for the function being fitted. Errors for the Monte
Carlo spectra include all statistical plus all systematical errors
as explained previously. The Monte Carlo errors do not include the
errors for the parameters of the input open charm corss sections.
All data spectra have statistical errors only. Figure
\ref{cu-w-light} is the same as figure \ref{cuf36} except the
open charm and light contributions are included.

\renewcommand{\arraystretch}{1.5}

\begin{table}[ht]
\caption[Parameter Values For \nsfuncexp]{Parameter values from
fits to the data using \funcexp as the input shape of the open
charm differential cross section. Values for $D$ are top line
and $\overline{D}$ are the bottom line. Errors
are the 70 percent confidence level errors as described in Chapter
\ref{errorcode}.}
\label{exp-table}
\begin{center}
\begin{tabular}
[c]{|clclcl|c|c|}\hline

\multicolumn{8}{|l|}{Copper Target Data}\\

\hline

\multicolumn{2}{|c}{$A_{1}$} &
 \multicolumn{2}{c}{$B$} &
 \multicolumn{2}{c|}{$N$} &
 \multicolumn{1}{c|}{\nsminchi} &
 \multicolumn{1}{c|}{\nspdfchi}\\

\hline

1.91 &
 $\latop{+ \, 0.32}{- \, 0.27}$ &
 1.91 &
 $\latop{+ \, 0.04}{- \, 0.04}$ &
  &
  &
  & \\

1.83 &
 $ \latop{+ \, 0.34}{- \, 0.28} $ &
 1.94 &
 $ \latop{+ \, 0.04}{- \, 0.04} $ &
 \raisebox{2.0ex}[0pt]{22.48} &
 \raisebox{2.0ex}[0pt]{$ \latop{0.23}{0.47} $} &
 \raisebox{2.0ex}[0pt]{81.4} &
 \raisebox{2.0ex}[0pt]{1.3}\\

\hline

\multicolumn{8}{c}{}\\

\hline

\multicolumn{8}{|l|}{Beryllium Target Data}\\

\hline

\multicolumn{2}{|c}{$A_{1}$} &
 \multicolumn{2}{c}{$B$} &
 \multicolumn{2}{c|}{$N$} &
 \multicolumn{1}{c|}{\nsminchi} &
 \multicolumn{1}{c|}{\nspdfchi}\\

\hline

0.60 &
 $\latop{+ \, 0.14}{- \, 0.23}$ &
 2.11 &
 $\latop{+ \, 0.06}{- \, 0.12}$ &
  &
  &
  & \\

0.57 &
 $ \latop{+ \, 0.16}{- \, 0.21} $ &
 2.18 &
 $ \latop{+ \, 0.07}{- \, 0.12} $ &
 \raisebox{2.0ex}[0pt]{42.11} &
 \raisebox{2.0ex}[0pt]{$ \latop{0.10}{0.66} $} &
 \raisebox{2.0ex}[0pt]{62.9} &
 \raisebox{2.0ex}[0pt]{1.0}\\

\hline

\end{tabular}
\end{center}
\end{table}

\renewcommand{\arraystretch}{1.0}

\renewcommand{\arraystretch}{1.5}

\begin{table}[!ht]
\caption[Parameter Values For \nsfuncthree]{Parameter values
for $D$ (top line) and $\overline{D}$ (second line) from
fits to the data where the differential cross section
was input as \nsfuncthree. Top two lines of both target data sets
are the full float minimization. Errors are calculated for 70
percent confidence level as explained in Chapter \ref{errorcode}.}
\label{f3-table}
\begin{center}
\begin{tabular}
[c]{|clclccl|c|c|}\hline

\multicolumn{9}{|l|}{Copper Target Data}\\

\multicolumn{2}{|c}{$A_{2}$} &
 \multicolumn{2}{c}{$\alpha$} &
 \multicolumn{1}{c}{$n$} &
 \multicolumn{2}{c}{$N$} &
 \multicolumn{1}{c}{$\chi^{2}$} &
 \multicolumn{1}{c|}{\nspdfchi}\\

\hline

\multicolumn{2}{|c}{649710} &
 \multicolumn{2}{c}{4.14} &
 6.32 &
 \multicolumn{2}{c|}{} &
  & \\

\multicolumn{2}{|c}{286500} &
 \multicolumn{2}{c}{3.77} &
 6.19 &
 \multicolumn{2}{c|}{\raisebox{2.0ex}[0pt]{21.84}} &
 \raisebox{2.0ex}[0pt]{54.09} &
 \raisebox{2.0ex}[0pt]{0.92}\\

\hline

37.52 &
 $\latop{+ \, 1.62}{- \, 3.45}$ &
 0.07 &
 $\latop{+ \, 0.01}{- \, 0.07}$ &
  &
 26.03 &
 $\latop{+ \, 0.76}{- \, 0.68}$ &
  & \\

31.91 &
 $\latop{+ \, 1.27}{- \, 2.81}$ &
 0.06 &
 $\latop{+ \, 0.01}{- \, 0.06}$ &
 \raisebox{2.0ex}[0pt]{4.0} &
 26.03 &
 $\latop{+ \, 0.55}{- \, 0.55}$ &
 \raisebox{2.0ex}[0pt]{305.41} &
 \raisebox{2.0ex}[0pt]{5.01}\\

\hline

2334 &
 $\latop{+ \, 217}{- \, 205}$ &
 1.71 &
 $\latop{+ \, 0.12}{- \, 0.13}$ &
  &
 22.39 &
 $\latop{+ \, 0.17}{- \, 0.47}$ &
  & \\

1931 &
 $\latop{+ \, 196}{- \, 181}$ &
 1.66 &
 $\latop{+ \, 0.13}{- \, 0.14}$ &
 \raisebox{2.0ex}[0pt]{$5.0$} &
 22.39 &
 $\latop{+ \, 0.11}{- \, 0.35}$ &
 \raisebox{2.0ex}[0pt]{99.58} &
 \raisebox{2.0ex}[0pt]{1.63}\\

\hline

161765 &
 $\latop{+ \, 19378}{- \, 15265}$ &
 3.53 &
 $\latop{+ \, 0.17}{- \, 0.17}$ &
  &
 21.70 &
 $\latop{+ \, 0.11}{- \, 0.34}$ &
  & \\

130126 &
 $\latop{+ \, 16808}{- \, 15265}$ &
 3.44 &
 $\latop{+ \, 0.18}{- \, 0.19}$ &
 \raisebox{2.0ex}[0pt]{$6.0$} &
 21.70 &
 $\latop{+ \, 0.11}{- \, 0.23}$ &
 \raisebox{2.0ex}[0pt]{56.08} &
 \raisebox{2.0ex}[0pt]{0.92}\\

\hline

\multicolumn{9}{c}{}\\

\hline

\multicolumn{9}{|l|}{Beryllium Target Data}\\

\multicolumn{2}{|c}{$A_{2}$} &
 \multicolumn{2}{c}{$\alpha$} &
 \multicolumn{1}{c}{$n$} &
 \multicolumn{2}{c}{$N$} &
 \multicolumn{1}{c}{\nsminchi} &
 \multicolumn{1}{c|}{\nspdfchi}\\

\hline

\multicolumn{2}{|c}{19238} &
 \multicolumn{2}{c}{3.03} &
 6.06 &
 \multicolumn{2}{c|}{} &
  & \\

\multicolumn{2}{|c}{17013} &
 \multicolumn{2}{c}{2.92} &
 6.14 &
 \multicolumn{2}{c|}{\raisebox{2.0ex}[0pt]{42.19}} &
 \raisebox{2.0ex}[0pt]{60.14} &
 \raisebox{2.0ex}[0pt]{0.99}\\

\hline

6.50 &
 $\latop{+ \, 0.34}{- \, 0.83}$ &
 0.08 &
 $\latop{+ \, 0.01}{- \, 0.08}$ &
  &
 42.39 &
 $\latop{+ \, 0.67}{- \, 0.67}$ &
  & \\

4.88 &
 $\latop{+ \, 0.25}{- \, 0.65}$ &
 0.08 &
 $\latop{+ \, 0.01}{- \, 0.08}$ &
 \raisebox{2.0ex}[0pt]{$4.0$} &
 42.39 &
 $\latop{+ \, 0.66}{- \, 0.67}$ &
 \raisebox{2.0ex}[0pt]{167.32} &
 \raisebox{2.0ex}[0pt]{2.66}\\

\hline

289 &
 $\latop{+ \, 29}{- \, 75}$ &
 1.46 &
 $\latop{+ \, 0.13}{- \, 0.39}$ &
  &
 41.97 &
 $\latop{+ \, 0.20}{- \, 0.46}$ &
  & \\

225 &
 $\latop{+ \, 13}{- \, 100}$ &
 1.50 &
 $\latop{+ \, 0.08}{- \, 0.73}$ &
 \raisebox{2.0ex}[0pt]{$5.0$} &
 41.97 &
 $\latop{+ \, 0.24}{- \, 0.92}$ &
 \raisebox{2.0ex}[0pt]{84.44} &
 \raisebox{2.0ex}[0pt]{1.34}\\

\hline

15280 &
 $\latop{+ \, 3847}{- \, 3046}$ &
 2.94 &
 $\latop{+ \, 0.32}{- \, 0.30}$ &
  &
 42.13 &
 $\latop{+ \, 0.11}{- \, 0.44}$ &
  & \\

11280 &
 $\latop{+ \, 1546}{- \, 4055}$ &
 2.91 &
 $\latop{+ \, 0.18}{- \, 0.58}$ &
 \raisebox{2.0ex}[0pt]{$6.0$} &
 42.13 &
 $\latop{+ \, 0.11}{- \, 0.67}$ &
 \raisebox{2.0ex}[0pt]{62.19} &
 \raisebox{2.0ex}[0pt]{0.99}\\

\hline

\end{tabular}
\end{center}
\end{table}

\renewcommand{\arraystretch}{1.0}

\renewcommand{\arraystretch}{1.5}

\begin{table}[!ht]
\caption[Parameter Values For \nsfuncfour]{The \minchi parameter
values when the open charm differential cross section was input
in the form \nsfuncfour and fitted to the copper target data.
Top line is for $D$ and second line
is for $\overline{D}$. All fits had $n$ fixed to the values
shown. $p_{beam}=19.38$ \cmom for this analysis. Errors were
calculated holding $\alpha$ fixed at its \minchi
value (top pair of lines for each $n$) and holding $m$ fixed at
its \minchi value (bottom pair of lines for each $n$) because
of strong correlations. Errors are calculated for 70 percent
confidence level as explained in Chapter \ref{errorcode}.}
\label{f4-cu-table}
\begin{center}
\begin{tabular}
[c]{|clclcclcl|c|c|}\hline

\multicolumn{11}{|l|}{Copper Target Data}\\

\multicolumn{2}{|c}{$A_{2}$} &
 \multicolumn{2}{c}{$\alpha$} &
 \multicolumn{1}{c}{$n$} &
 \multicolumn{2}{c}{$m$} &
 \multicolumn{2}{c}{$N$} &
 \multicolumn{1}{c}{\nsminchi} &
 \multicolumn{1}{c|}{\nspdfchi}\\

\hline

1412 &
 $\latop{+ \, 180}{- \, 171}$ &
 2.75 &
  &
  &
 8.00 &
 $\latop{+ \, 0.47}{- \, 0.48}$ &
  &
 $\latop{+ \, 0.22}{- \, 0.33}$ &
  & \\

1616 &
 $\latop{+ \, 217}{- \, 220}$ &
 2.99 &
  &
  &
 8.80 &
 $\latop{+ \, 0.51}{- \, 0.57}$ &
  &
 $\latop{+ \, 0.11}{- \, 0.33}$ &
  & \\

1412 &
 $\latop{+ \, 158}{- \, 156}$ &
 2.75 &
 $\latop{+ \, 0.22}{- \, 0.24}$ &
  &
 8.00 &
  &
  &
 $\latop{+ \, 0.16}{- \, 0.23}$ &
  & \\

1616 &
 $\latop{+ \, 196}{- \, 206}$ &
 2.99 &
 $\latop{+ \, 0.25}{- \, 0.29}$ &
 \raisebox{6.1ex}[0pt]{$4.0$} &
 8.80 &
  &
 \raisebox{6.1ex}[0pt]{21.02} &
 $\latop{+ \, 0.10}{- \, 0.22}$ &
 \raisebox{6.1ex}[0pt]{58.48} &
 \raisebox{6.1ex}[0pt]{0.99} \\

\hline

15959 &
 $\latop{+ \, 2120}{- \, 1774}$ &
 3.30 &
  &
  &
 4.17 &
 $\latop{+ \, 0.47}{- \, 0.44}$ &
 &
 $\latop{+ \, 0.22}{- \, 0.46}$ &
  & \\

10804 &
 $\latop{+ \, 1520}{- \, 1283}$ &
 3.02 &
  &
  &
 3.81 &
 $\latop{+ \, 0.51}{- \, 0.51}$ &
  &
 $\latop{+ \, 0.10}{- \, 0.24}$ &
  & \\

15959 &
 $\latop{+ \, 1940}{- \, 1740}$ &
 3.30 &
 $\latop{+ \, 0.20}{- \, 0.21}$ &
  &
 4.17 &
  &
  &
 $\latop{+ \, 0.12}{- \, 0.35}$ &
  & \\

10804 &
 $\latop{+ \, 1371}{- \, 1243}$ &
 3.02 &
 $\latop{+ \, 0.21}{- \, 0.21}$ &
 \raisebox{6.1ex}[0pt]{$5.0$} &
 3.81 &
  &
 \raisebox{6.1ex}[0pt]{21.95} &
 $\latop{+ \, 0.12}{- \, 0.23}$ &
 \raisebox{6.1ex}[0pt]{54.61} &
 \raisebox{6.1ex}[0pt]{0.93} \\

\hline

300220 &
 $\latop{+ \, 39248}{- \, 33390}$ &
 4.07 &
  &
  &
 1.29 &
 $\latop{+ \, 0.46}{- \, 0.43}$ &
  &
 $\latop{+ \, 0.20}{- \, 0.36}$ &
  & \\

137891 &
 $\latop{+ \, 19998}{- \, 8836}$ &
 3.48 &
  &
  &
 0.13 &
 $\latop{+ \, 0.52}{- \, 0.13}$ &
 &
 $\latop{+ \, 0.34}{- \, 0.24}$ &
  & \\

300220 &
 $\latop{+ \, 39002}{- \, 34736}$ &
 4.07 &
 $\latop{+ \, 0.20}{- \, 0.20}$ &
  &
 1.29 &
  &
  &
 $\latop{+ \, 0.11}{- \, 0.34}$ &
  & \\

137891 &
 $\latop{+ \, 18010}{- \, 16242}$ &
 3.48 &
 $\latop{+ \, 0.19}{- \, 0.19}$ &
 \raisebox{6.1ex}[0pt]{$6.0$} &
 0.13 &
  &
 \raisebox{6.1ex}[0pt]{21.75} &
 $\latop{+ \, 0.11}{- \, 0.23}$ &
 \raisebox{6.1ex}[0pt]{54.23} &
 \raisebox{6.1ex}[0pt]{0.92} \\

\hline

\end{tabular}
\end{center}
\end{table}

\renewcommand{\arraystretch}{1.0}

\renewcommand{\arraystretch}{1.5}

\begin{table}[!ht]
\caption[Parameter Values For \nsfuncfour]{The \minchi parameter
values when the open charm differential cross section was input
in the form \nsfuncfour and fitted to the beryllium target data.
Top line is for $D$ and second line
is for $\overline{D}$. All fits had $n$ fixed to the values
shown. $p_{beam}=19.38$ \cmom for this analysis. Errors were
calculated holding $\alpha$ fixed at its \minchi
value (top pair of lines for each $n$) and holding $m$ fixed at
its \minchi value (bottom pair of lines for each $n$) because
of strong correlations. Errors are calculated for 70 percent
confidence level as explained in Chapter \ref{errorcode}.}
\label{f4-be-table}
\begin{center}
\begin{tabular}
[c]{|clclcclcl|c|c|}\hline

\multicolumn{11}{|l|}{Beryllium Target Data}\\

\multicolumn{2}{|c}{$A_{2}$} &
 \multicolumn{2}{c}{$\alpha$} &
 \multicolumn{1}{c}{$n$} &
 \multicolumn{2}{c}{$m$} &
 \multicolumn{2}{c}{$N$} &
 \multicolumn{1}{c}{\nsminchi} &
 \multicolumn{1}{c|}{\nspdfchi}\\

\hline

508 &
 $\latop{+ \, 153}{- \, 117}$ &
 3.34 &
  &
  &
 11.23 &
 $\latop{+ \, 1.17}{- \, 1.15}$ &
  &
 $\latop{+ \, 0.10}{- \, 0.67}$ &
  & \\

402 &
 $\latop{+ \, 136}{- \, 97}$ &
 3.15 &
  &
  &
 11.87 &
 $\latop{+ \, 1.33}{- \, 1.24}$ &
  &
 $\latop{+ \, 0.11}{- \, 0.44}$ &
  & \\

508 &
 $\latop{+ \, 142}{- \, 119}$ &
 3.34 &
 $\latop{+ \, 0.57}{- \, 0.57}$ &
  &
 11.23 &
  &
  &
 $\latop{+ \, 0.07}{- \, 0.44}$ &
  & \\

402 &
 $\latop{+ \, 126}{- \, 100}$ &
 3.15 &
 $\latop{+ \, 0.60}{- \, 0.58}$ &
 \raisebox{6.1ex}[0pt]{$4.0$} &
 11.87 &
  &
 \raisebox{6.1ex}[0pt]{42.18} &
 $\latop{+ \, 0.04}{- \, 0.44}$ &
 \raisebox{6.1ex}[0pt]{59.30} &
 \raisebox{6.1ex}[0pt]{0.97} \\

\hline

4151 &
 $\latop{+ \, 1228}{- \, 903}$ &
 3.46 &
  &
  &
 6.50 &
 $\latop{+ \, 1.15}{- \, 1.06}$ &
  &
 $\latop{+ \, 0.10}{- \, 0.45}$ &
  & \\

3483 &
 $\latop{+ \, 1157}{- \, 810}$ &
 3.37 &
  &
  &
 7.25 &
 $\latop{+ \, 1.30}{- \, 1.18}$ &
  &
 $\latop{+ \, 0.10}{- \, 0.44}$ &
  & \\

4151 &
 $\latop{+ \, 1167}{- \, 946}$ &
 3.46 &
 $\latop{+ \, 0.46}{- \, 0.45}$ &
  &
 6.50 &
  &
  &
 $\latop{+ \, 0.11}{- \, 0.45}$ &
  & \\

3483 &
 $\latop{+ \, 1105}{- \, 860}$ &
 3.37 &
 $\latop{+ \, 0.50}{- \, 0.47}$ &
 \raisebox{6.1ex}[0pt]{$5.0$} &
 7.25 &
  &
 \raisebox{6.1ex}[0pt]{42.18} &
 $\latop{+ \, 0.05}{- \, 0.44}$ &
 \raisebox{6.1ex}[0pt]{58.44} &
 \raisebox{6.1ex}[0pt]{0.96} \\

\hline

67537 &
 $\latop{+ \, 19855}{- \, 14577}$ &
 4.11 &
  &
  &
 3.31 &
 $\latop{+ \, 1.13}{- \, 1.05}$ &
  &
 $\latop{+ \, 0.19}{- \, 0.45}$ &
  & \\

68581 &
 $\latop{+ \, 22785}{- \, 16101}$ &
 4.19 &
  &
  &
 4.49 &
 $\latop{+ \, 1.32}{- \, 1.20}$ &
  &
 $\latop{+ \, 0.21}{- \, 0.45}$ &
  & \\

67537 &
 $\latop{+ \, 20187}{- \, 16248}$ &
 4.11 &
 $\latop{+ \, 0.43}{- \, 0.43}$ &
  &
 3.31 &
  &
  &
 $\latop{+ \, 0.10}{- \, 0.67}$ &
  & \\

68581 &
 $\latop{+ \, 23679}{- \, 18335}$ &
 4.19 &
 $\latop{+ \, 0.49}{- \, 0.48}$ &
 \raisebox{6.1ex}[0pt]{$6.0$} &
 4.49 &
  &
 \raisebox{6.1ex}[0pt]{42.18} &
 $\latop{+ \, 0.05}{- \, 0.44}$ &
 \raisebox{6.1ex}[0pt]{58.25} &
 \raisebox{6.1ex}[0pt]{0.95} \\

\hline

\end{tabular}
\end{center}
\end{table}

\renewcommand{\arraystretch}{1.0}

\clearpage

\section{The Open Charm And Anti-charm Cross Sections}

\begin{center}
\begin{figure}[!ht]
\resizebox{5.9in}{6.13in}
{\includegraphics[64,202][550,707]{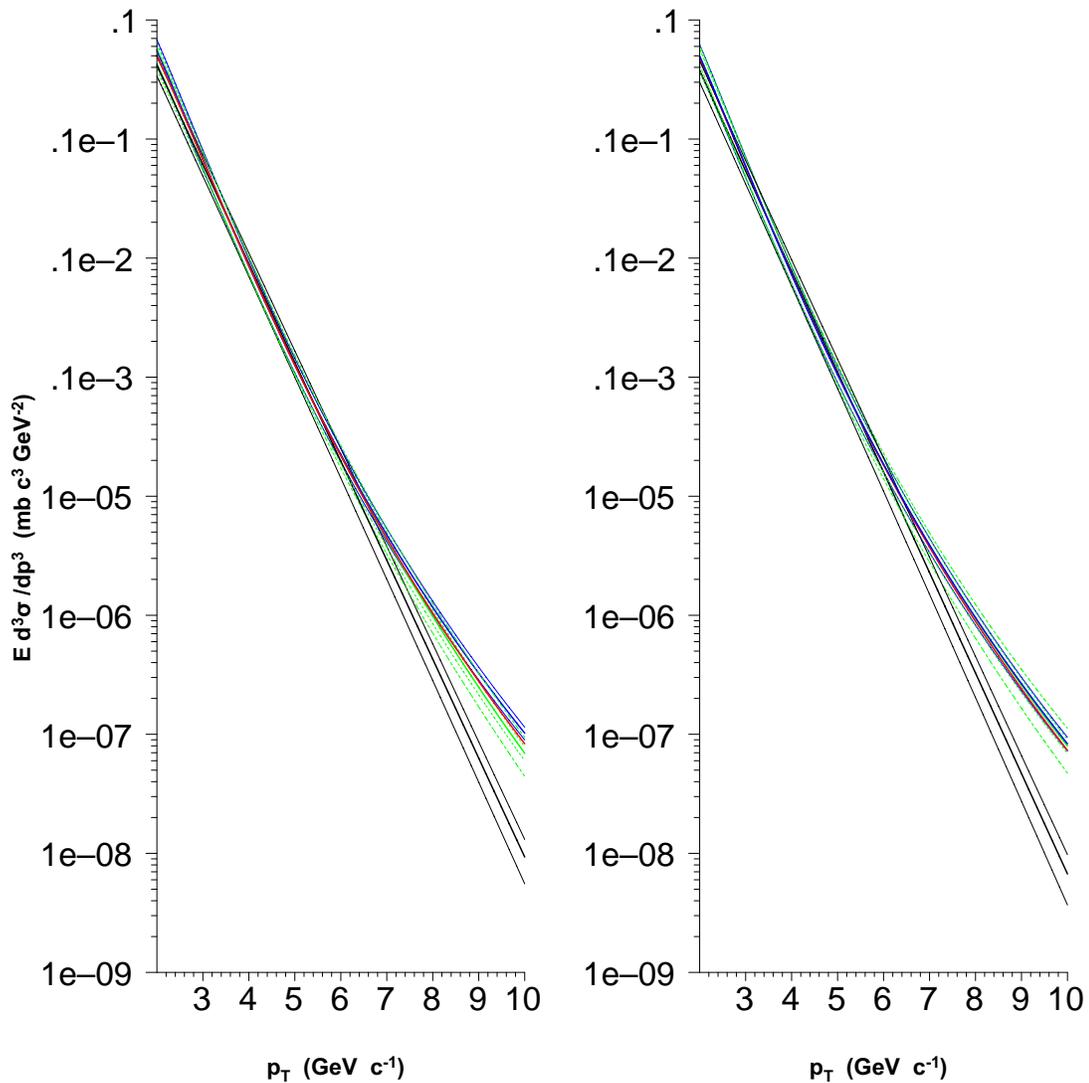}}
\caption[Open Charm \pcu Cross Sections]{The differential cross
sections for $D$ (left) and $\overline{D}$ (right) determined
from fits to the copper data for the various functions:
exponential (black), 3
parameter with $n$ a free parameter
(red), the 3 parameter with $n$ fixed to $n=6$ (blue) and
the 4 parameter with $n=6$ (green). Errors shown include
the error on $N$ added in quadrature.}
\label{cuallwe}
\end{figure}
\end{center}

\begin{center}
\begin{figure}[!ht]
\resizebox{5.8in}{6.23in}
{\includegraphics[66,196][552,718]{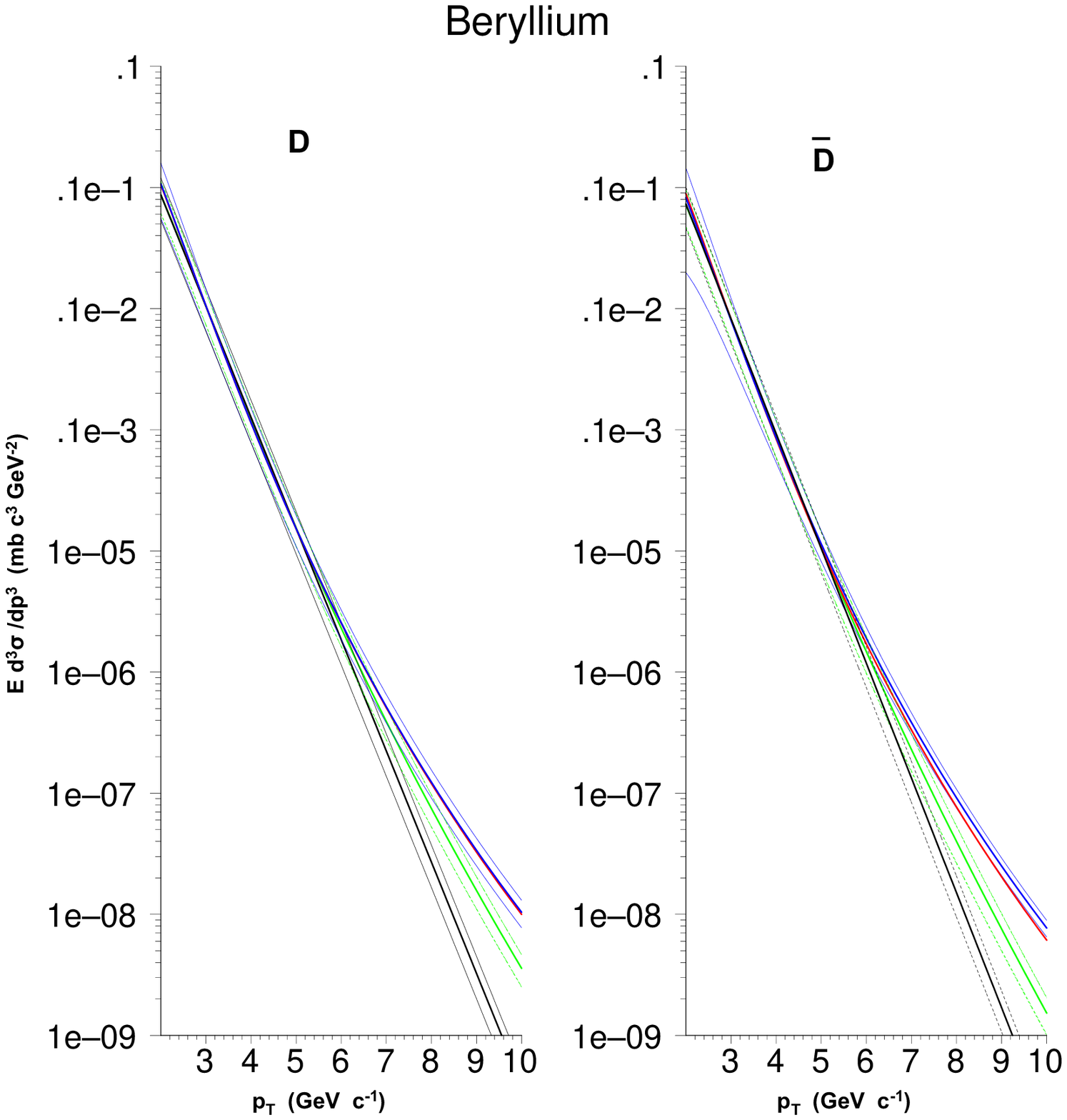}}
\caption[Open Charm \pbe Cross Sections]{The differential cross
sections for $D$ (left) and $\overline{D}$ (right) determined
from fits to the beryllium data for the various function:
exponential (black), 3
parameter with $n$ a free parameter
(red), the 3 parameter with $n$ fixed to $n=6$ (blue) and
the 4 parameter with $n=6$ (green). Errors shown include
the error on $N$ added in quadrature.}
\label{beallwe}
\end{figure}
\end{center}

\begin{center}
\begin{figure}[!ht]
\resizebox{5.6in}{6.46in}
{\includegraphics[83,181][551,721]{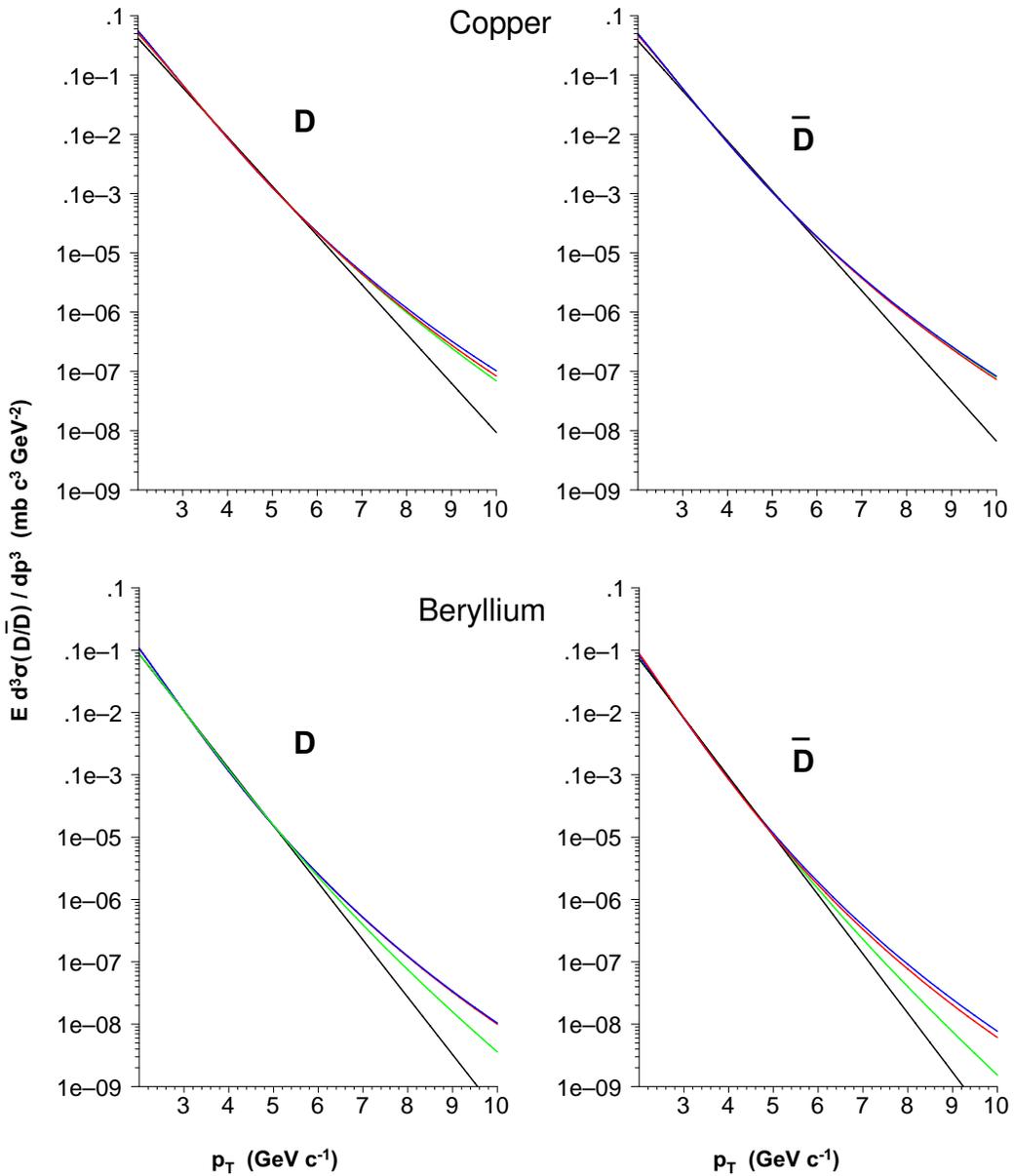}}
\caption[The \pcu And \pbe Cross Sections Without Errors]{The \pcu
(top) and \pbe (bottom) $D$ (left) and $\overline{D}$ (right)
differential
cross sections from fits to the data for the various functions:
exponential (black), 3 parameter with $n$ a free parameter (red),
3 parameter with $n=6$ (blue) and the 4 parameter with $n=6$
(green). No errors are shown for clarity. }
\label{bothwoe}
\end{figure}
\end{center}

\begin{center}
\begin{figure}[!ht]
\resizebox{5.4in}{6.23in}
{\includegraphics[72,180][540,720]{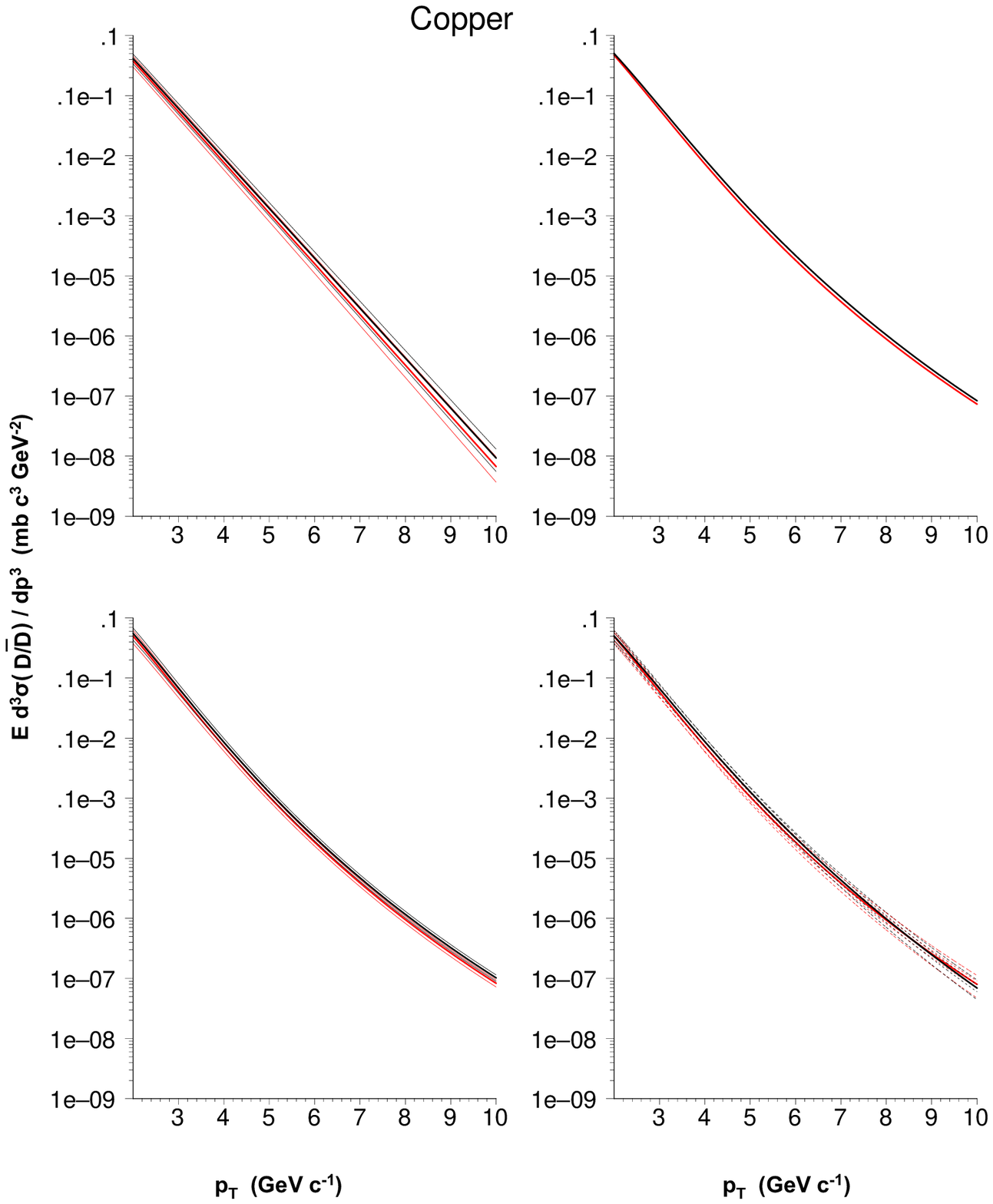}}
\caption[The $D/\overline{D}$ \pcu Cross Sections By
Function]{The \pcu $D$ (black) and $\overline{D}$ (red) differential
cross sections for the various functions: exponential (top left),
3 parameter with $n$ a free parameter (top right), 3 parameter with
$n$ fixed at $n=6$ (bottom left) and the 4 parameter function with
$n$ fixed at $n=6$. Errors include the error on the parameter $N$
added in quadrature.}
\label{cupm}
\end{figure}
\end{center}

\begin{center}
\begin{figure}[!ht]
\resizebox{5.6in}{6.55in}
{\includegraphics[80,180][548,727]{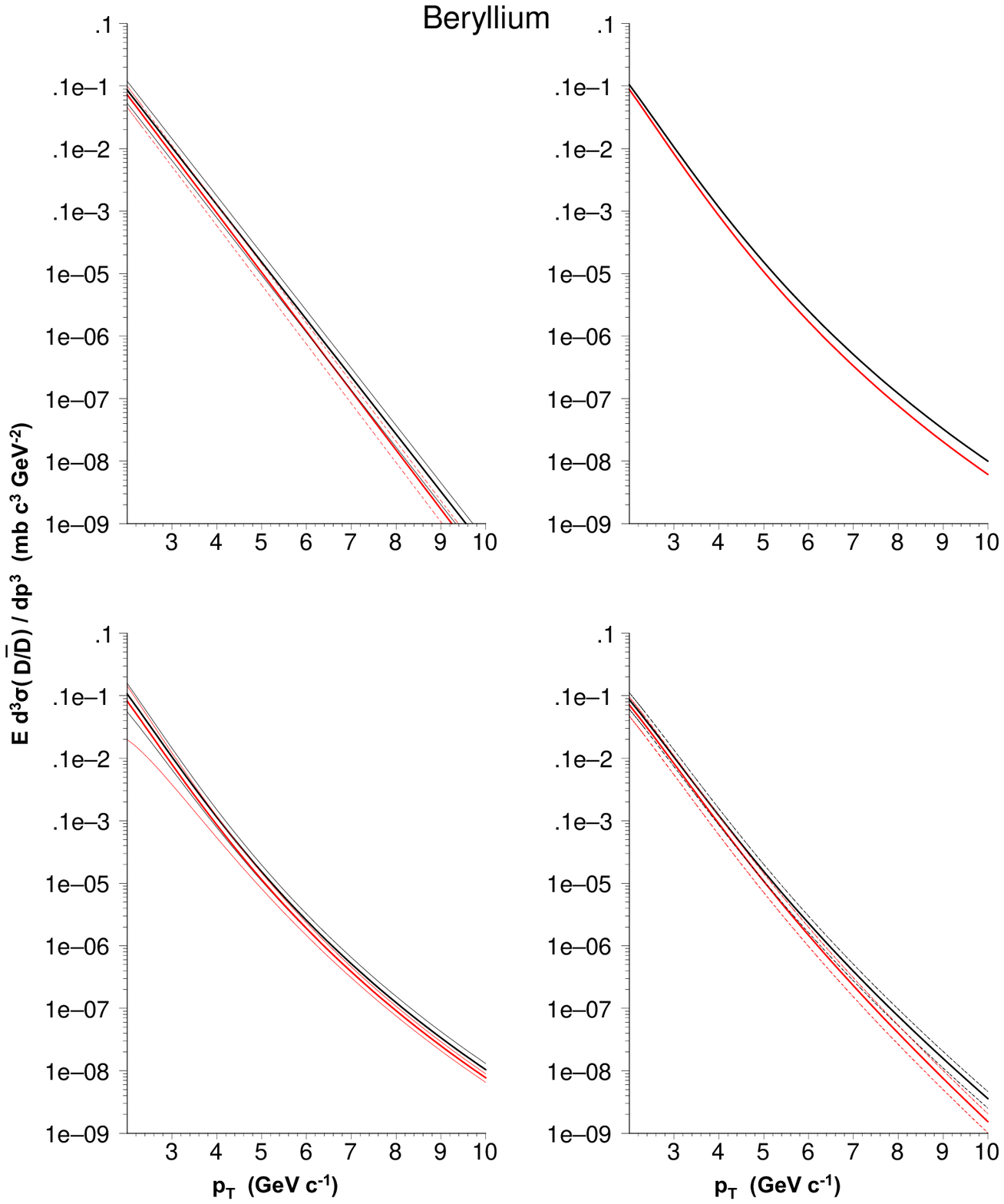}}
\caption[The $D/\overline{D}$ \pbe Cross Sections By
Function]{The \pbe $D$ (black) and $\overline{D}$ (red) differential
cross sections for the various functions: exponential (top left),
3 parameter with $n$ a free parameter (top right), 3 parameter with
$n$ fixed at $n=6$ (bottom left) and the 4 parameter function with
$n$ fixed at $n=6$. Errors include the error on the parameter $N$
added in quadrature.}
\label{bepm}
\end{figure}
\end{center}

\begin{center}
\begin{figure}[!ht]
\resizebox{5.8in}{6.1in}
{\includegraphics[80,157][505,607]{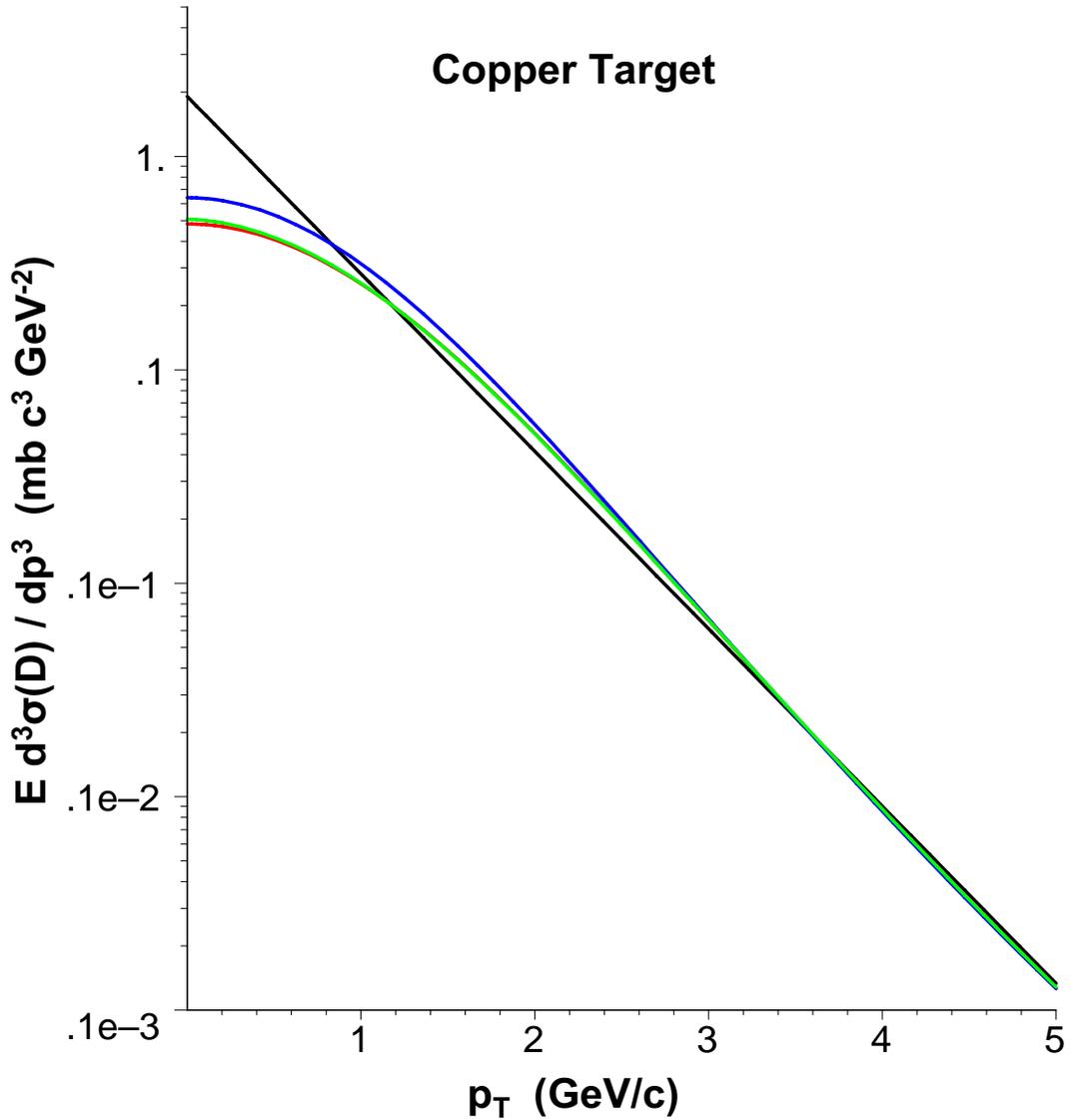}}
\caption[The \pcu $D$ Cross Sections At Low \nspt]{The \pcu $D$
differential cross section projected to low \pt for the
various functions: exponential (black),
3 parameter with $n$ a free parameter (red), 3 parameter with
$n$ fixed at $n=6$ (blue) and the 4 parameter function with
$n$ fixed at $n=6$ (green). No errors are shown.}
\label{cuplowpt}
\end{figure}
\end{center}

\begin{center}
\begin{figure}[!ht]
\resizebox{5.9in}{6.0in}
{\includegraphics[58,228][505,685]{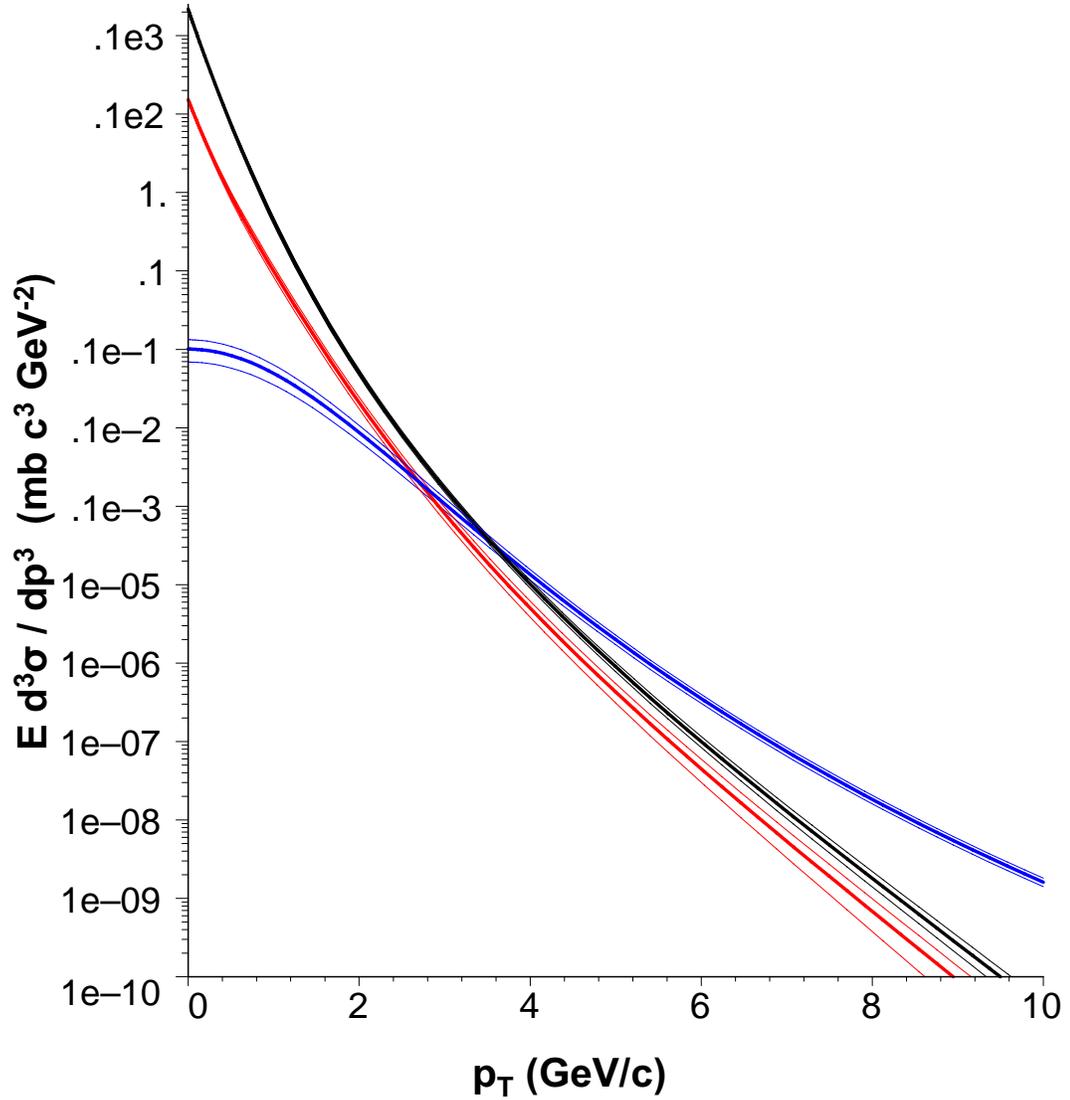}}
\caption[The \pcu $D$, \pluspi and \plusk Cross Sections]{The
\pcu $D$ differential cross section projected to low \pt for the
3 parameter function with $n=6$, and scaled down by $A=63.546$
(blue). Black is the \pluspi and red is the \plusk differential
\ppp cross sections as given in \cite{nucl-phys-b-100-237}.}
\label{cu-pi-k}
\end{figure}
\end{center}

The differential \pcu  and \pbe open charm cross sections,
determined from fits to the data shown in figures \ref{cuexp}
through \ref{bef46} are shown in figures \ref{cuallwe} and
\ref{beallwe}. Errors were calculated using the error propagation
equation as described in \ref{chijerrors} for the parameters in the
function used to describe the cross section added in quadrature
to the error of the scaling variable $N$. The error shown
represents a 70 percent confidence level that if
the value of the first parameter was held within the stated range
in values, the cross section would
lie between the upper and lower error bars. Errors for the 4
parameter function were performed twice because of large
correlations; one set of errors were calculated holding the value
of $\alpha$ fixed (solid lines), and one set with $m$ fixed
(dashed lines). Fits to the 3 parameter function with $n$ a free
parameter had large correlations as well, and no errors were
calculated for those fits. For clarity, the same cross sections
are plotted without errors in figure \ref{bothwoe}. The \pcu and
\pbe $D$ and $\overline{D}$ differential cross
sections are plotted together for the four selected functions
in figures \ref{cupm}, and \ref{bepm}. The four fits selected
appear to be well constrained, even when
projected to $\nspt=0$ \nscmom, as shown in figure \ref{cuplowpt}.
Figure \ref{cu-pi-k} shows the $D$ differential cross section
scaled by $A=63.546$ plotted with the \pluspi and \plusk
cross sections given in \cite{nucl-phys-b-100-237}.

\afterpage{\clearpage}

\subsection{Comparison To Other Experiments}

The open charm cross sections from this analysis were extracted
from single muon events from 2.25 to $\sim$7 \nscmom. Extensive
Monte Carlo studies have determined that contributions from open
charm hadrons to the single muon spectra had
$0.2 \lesssim x_{F} \lesssim 0.8$ and
$2.25 \lesssim \nspt \lesssim 7.0$
(\nscmom). The open charm cross sections were extracted using
3 functions to describe the shape of the cross section: \nsfuncexp,
\nsfuncthree and \nsfuncfour. Comparison to other \ppp or \pA
results at similar energies is problematical since
all fit their data to functions dissimilar to
those chosen here. The increased range in transverse-momentum from
this analysis is far beyond other \pA experiments at this energy,
covering a much larger range in $x_{F}$ as well.

The LEBC-MPS Collaboration (E743) \cite{prl-61-2185} fitted their
800 \cmom \ppp data with the function
$$ \left( 1 - | \, x_{F} \, | \right) ^{n} e^{-a \, \nspt^{2}} $$ and
E789 \cite{prl-72-2542} fitted their results from 800 \cmom \pA
data with the function
$$ \nspt \, e^{-n \, \nspt^{2}} $$ The exponential using the square
of the transverse-momentum failed to adequately describe the data
from this analysis because it falls too steeply over the range in
transverse-momentum.

\begin{center}
\begin{figure}[ph]
\resizebox{5.9in}{6.0in}
{\includegraphics[61,220][547,716]{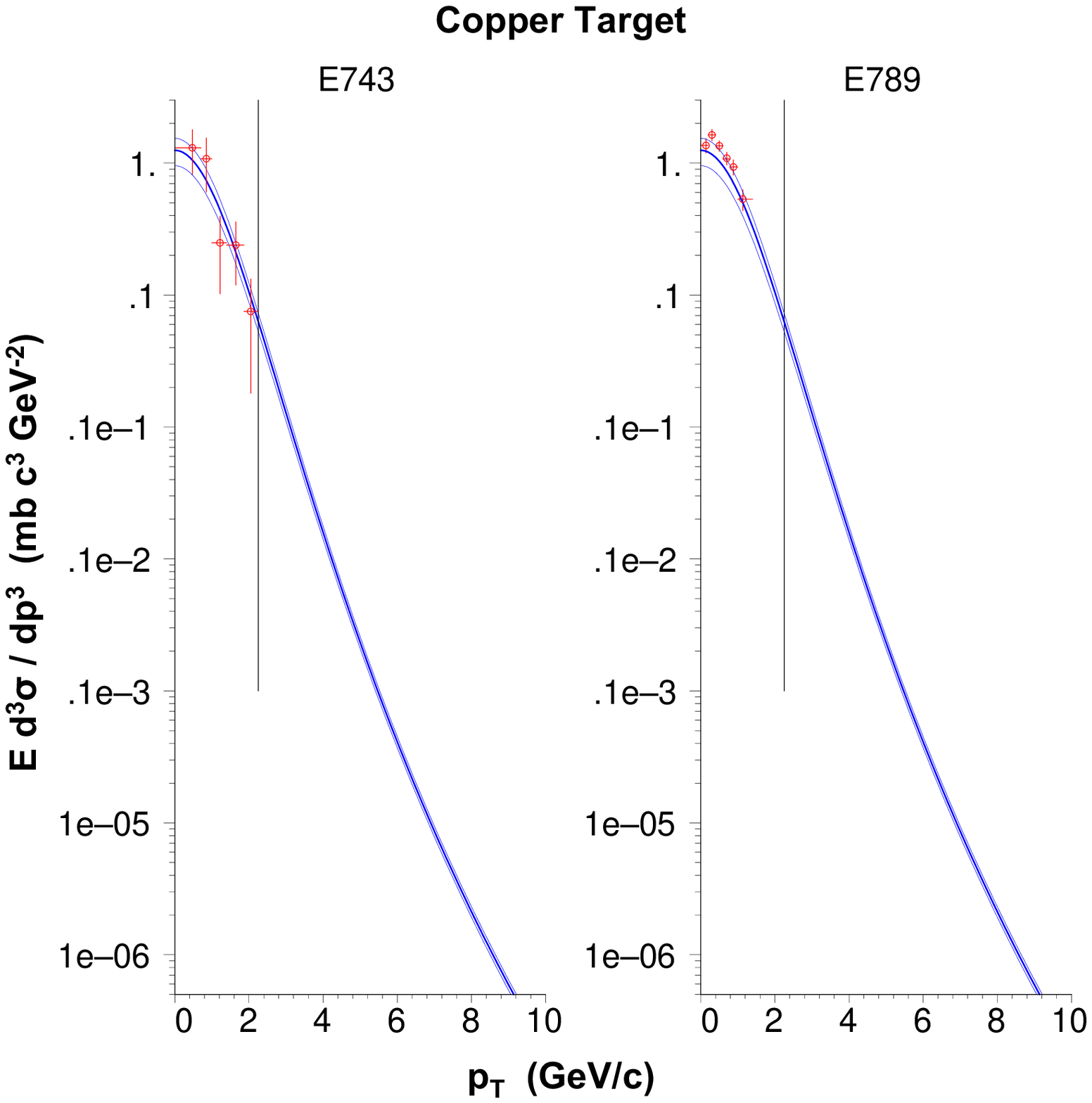}}
\caption[Comparison Of \pcu To E743 And E789.]
{Comparison of the results from this analysis from fitting
the \pcu data using the 3 parameter function with $n=6$ to the
E743\cite{prl-61-2185} (left) and E789\cite{prl-72-2542} (right)
data. The cross section from this analysis is the sum of the
$D$ and $\overline{D}$ cross sections. The invariant cross sections
for E743 and E789 were determined by dividing their published
results by $\int^{1}_{-1} \, \left( 1 - |x_{F}| \right)^{8.6} \times
2 \times \pi$. The data were then scaled by $A_{Cu}=63.546$. Black
vertical line indicates the minimium \mupt data used by
this analysis. The reader is cautioned that the data from E743
and E789 are evaluated at $x_{F} = y = 0$ and compared to the cross
section from this analysis projected to lower momentum.}
\label{cu-outdata}
\end{figure}
\end{center}

\begin{center}
\begin{figure}[!ht]
\resizebox{5.9in}{6.0in}
{\includegraphics[61,221][547,718]{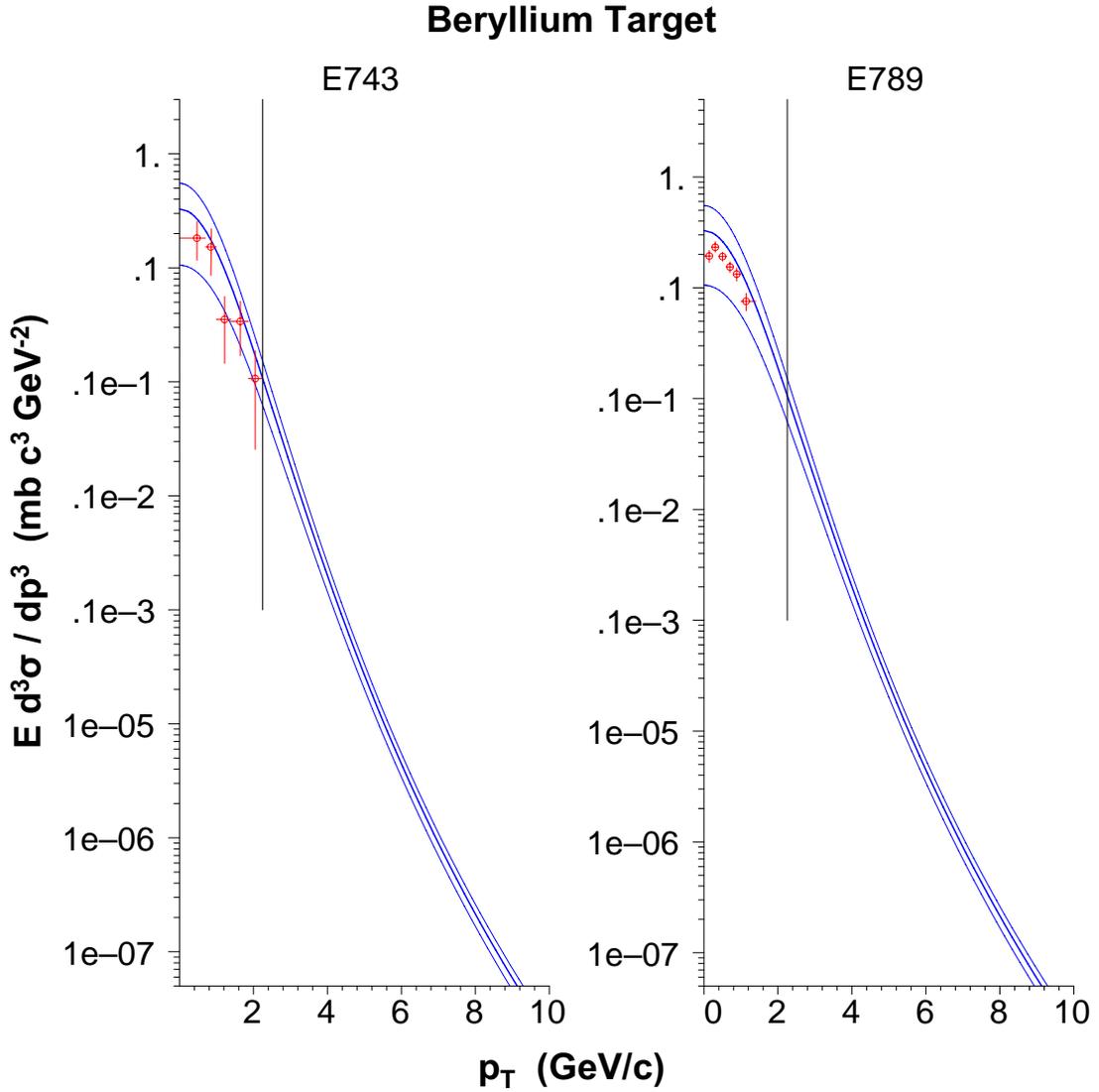}}
\caption[Comparison Of \pbe To E743 And E789.]
{Comparison of the results from this analysis from fitting
the \pbe data using the 3 parameter function with $n=6$ to the
E743\cite{prl-61-2185} (left) and E789\cite{prl-72-2542} (right)
data. The cross section from this analysis is the sum of the
$D$ and $\overline{D}$ cross sections. The invariant cross sections
for E743 and E789 were determined by dividing their published
results by $\int^{1}_{-1} \, \left( 1 - |x_{F}| \right)^{8.6} \times
2 \times \pi$. The data were then scaled by $A_{Be}=9.012182$. Black
vertical line indicates the minimium \mupt data used by
this analysis. The reader is cautioned that the data from E743
and E789 are evaluated at $x_{F} = y = 0$ and compared to the cross
section from this analysis projected to lower momentum.}
\label{be-outdata}
\end{figure}
\end{center}

Figure \ref{prl-72-2542-pt} shows the per nucleon total inclusive
neutral open charm cross section from E789, and Figure
\ref{prl-61-2185-pt2} shows the \ppp total inclusive open charm
cross section from E743. For comparison to the results from this
analysis, the E743 data was divided by 2 since their results were
the total open charm cross section which is assumed to be twice the
neutral cross sections determined by E789 and twice the sum of the
$D$ and $\overline{D}$ cross sections found by this analysis, and
both the E743 and E789 results were scaled by $A_{Cu}=63.546$ and
$A_{Be}=9.012182$. The sum of the $D$ and $\overline{D}$ cross
sections determined from fitting the 3 parameter function with $n=6$
to the \pcu
data is plotted with the scaled results from E743 (left) and E789
(right) in Figure \ref{cu-outdata}, and the result from fitting
the same function to the \pbe data is shown in Figure
\ref{be-outdata}. The reader is cautioned that the cross sections
shown from this analysis below 2.25 \cmom are the \textit{projected}
cross sections, since the analysis used no data below that momentum.
The reader is also cautioned that both the E743 and E789 data are
scaled to $A$ assuming $\alpha(\nspt)=1.0$ (this is a result
presented in E789). These comparisons have also ignored the
ratio of charged to neutral production. These figures show the
results from this analysis lie between the
two previous measurements. This confirms that the use of single
muon spectra and the methods adopted by this analysis to extract
the differential cross sections provided reasonable results.

Table \ref{comp-params} provides results from selected
experiments. It lists the type of data,
\pt and $x_{F}$ ranges covered, the function(s) and values of
the parameters from fits to their data
and results from this analysis. From figures \ref{cu-outdata},
\ref{be-outdata} and the table the author concludes:
\begin{itemize}

\item The open charm cross sections from this analysis reproduce
 the open charm \pt distributions from both previous 800 \cmom
 \ppp and \pA experiments, when projected to low hadron \nspt.

\item Data from 250 \cmom \pA interactions (E769 \cite{prl-77-2392})
 was fit to the exponential function. The value of $B$ reported was
 3.0 $\pm$ 0.3, while the value of $B$ found by this analysis
 was 1.91 $\pm$ 0.04 (copper) and 2.11 $\pm$ 0.12 (beryllium). This
 is reasonable since the E769 data was from
 $0.0 \le \pt \le \sqrt{10}$ \nscmom, while the data from this
 analysis is at higher \nspt. The slope of the
 cross section becomes smaller with increasing \pt, as
 evidenced by the results from this experiment.

\item 250 \cmom $\pi$\textit{-A} data from E769 was fitted with
 the 3 parameter function, where $\alpha=1.4 \pm 0.3$ and
 $n=5.0 \pm 0.6$. This analysis determined that $n=6$ and
 $\alpha \sim 3.5$ (copper) and $\alpha \sim 2.9$ (beryllium),
 indicating that open charm production, as a function of
 \nspt, is softer for proton interactions than meson interactions
 (see figure \ref{prl-77-2392-pt2}).

\end{itemize}

\renewcommand{\arraystretch}{1.5}

\begin{table}[!ph]
\caption[Comparison Of Parameter Values To Other Experiments]
{Top is the experiment, interaction, beam momentum and
range in \pt and $x_{F}$ covered. Bottom table gives the
parameterizations used to fit the data by the experiment and the
values of the parameters reported. Note that the
value of $B^{\prime}$ from E789 does not include all \pt
dependencies of the function used by that experiment (see text).
Values given for this analysis (E866) are for fits to determine
the $D$ cross section only. N/A indicates that the parameterization
was attempted but resulted in very large \minchi and the values of
the parameters are not shown. Comparison of the cross sections from
this analysis and the E743 and E789 experiments are shown in figures
\ref{cu-outdata} and \ref{be-outdata}.}
\label{comp-params}
\begin{center}
\begin{tabular}[c]{|l|c|c|c|c|}

\hline

\multicolumn{5}{|c|}{Experiment And Data}\\

\multicolumn{1}{|c}{} &
 \multicolumn{1}{c}{} &
 \multicolumn{1}{c}{$p_{b}$} &
 \multicolumn{1}{c}{\nspt} & \\

\multicolumn{1}{|c}{} &
 \multicolumn{1}{c}{} &
 \multicolumn{1}{c}{\nscmom} &
 \multicolumn{1}{c}{\nscmom} &
 \multicolumn{1}{c|}{\raisebox{2.0ex}[0pt]{$x_{F}$}}\\

\hline

  &
 $\pi^{\pm}$\textit{-A} &
 250 &
 0 - 4 &
 -0.1 - 0.8\\

\raisebox{2.0ex}[0pt]{E769 \cite{prl-77-2392}} &
 \pA &
 250 &
 0 - $\sqrt{10}$ &
 -0.1 - 0.5\\

\hline

E743 \cite{prl-61-2185} &
 \ppp &
 800 &
 0 - $\sqrt{5}$ &
 -0.1 - 0.4\\

\hline

E789 \cite{prl-72-2542} &
 \pA &
 800 &
 0 - 1.1 &
 0.00 - 0.08\\

\hline

  &
 \pcu &
  &
  & \\

\raisebox{2.0ex}[0pt]{E866} &
 \pbe &
 \raisebox{2.0ex}[0pt]{800} &
 \raisebox{2.0ex}[0pt]{2.25 $\sim$ 7.0} &
 \raisebox{2.0ex}[0pt]{0.2 - 0.8}\\

\hline
\hline

\multicolumn{5}{|c|}{Parameterizations And Values}\\

\multicolumn{1}{|c}{} &
 \multicolumn{1}{c}{$ e^{-B^{\prime} \, \nspt^{2}}$} &
 \multicolumn{1}{c}{$ e^{-B \, \nspt} $ } &
 \multicolumn{2}{c|}{$ \left(
 \nspt^{2}+ \alpha \, m^{2}_{c} \right)^{-n}$}\\

\multicolumn{1}{|c}{} &
 \multicolumn{1}{c}{$B^{\prime}$} &
 \multicolumn{1}{c}{$B$} &
 \multicolumn{1}{c}{$\alpha$} &
 \multicolumn{1}{c|}{$n$}\\

\hline

  &
 1.08 $\pm$ 0.05 &
 2.74 $\pm$ 0.09 &
 1.4 $\pm$ 0.3 &
 5.0 $\pm$ 0.6\\

\raisebox{2.0ex}[0pt]{E769} &
 1.08 $\pm$ 0.05 &
 3.0 $\pm$ 0.3 &
  & \\

\hline

E743 &
 0.8 $\pm$ 0.2 &
  &
  & \\

\hline

E789 &
 0.91 $\pm$ 0.12 &
  &
  & \\

\hline

  &
  &
 1.91 $\latop{+ \, 0.04}{- \, 0.04}$ &
 3.53 $\latop{+ \, 0.17}{- \, 0.17}$ &
 6\\

\raisebox{2.0ex}[0pt]{E866} &
 \raisebox{2.0ex}[0pt]{N/A} &
 2.11 $\latop{+ \, 0.06}{- \, 0.12}$ &
 2.94 $\latop{+ \, 0.32}{- \, 0.30}$ &
 6\\

\hline

\end{tabular}
\end{center}
\end{table}

\renewcommand{\arraystretch}{1.0}

\afterpage{\clearpage}

\subsection{The Ratio Of Charged To Neutral Production}
\label{ratcton}

Theory suggests that
$$ R_{CN} = \frac{ \sigma \left( D^{+} \right) }{ \sigma \left(
D^{0} \right) } \sim 0.32 $$ where $R_{CN}$ is the ratio
of charged to neutral open charm production \cite{hep-ph-9702287}
\cite{hep-ph-0306212}. Prior to 2000,
experimental results from proton induced charm production showed
the ratio to be consistent with unity \cite{prl-61-2185}
\cite{phys-lett-b-263-573}. Results from the production of open
charm via pion-nucleon interactions show the ratio
from 0.27 to 0.50.\footnote{These results are presented in tabular
form in \cite{hep-ph-9702287}} The discrepancy between the ratios
from meson interactions versus proton interactions, which should
be the same, has generated some interest
in the community \cite{hep-ph-9702287}. A more recent study by the
HERA-B collaboration \cite{hep-ex-0408110} found the ratio to be
0.54 $\pm$ 0.11 $\pm$ 0.14 in 920 \cmom \pA collisions. Table
\ref{ctonrat} shows four experimentally determined charged to
neutral production ratios for open charm from \ppp and \pA
interactions.

\renewcommand{\arraystretch}{1.5}

\begin{table}[!ht]
\caption[Charged To Neutral Production Ratios From Other
Experiments]{The charged to neutral production ratio, \nscton,
from other \ppp or \pA experiments.}
\label{ctonrat}
\begin{center}
\begin{tabular}
[c]{|lccc|}\hline

  &
  &
 Momentum &\\

\raisebox{1.5ex}[0pt]{Experiment} &
  &
 \cmom &
 \raisebox{1.5ex}[0pt]{Ratio}\\

\hline

E783 \cite{prl-61-2185} &
 \ppp &
 800 &
 1.2 $\pm$ 0.6\\

E653 \cite{phys-lett-b-263-573} &
 \pA &
 800 &
 1.0 $\pm$ 0.6\\

LEBC-EHS \cite{z-phys-c-40-321} &
 \ppp &
 400 &
 0.7 $\pm$ 0.1\\

HERA-B \cite{hep-ex-0408110} &
 \pA &
 920 &
 0.54 $\pm$ 0.18\\

\hline
\hline

\multicolumn{3}{|c}{Weighted Average} &
 0.7 $\pm$ 0.1\\

\hline

\end{tabular}
\end{center}
\end{table}

\renewcommand{\arraystretch}{1.0}

This analysis lacks the sensitivity required to measure the
charged to neutral production ratio, \nscton. This is due to the
large transverse momentum shift from the parent hadron to the
decay muon. This shift resulted in the shapes of the muon spectra
being very similar for all open charm hadrons. During
fitting the contributions from both the charged and neutral
open charm mesons were added together, to give the total
open charm Monte Carlo contribution. The
analysis used $\nscton=1$ for all fits, so the two open
charm cross sections determined from each fit were
designated as \dcross and \nsadcross. The $D$ cross section was
then the cross section for either \plusd or \nsdzero, and the
$\overline{D}$ cross section was the cross section for either the
\minusd or \nsadzero.

By using a charged to neutral production ratio other than 1,
fitting would then be to either the charged cross section, or
the neutral cross section, depending on how the ratio was introduced
into calculating the open charm and open anti-charm Monte Carlo
spectra. An open charm Monte Carlo spectrum is used as an example.
The open charm contribution was determined from
$$ W^{\mu}_{j} = \sum_{i=1}^{40}
\left[ \left( W^{\sigma_{D}}_{x,i} \,
\mathbf{W}^{\nsplusd \rightarrow \nsmuplus}_{i \, j}
\right) +
\left( W^{\sigma_{D}}_{x,i} \,
\mathbf{W}^{\nsdzero \rightarrow \nsmuplus}_{i \, j}
\right) \right] $$

Choosing $\sigma_{\nsplusd} = R \, \sigma_{\nsdzero}$ the equation
above may be re-written as
$$ W^{\mu}_{j} = \sum_{i=1}^{40}
\left[ \left( R \, W^{\sigma_{\nsdzero}}_{x,i} \,
\mathbf{W}^{\nsplusd \rightarrow \nsmuplus}_{i \, j}
\right) +
\left( W^{\sigma_{\nsdzero}}_{x,i} \,
\mathbf{W}^{\nsdzero \rightarrow \nsmuplus}_{i \, j}
\right) \right] $$ Further, assuming the two contributions are
related by (for this anlysis, the open charm contributions had
virtually identical shapes, though different magnitudes) 
$$ \mathbf{W}^{\nsplusd \rightarrow \nsmuplus}_{i \, j} =
s \, \mathbf{W}^{\nsdzero \rightarrow \nsmuplus}_{i \, j} $$ where
$s$ is a constant, the open charm contribution to the total Monte
Carlo spectra can be expressed as

\begin{equation}
W^{\mu}_{j} = \left( s \, R + 1 \right) \: \sum_{i=1}^{40}
\left[ W^{\sigma_{\nsdzero}}_{x,i} \,
\mathbf{W}^{\nsdzero \rightarrow \nsmuplus}_{i \, j} \right]
\label{effcton}
\end{equation}

While it is possible to include the ratio as a free parameter
in the fitting routines in much the same manner as the
free parameter $N$, the similarity in the shapes of
the muon spectra from \plusd and \dzero production make the
sensitivity to \cton very low. The free parameter $N$ is constrained
by the light contribution which
is different in shape to the open charm contribution.

Effects of $\nscton \ne 1$ can be examined by using Equation
\ref{effcton} above. Values of \cton less than one result in a
smaller open charm contribution, which results in a \dzero cross
section larger than the one determined with $\nscton = 1$, and
values of \cton greater than one result in \dzero cross sections
that are smaller. The amount they would differ is given by
$$ E \, \frac{ d^{3} \, \sigma \left( D^{0} \right) }
{ d \, p^{3} } \left( p_{T} \right) =
\frac{ s + 1 }{ \left( s \, R + 1 \right)} \;
 E \,
\frac{ d^{3} \, \sigma \left( D \right) }
{ d \, p^{3} } \left( p_{T} \right)$$

\begin{center}
\begin{figure}[!ht]
\resizebox{5.8in}{6.2in}
{\includegraphics[80,250][512,711]{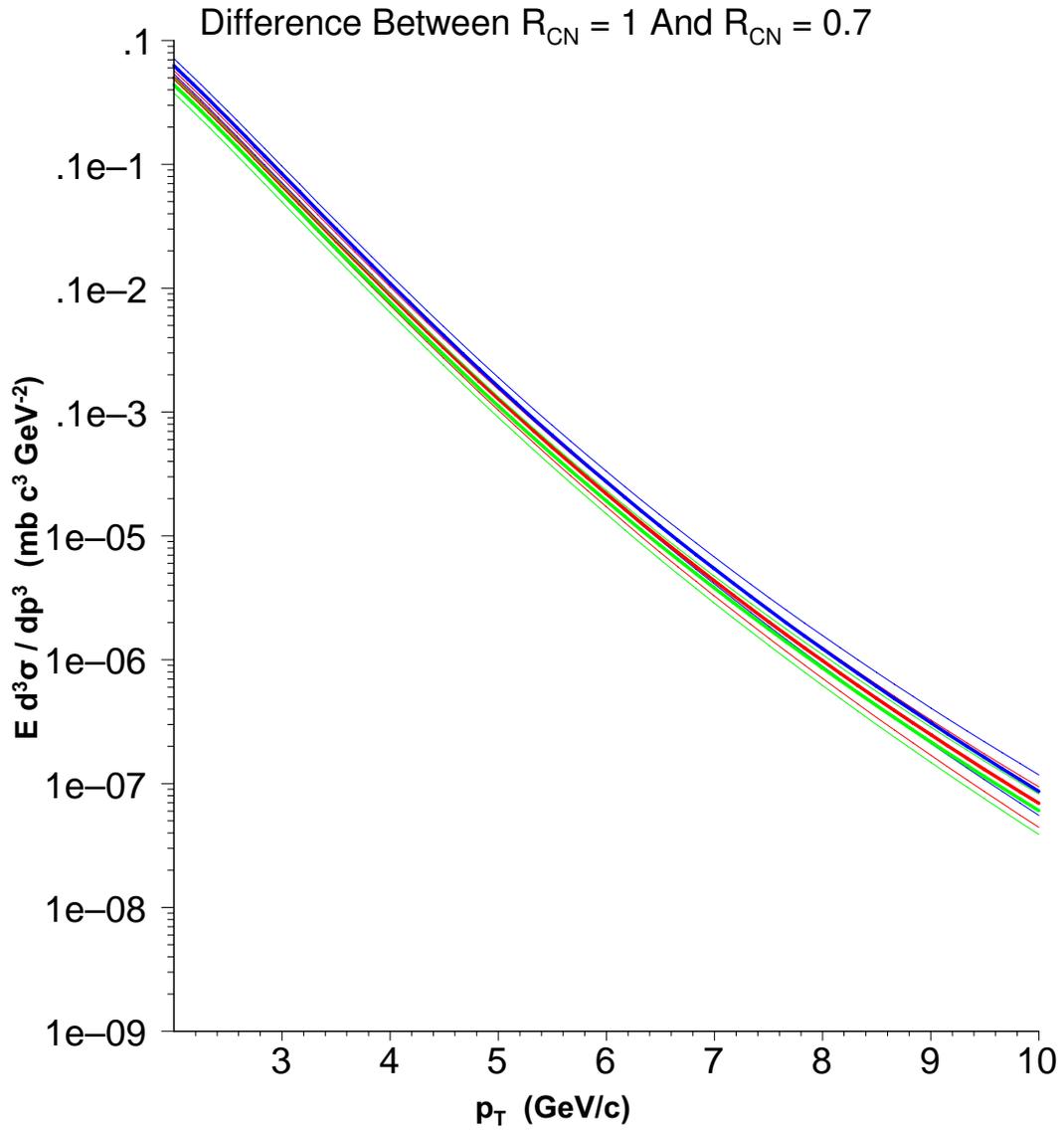}}
\caption[The \dzero And \plusd Cross Sections For $\nscton=0.7$.]
{The $D$ cross section determined by fitting the 4 variable
function with $n$ fixed at $n=6$ and $\nscton=1$ to the \pcu data
(red), and the \dzero (blue) and \plusd (green) cross sections that
would result for $\nscton = 0.7$. All errors are
calculated with the value of $\alpha$ held fixed to its \minchi
value.}
\label{rdiff}
\end{figure}
\end{center}

Figure \ref{rdiff} shows the $D$ cross section from the fit
to the \pcu data where the shape of the cross section was assumed
to be given by the 4 parameter function with $n$ fixed at $n=6$
(red). The errors shown are
those calculated holding $\alpha$ fixed at its \minchi value. The
blue line is the \dzero cross section, and the green line is the
\plusd cross section that would be found for $\nscton=0.7$.

\clearpage

\subsection{Nuclear Dependency}

\begin{center}
\begin{figure}[!ht]
\resizebox{4.7in}{3.13in}
{\includegraphics[274,300][560,486]{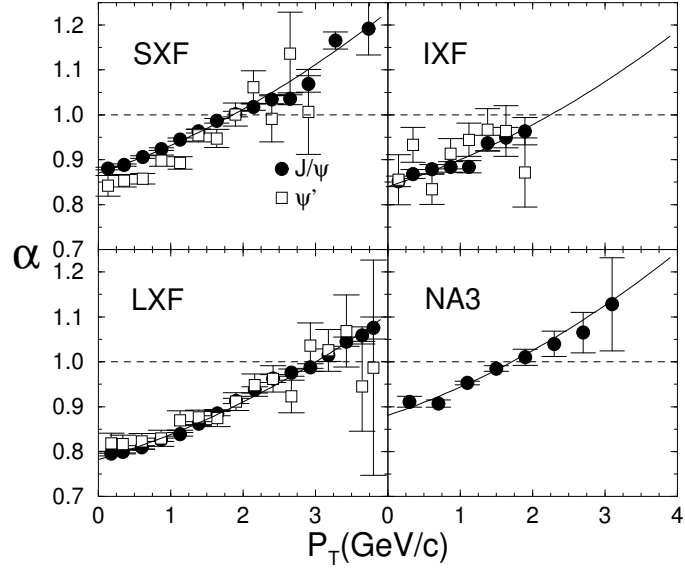}}
\caption[$\alpha$ Versus \pt From E866 Dimuon Analses]
{$\alpha$ versus \pt for $J/\Psi$ (solid circles) and
$\Psi^{\prime}$
(open boxes) production by 800 \cmom protons. Results are shown
for the three data sets \-- SXF, IXF, and LXF (see text) \--
which have $\langle x_{F} \rangle =$ 0.055, 0.308 and 0.480
respectively. Only statistical uncertainties are shown. An additional
systematic uncertainty of 0.5\% is not included. Also shown are the
NA3 results at 200 \cmom \cite{z-phys-c-20-101}. The solid curves
represent the parameterization $\alpha(\nspt) = A_{i} (1 + 0.0604
\nspt + 0.0107 p^{2}_{T})$, where $A_{i} =$ 0.870, 0.840,
0.782, and 0.881 for the SXF, IXF, LXF data sets,
and the NA3 data, respectively. Taken from \cite{prl-84-3256}}
\label{prl-84-3256-alpha}
\end{figure}
\end{center}

Nuclear dependency is defined by
$$ \sigma_{A} = A^{\alpha} \, \sigma_{N}$$ where $\alpha$ is a
function of either $x_{F}$, \pt or both, $A$ is the atomic weight
of the material and $N$ is nucleon. E789 studied the production
of neutral open charm produced in 800 \cmom \pA collisions near
$x_{F}=0$. The nuclear dependency reported by this
study was $\alpha = 1.02 \pm 0.03 \pm 0.02$, implying that the
cross sections scaled as the number of nucleons \cite{prl-72-2542}.
In contrast, E866  \cite{prl-84-3256}
measured the power $\alpha$
for hidden charm for three regions in $x_{F}$,
SXF ($-0.1 \le x_{F} \le 0.3$), IXF($0.2 \le x_{F} 0.6$) and
LXF ($0.3 \le x_{F} \le 0.93$). For this analysis,
$0.2 \le x_{F} \le 0.8$, which
lies between the IXF and LXF data from E866.
Their results for the power $\alpha(\nspt)$ are shown in Figure
\ref{prl-84-3256-alpha}.

\begin{center}
\begin{figure}[!ht]
\resizebox{5.6in}{6.3in}
{\includegraphics[81,143][513,629]{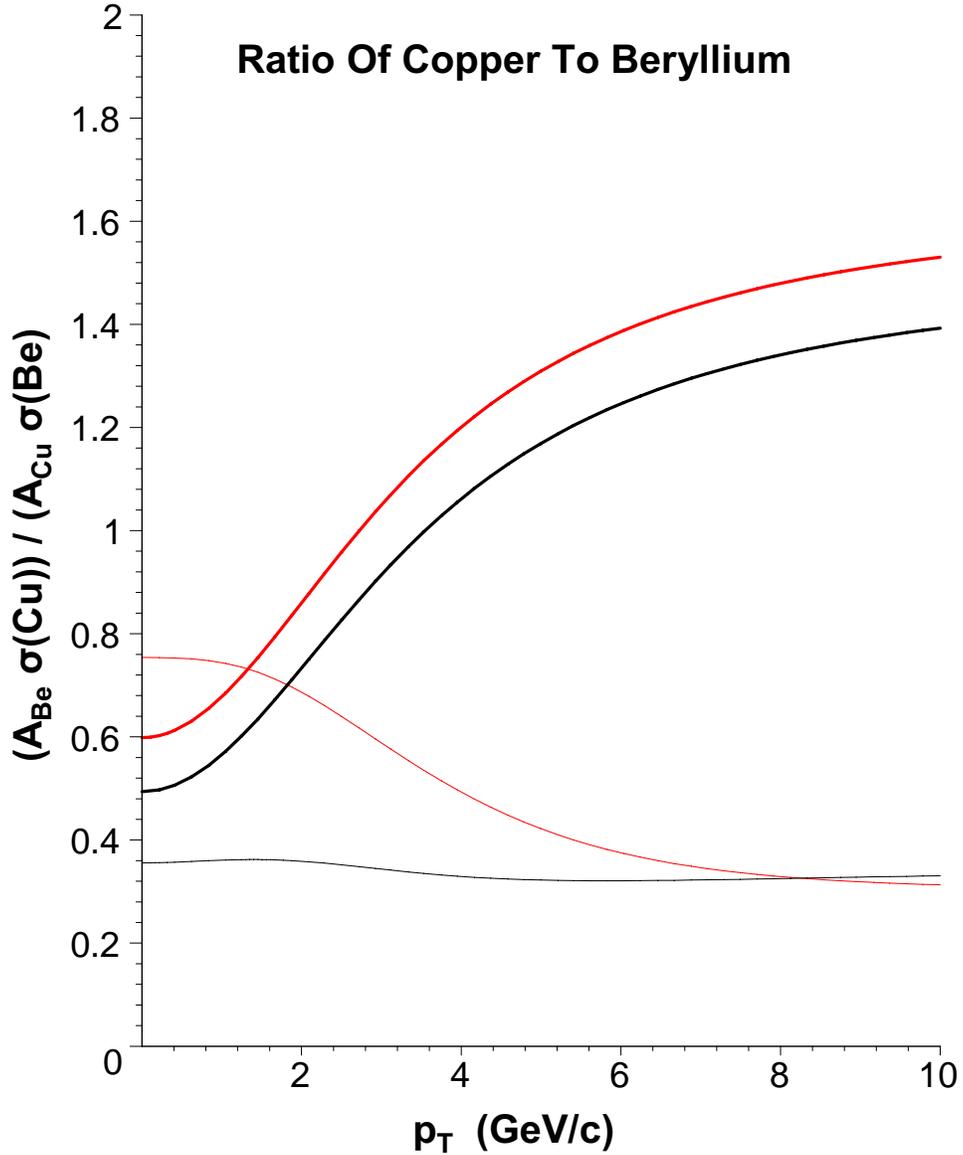}}
\caption[Ratio Of $\sigma^{\mbox{Cu}}(D) / \sigma^{\mbox{Be}}(D)$]
{The ratio of the copper and beryllium $D$ cross sections (black),
and the copper and beryllium $\overline{D}$ cross sections (red).
Both ratios are normalized to $A_{Be}/A_{Cu}$ and use the results
of the fits to the data of the 3 parameter polynomial function
with $n=6$. Thin black (red) lines are the scale of the errors of
the ratios. Errors include the error of $N$ added in quadrature
to the statistical and systematic errors of the other parameters.}
\label{rxsects}
\end{figure}
\end{center}

The ratio
$$ R_{\sigma} = \frac{A_{Be} \, \nsbothcucross}
{A_{Cu} \, \nsbothbecross}$$ for the 3 parameter $n=6$ results
from this analysis are shown in Figure \ref{rxsects}. The power
$\alpha$ can be determined from the ratio of the cross sections,
$r=\sigma_{Cu} / \sigma_{Be}$, and the relations
$\sigma_{Cu} = A_{Cu}^{\alpha} \, \sigma_{N}$ and
$\sigma_{Be} = A_{Be}^{\alpha} \, \sigma_{N}$:
$$ \alpha(\nspt) = \frac{1}
{\mbox{ln}(A_{Cu}) - \mbox{ln}(A_{Be})} \,
\mbox{ln} \left( \frac{1}{r} \right) $$

\begin{center}
\begin{figure}[!ht]
\resizebox{5.9in}{3.52in}
{\includegraphics[88,347][510,599]{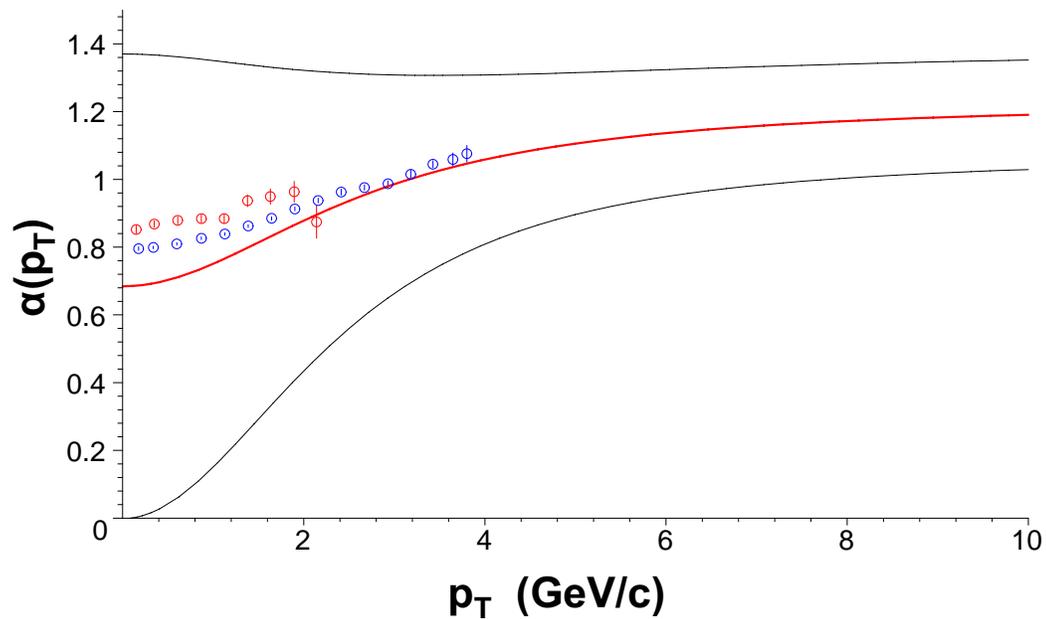}}
\caption[The Power $\alpha(\nspt)$ For Open Charm/Anti-Charm]
{The power $\alpha(\nspt)$ determined by this analysis for
$D + \overline{D}$, where the results from fits using
the 3 parameter polynomial with $n=6$ were used as the open
charm/anti-charm copper and beryllium cross sections. Red open
circles are the IXF data, and the blue open circles are the LXF
data taken from \cite{prl-84-3256}. Thin black lines are
the errors for $\alpha$ from this analysis. Errors
include the error for the parameter $N$ added in quadrature to the
error of the remaining parameters.}
\label{alpha}
\end{figure}
\end{center}

This analysis has extracted $\alpha(\nspt)$ for
both ratios ($r(D)$ and $r(\overline{D})$), and the resulting
$\alpha(\nspt)$ are shown with the IXF and LXF
$J/\Psi$ data from \cite{prl-84-3256}
in Figure \ref{alpha}. Both ratios used the fits of the 3 parameter
$n=6$ to the copper and beryllium data. Based on these results,
this analysis concludes that the cross sections, as a function of
\nspt, do not simply scale by the atomic weight $A$ (admittedly
within large errors, though).

\clearpage

\section{Summary}
\label{summary}

The author takes the position that determination of the 'shape' of
the open charm cross sections is more important than determination of
the absolute magnitude. Clearly both would be optimal, but the use
of secondary spectra required making several assumptions such
as the ratio of charged to neutral open charm production. These
assumptions provided uncertainties in the absolute magnitudes of the
open charm and open anti-charm cross sections that could not be
removed.

Single muon data from 800 \cmom \pcu and \pbe interactions were used
to extract the differential open charm/anti-charm cross sections
as a function of \nspt. Several functions describing theoretical
predictions of the shapes of the cross sections were fit to the
data. Monte Carlo studies indicated that the open charm
contributions were from open charm hadrons produced with
$2.25 \le \pt \lesssim 7.0$ (\nscmom),
$0 \lesssim y_{c.m.} \lesssim 2.0$ and
$0.2 \lesssim x_{F} \lesssim 0.8$. Tables \ref{exp-table},
\ref{f3-table},
\ref{f4-cu-table} and
\ref{f4-be-table} show that the 3 parameter (with $n$ a free
parameter or fixed at the integer value 6) and 4 parameter (with
the parameter $n$ fixed at the integer value 6) polynomial
functions resulted in the smallest \pdfchi for both the \pcu and
\pbe data. Extrapolation of the results
from the 3 parameter function with $n=6$ show good agreement with
previous experiments. Fits of the three functions to both the
\pcu and \pbe data indicate that the parameter $n$ is very close
to 6.0. Due to large correlations introduced by the scaling factor
$N$ and loss of data below 2.25 \nscmom, errors for the parameters
from fits using the 3 variable function with $n$ a free parameter
cannot be determined accurately. In summary:

\begin{itemize}

\item The production of open charm hadrons, as a function of the
 hadron \nspt, is well represented by the function:
$$ \nsfuncthree $$ with $n=6$. Previous fits using this function
 have only been reported for $\pi$\textit{-A} data, where it was
 found that the same function fit the data well, and $n=5.0$.

\item The cross sections can also be well described using the simple
 exponential
$$ \nsfuncexp $$ for $2.25 \le \nspt \lesssim 8.0$ \nscmom.

\item The weighted average value (over $D$ and $\overline{D}$)
 of $\alpha$ for the 3 parameter
 function and $n=6$ for \pcu production is 3.49 $\pm$ 0.01, while
 the
 weighted average for \pbe production is 2.92 $\pm$ 0.02. The
 difference indicates that the production via \pcu has a steeper
 slope than \pbe production for the region in \pt covered by
 this analysis.

\item Extraction of the power $\alpha$ used to relate proton-nucleon
 cross sections to proton-nucleus cross sections, often referred to
 as the nuclear dependency, indicates a structure similar to that
 found for the production of $J/\Psi$, though large uncertainties
 at low \pt make it difficult to claim that such a structure was
 found with certainty. Regardless, the author claims that, for
 the region in \pt for this analysis, the power $\alpha$
 continuously rises even if the large errors
 are taken into account at $\pt \sim 7.0$ \nscmom. The conclusion
 reached for this analysis is that the cross sections do not simply
 scale by the nuclear weight $A$.

\end{itemize}

\newpage

\addcontentsline{toc}{chapter}{\bibname}


\newpage

\addcontentsline{toc}{chapter}{Appendices}
\appendix

\chapter{Miscellaneous Apparatus Information}

\renewcommand{\arraystretch}{1.5}

\begin{table}[ht]
\caption[Drift Chamber Specifications]{Drift chamber information.
$z$ is given for the average of the two planes for each pair.
Chambers are listed in ascending distance along the $z$ axis. All
lengths are in inches.}
\label{wireplanes}
\begin{center}
\begin{tabular}
[c]{|ccccccc|}
\hline

Detector &
 Angle &
 Number &
 Wire &
  &
 Aperture &
 Operating\\

Pair &
 (deg) &
 of Wires &
 Spacing &
 \raisebox{2.0ex}[0pt]{$z$} &
 ($x\times y$) &
 Voltage\\

\hline

V1-V1$^{\prime}$ &
 -14 &
 200 &
 0.250 &
 724.69 &
 48.0 $\times$ 50.0 &
 +1700\\

Y1-Y1$^{\prime}$ &
 0 & 160 &
 0.250 &
 748.81 &
 48.0 $\times$ 40.0 &
 +1700\\

U1-U1$^{\prime}$ &
 +14 &
 200 &
 0.250 &
 755.48 &
 48.0 $\times$ 50.0 &
 +1700\\

\hline

U2-U2$^{\prime}$ &
 +14 &
 160 &
 0.388 &
 1083.40 &
 66.0 $\times$ 62.1 &
 -1950\\

Y2-Y2$^{\prime}$ &
 0 &
 128 &
 0.400 &
 1093.21 &
 66.0 $\times$ 51.2 &
 -2000\\

V2-V2$^{\prime}$ &
 -14 & 160 &
 0.388 &
 1103.25 &
 66.0 $\times$ 62.1 &
 -2000\\

\hline

U3-U3$^{\prime}$ &
 +14 &
 144 &
 0.796 &
 1790.09 &
 106.0 $\times$ 114.7 &
 -2200\\

Y3-Y3$^{\prime}$ &
 0 &
 112 &
 0.820 &
 1800.20 &
 106.0 $\times$ 91.8 &
 -2200\\

V3-V3$^{\prime}$ &
 -14 &
 144 &
 0.796 &
 1810.24 &
 106.0 $\times$ 114.7 &
 -2200\\

\hline

\end{tabular}
\end{center}
\end{table}

\renewcommand{\arraystretch}{1.0}
\renewcommand{\arraystretch}{1.5}

\begin{center}
\begin{table}[ht]
\caption[Hodoscope Specifications]{Information on the banks of
hodoscopes. Planes are presented in the order in which a particle
would traverse them. Minimum and maximum values of
$| \nsxmom/\nszmom | $ for each $x$ plane are
presented for reference, and are based on disconnecting the high
voltage supplies to the middle half of each bank as shown in
figures \ref{hodos} and \ref{xplane}. All lengths are in inches.}
\label{hodoplanes}
\begin{tabular}[c]{|ccccccc|}
\hline

  &
  &
 Number &
 Scint. &
 Aperture &
 \tantx &
 \tantx\\

 \raisebox{2.0ex}[0pt]{Det.} &
 \raisebox{2.0ex}[0pt]{$z$} &
 Scint. &
 Width &
 $x\times y$ &
 Min &
 Max\\

\hline

Y1 &
 769.78 &
 16 &
 2.5 &
 47.50 $\times$ 40.75 &
  & \\

X1 &
 770.72 &
 12 & 4.0 &
 47.53 $\times$ 40.78 &
 0.015 &
 0.030\\

Y2 &
 1114.94 &
 16 &
 3.0 &
 64.625 $\times$ 48.625 &
  & \\

X3 &
 1822.00 &
 12 &
 8.68 &
 105.18 $\times$ 92.0 &
 0.014 &
 0.030\\

Y3 &
 1832.00 &
 13 &
 7.5 &
 104.00 $\times$ 92.0 &
  & \\

Y4 &
 2035.50 &
 14 &
 8.0 &
 116.00 $\times$ 100.00 &
  & \\

X4 &
 2131.12 &
 16 & 7.125 &
 126.00 $\times$ 114.00 &
 0.015 &
 0.030\\

\hline
\hline
\end{tabular}
\end{table}
\end{center}

\renewcommand{\arraystretch}{1.0}
\renewcommand{\arraystretch}{1.5}

\begin{table}[ht]
\caption[Station Four Specifications]{Information for the three
layers of proportional tubes in Station 4 as used in taking data
for this analysis. Layers are listed in ascending distance from
the origin. All lengths are in inches.}
\label{ptubes}
\begin{center}
\begin{tabular}
[c]{|cccccc|}
\hline

  &
 Number &
 Cell &
 Aperture &
  &
 Operating\\

\raisebox{2.0ex}[0pt]{Bank} &
 of Tubes &
 ($x\times y$) &
 ($x\times y$) &
 \raisebox{2.0ex}[0pt]{$z$} &
 Voltage\\

\hline

PT-Y1 &
 120 &
 1.00 $\times$ 1.00 &
 117.0 $\times$ 120.0 &
 2041.75 &
 +2500\\

PT-X &
 135 &
 1.00 $\times$ 1.00 &
 135.4 $\times$ 121.5 &
 2135.88 &
 +2500\\

PT-Y2 &
 143 &
 1.00 $\times$ 1.00 &
 141.5 $\times$ 143.0 &
 2200.75 &
 +2500\\

\hline

\end{tabular}
\end{center}
\end{table}

\renewcommand{\arraystretch}{1.0}

\begin{center}
\begin{figure}[ht]
\resizebox{4.8in}{7.7in}
{\includegraphics[165,115][455,675]{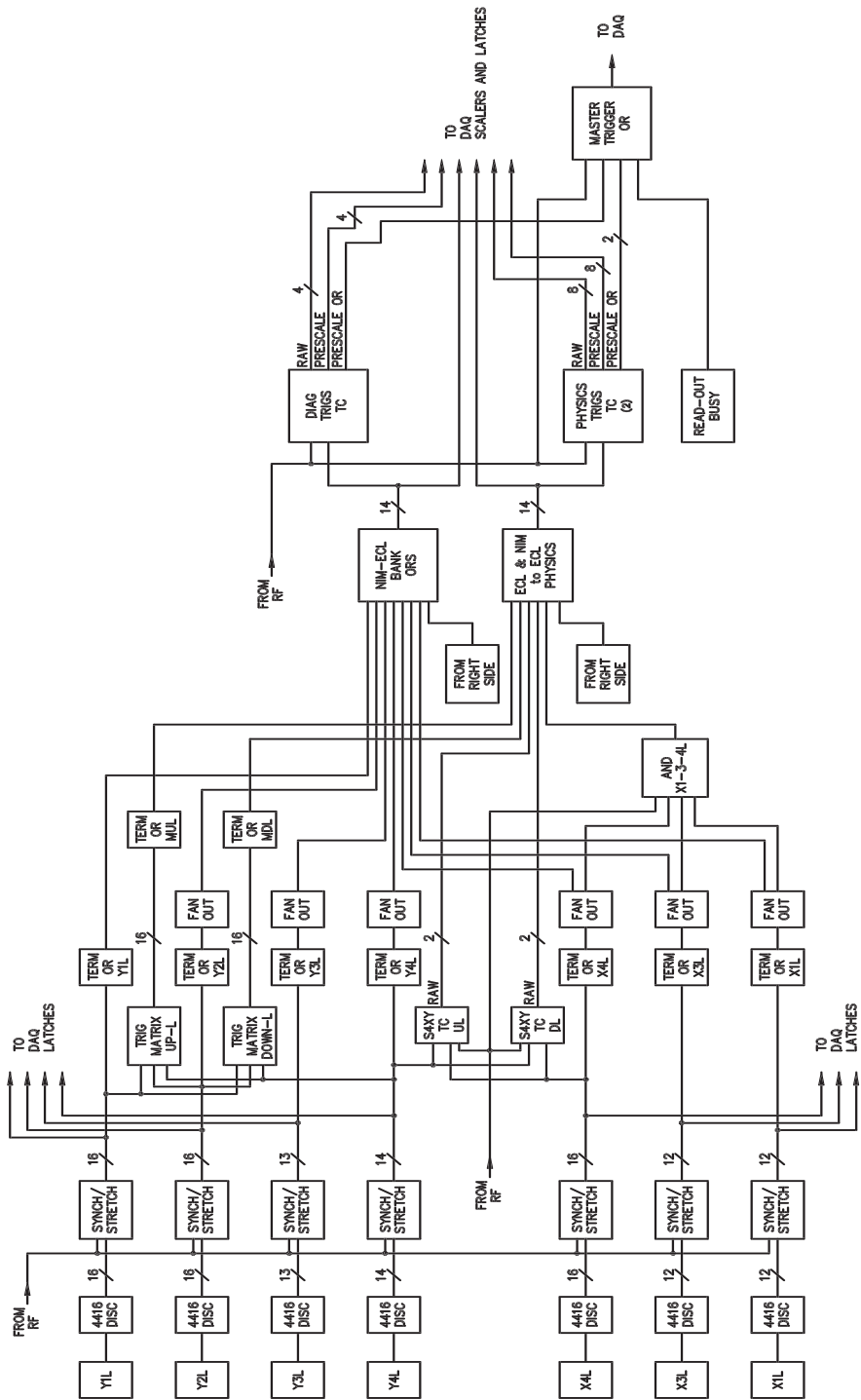}}
\caption[The E866 Trigger]{Block diagram of the E866 trigger system
for the left hand side. Taken from \cite{hawker}}
\label{trig_nim}
\end{figure}
\end{center}

\chapter{Analysis}

\begin{table}[ht]
\caption[Fraction of Dump Events in a Target Analysis]{Estimated
fraction of events in the target analysis of Run 2751 that
originated in the dump. Dump IC3SB is the fraction of the luminosity
that survived the target and interacted in the dump. Totals are for
the three targets only.}
\label{destimate}
\begin{center}
\begin{tabular}[c]{|r|c|c|c|c|c|}
\hline

Target &
 Total &
 Dump &
 Estimated &
 Number &
 Fraction\\

  &
 IC3SB &
 IC3SB &
 Dump &
 Events &
 Dump\\

  &
  &
  &
 Events &
  &
 Events\\

\hline

0 &
 40661 &
 40661 &
 143 $\pm$ 12 &
 143 &
 1.000\\

\hline

1 &
 53013 &
 48708 &
 170 $\pm$ 15 &
  & \\

2 &
 52872 &
 46663 &
 163 $\pm$ 14 &
  &
 \\

3 &
 27225 &
 22983 &
 80 $\pm$ 7 &
  &
 \\

\hline

Total &
 133110 &
 118324 &
 413 $\pm$ 22 &
 6350 $\pm$ 80 &
 0.065 $\pm$
0.004\\

\hline\hline
\end{tabular}
\end{center}
\end{table}

\begin{table}[ht]
\caption[Magnetic Field Calibrations.]{Spectrometer calibrations
for data taken with the magnetic fields in SM12 and SM3 parallel.
All target analyses used $Z_{tgt}=-24.0$ and $Z_{scat}=200.0$
inches. All dump analyses used $Z_{tgt}=85.1$ and $Z_{scat}=235.0$
inches.}
\label{lscalibrations}
\begin{center}
\begin{tabular}[c]{|r|cccccc|}
\hline

Run &
  &
 $TWEE$ &
 $X0$ &
 $Y0$ &
 $\theta_{x}^{\prime}$ &
 $\theta_{y}^{\prime}$\\

  &
  &
  &
 (in) &
 (in) &
 (rad) &
 (rad)\\

\hline

2748 &
 Target &
 0.993 &
 0.040 &
 -0.525 &
 0.00015 &
 -0.00025\\

  &
 Dump &
 0.995 &
 0.046 &
 -0.560 &
 0.00008 &
 -0.00015\\

\hline

2749 &
 Target &
 0.995 &
 0.051 &
 -0.538 &
 0.00010 &
 -0.00035\\

  &
 Dump &
 0.997 &
 0.063 &
 -0.572 &
 0.00005 &
 0.00000\\

\hline

2750 &
 Target &
 0.990 &
 0.070 &
 -0.540 &
 0.00010 &
 -0.00035\\

  &
 Dump &
 0.992 &
 0.081 &
 -0.572 &
 0.00008 &
 0.00004\\

\hline

2751 &
 Target &
 0.990 &
 0.007 &
 -0.563 &
 0.00005 &
 -0.00035\\

  &
 Dump &
 0.992 &
 0.024 &
 -0.585 &
 0.00005 &
 0.00004\\

\hline

2752 &
 Target &
 0.990 &
 0.032 &
 -0.553 &
 0.00005 &
 -0.00035\\

  &
 Dump &
 0.992 &
 0.051 &
 -0.575 &
 0.00000 &
 0.00004\\

\hline

2753 &
 Target &
 0.990 &
 0.050 &
 -0.555 &
 0.00005 &
 -0.00035\\

  &
 Dump &
 0.992 &
 0.056 &
 -0.578 &
 0.00000 &
 0.00004\\

\hline

2754 &
 Target &
 0.988 &
 0.035 &
 -0.549 &
 0.00005 &
 -0.00035\\

  &
 Dump &
 0.990 &
 0.054 &
 -0.583 &
 0.00000 &
 0.00000\\

\hline

2755 &
 Target &
 0.986 &
 0.040 &
 -0.569 &
 0.00005 &
 -0.00035\\

  &
 Dump &
 0.990 &
 0.049 &
 -0.593 &
 0.00005 &
 -0.00015\\

\hline

2756 &
 Target &
 0.986 &
 0.049 &
 -0.545 &
 0.00001 &
 -0.00035\\

  &
 Dump &
 0.990 &
 0.072 &
 -0.581 &
 0.00000 &
 -0.00005\\

\hline

2757 &
 Target &
 0.986 &
 0.045 &
 -0.547 &
 0.00001 &
 -0.00035\\

  &
 Dump &
 0.990 &
 0.066 &
 -0.581 &
 -0.00002 &
 -0.00008\\

\hline

\end{tabular}
\end{center}
\end{table}

\chapter{\chisquare From The Minimizations}

\section{Sub-Total And Total \nschisquare}
\label{subtotals}

\renewcommand{\arraystretch}{1.25}

\begin{table}[th]
\caption[Sub Total \chisquare For Minimizations To The \pcu Data]
{The eight sub-total, total, number of degrees of freedom and
reduced \nschisquare, \pdfchi for all minimizations to the
copper target data. Top line gives the individual spectra sub-totals,
and the second line gives the sub-totals by charge and production
region.}
\label{indivcu}
\begin{center}
\begin{tabular}
[c]{|l|c|cccc|ccc|}\hline

\multicolumn{9}{|c|}{Copper Data}\\

\hline
\hline

  &
  &
 \nstpluschi &
 \nstminuschi &
 \nsdpluschi &
 \nsdminuschi &
  &
  & \\

\raisebox{2.0ex}[0pt]{Function} &
 \raisebox{2.0ex}[0pt]{$n$} &
 \nspluschi &
 \nsminuschi &
 \nstgtchi &
 \nsdmpchi &
 \raisebox{2.0ex}[0pt]{\nschisquare} &
 \raisebox{2.0ex}[0pt]{$ndf$} &
 \raisebox{2.0ex}[0pt]{\nspdfchi}\\

\hline
\hline

  &
  &
 18.43 &
 21.68 &
 16.90 &
 24.41 &
  &
  & \\

\raisebox{2.0ex}[0pt]{\nsfuncexp} &
 \raisebox{2.0ex}[0pt]{NA} &
 35.33 &
 46.09 &
 40.11 &
 41.31 &
 \raisebox{2.0ex}[0pt]{81.42} &
 \raisebox{2.0ex}[0pt]{61} &
 \raisebox{2.0ex}[0pt]{1.33}\\

\hline
\hline

  &
  &
 12.28 &
 18.40 &
 10.65 &
 12.25 &
  &
  & \\

  &
 \raisebox{2.0ex}[0pt]{FF} &
 22.94 &
 31.16 &
 31.19 &
 22.91 &
 \raisebox{2.0ex}[0pt]{54.09} &
 \raisebox{2.0ex}[0pt]{59} &
 \raisebox{2.0ex}[0pt]{0.92}\\

\cline{2-9}

  &
  &
 39.72 &
 48.38 &
 133.36 &
 83.95 &
  &
  & \\

  &
 \raisebox{2.0ex}[0pt]{4} &
 173.08 &
 132.32 &
 88.10 &
 217.31 &
 \raisebox{2.0ex}[0pt]{305.41} &
 \raisebox{2.0ex}[0pt]{61} &
 \raisebox{2.0ex}[0pt]{5.01}\\

\cline{2-9}

  &
  &
 15.04 &
 23.89 &
 35.44 &
 25.21 &
  &
  & \\

  &
 \raisebox{2.0ex}[0pt]{5} &
 50.48 &
 49.10 &
 38.93 &
 60.65 &
 \raisebox{2.0ex}[0pt]{99.58} &
 \raisebox{2.0ex}[0pt]{61} &
 \raisebox{2.0ex}[0pt]{1.63}\\

\cline{2-9}

\raisebox{9.0ex}[0pt]{\functhreedis} &
  &
 11.98 &
 18.96 &
 12.80 &
 12.33 &
  &
  & \\

  &
 \raisebox{2.0ex}[0pt]{6} &
 24.78 &
 31.29 &
 30.94 &
 25.14 &
 \raisebox{2.0ex}[0pt]{56.08} &
 \raisebox{2.0ex}[0pt]{61} &
 \raisebox{2.0ex}[0pt]{0.92}\\

\hline
\hline

  &
  &
 13.71 &
 18.31 &
 11.59 &
 14.87 &
  &
  & \\

  &
 \raisebox{2.0ex}[0pt]{4} &
 25.30 &
 33.18 &
 32.03 &
 26.46 &
 \raisebox{2.0ex}[0pt]{58.48} &
 \raisebox{2.0ex}[0pt]{59} &
 \raisebox{2.0ex}[0pt]{0.99}\\

\cline{2-9}

  &
  &
 12.32 &
 19.14 &
 10.44 &
 12.70 &
  &
  & \\

  &
 \raisebox{2.0ex}[0pt]{5} &
 22.76 &
 31.85 &
 31.47 &
 23.15 &
 \raisebox{2.0ex}[0pt]{54.61} &
 \raisebox{2.0ex}[0pt]{59} &
 \raisebox{2.0ex}[0pt]{0.93}\\

\cline{2-9}

  &
  &
 12.52 &
 18.94 &
 10.45 &
 12.32 &
  &
  & \\

\raisebox{9.0ex}[0pt]{\funcfourdis} &
 \raisebox{2.0ex}[0pt]{6} &
 22.96 &
 31.26 &
 31.46 &
 22.76 &
 \raisebox{2.0ex}[0pt]{54.23} &
 \raisebox{2.0ex}[0pt]{59} &
 \raisebox{2.0ex}[0pt]{0.92}\\

\hline

\end{tabular}
\end{center}
\end{table}

\renewcommand{\arraystretch}{1.0}

\renewcommand{\arraystretch}{1.25}

\begin{table}[th]
\caption[Sub Total \chisquare For Minimizations To The \pbe Data]
{The eight sub-total, total, number of degrees of freedom and
reduced \nschisquare, \pdfchi for all minimizations to the
beryllium target data. Top line gives the individual spectra
sub-totals and the second line gives the sub-totals by charge and
production region.}
\label{indivbe}
\begin{center}
\begin{tabular}
[c]{|l|c|cccc|ccc|}\hline

\multicolumn{9}{|c|}{Beryllium Data}\\

\hline
\hline

  &
  &
 \nstpluschi &
 \nstminuschi &
 \nsdpluschi &
 \nsdminuschi &
  &
  & \\

\raisebox{2.0ex}[0pt]{Function} &
 \raisebox{2.0ex}[0pt]{$n$} &
 \nspluschi &
 \nsminuschi &
 \nstgtchi &
 \nsdmpchi &
 \raisebox{2.0ex}[0pt]{\nschisquare} &
 \raisebox{2.0ex}[0pt]{$ndf$} &
 \raisebox{2.0ex}[0pt]{\nspdfchi}\\

\hline
\hline

  &
  &
 9.55 &
 7.44 &
 12.71 &
 33.18 &
  &
  & \\

\raisebox{2.0ex}[0pt]{\nsfuncexp} &
 \raisebox{2.0ex}[0pt]{NA} &
 22.26 &
 40.62 &
 16.99 &
 45.89 &
 \raisebox{2.0ex}[0pt]{62.88} &
 \raisebox{2.0ex}[0pt]{63} &
 \raisebox{2.0ex}[0pt]{1.00}\\

\hline
\hline

  &
  &
 7.58 &
 6.69 &
 12.94 &
 32.93 &
  &
  & \\

  &
 \raisebox{2.0ex}[0pt]{FF} &
 20.52 &
 39.62 &
 14.27 &
 45.87 &
 \raisebox{2.0ex}[0pt]{60.14} &
 \raisebox{2.0ex}[0pt]{61} &
 \raisebox{2.0ex}[0pt]{0.95}\\

\cline{2-9}

  &
  &
 65.08 &
 56.19 &
 13.65 &
 32.40 &
  &
  & \\

  &
 \raisebox{2.0ex}[0pt]{4} &
 78.73 &
 88.60 &
 121.27 &
 46.06 &
 \raisebox{2.0ex}[0pt]{167.32} &
 \raisebox{2.0ex}[0pt]{63} &
 \raisebox{2.0ex}[0pt]{2.66}\\

\cline{2-9}

  &
  &
 17.59 &
 20.76 &
 12.34 &
 33.73 &
  &
  & \\

  &
 \raisebox{2.0ex}[0pt]{5} &
 29.94 &
 54.40 &
 38.36 &
 46.08 &
 \raisebox{2.0ex}[0pt]{84.44} &
 \raisebox{2.0ex}[0pt]{63} &
 \raisebox{2.0ex}[0pt]{1.34}\\

\cline{2-9}

\raisebox{9.0ex}[0pt]{\functhreedis} &
  &
 7.76 &
 8.54 &
 12.76 &
 33.12 &
  &
  & \\

  &
 \raisebox{2.0ex}[0pt]{6} &
 20.52 &
 41.66 &
 16.30 &
 45.88 &
 \raisebox{2.0ex}[0pt]{62.19} &
 \raisebox{2.0ex}[0pt]{63} &
 \raisebox{2.0ex}[0pt]{0.99}\\

\hline
\hline

  &
  &
 7.28 &
 6.15 &
 12.91 &
 32.96 &
  &
  & \\

  &
 \raisebox{2.0ex}[0pt]{4} &
 20.18 &
 39.12 &
 13.43 &
 45.87 &
 \raisebox{2.0ex}[0pt]{59.30} &
 \raisebox{2.0ex}[0pt]{61} &
 \raisebox{2.0ex}[0pt]{0.97}\\

\cline{2-9}

  &
  &
 6.69 &
 5.88 &
 12.91 &
 32.96 &
  &
  & \\

  &
 \raisebox{2.0ex}[0pt]{5} &
 19.60 &
 38.84 &
 12.57 &
 45.87 &
 \raisebox{2.0ex}[0pt]{58.44} &
 \raisebox{2.0ex}[0pt]{61} &
 \raisebox{2.0ex}[0pt]{0.96}\\

\cline{2-9}

  &
  &
 6.59 &
 5.79 &
 12.90 &
 32.97 &
  &
  & \\

\raisebox{9.0ex}[0pt]{\funcfourdis} &
 \raisebox{2.0ex}[0pt]{6} &
 19.49 &
 38.76 &
 12.38 &
 45.87 &
 \raisebox{2.0ex}[0pt]{58.25} &
 \raisebox{2.0ex}[0pt]{61} &
 \raisebox{2.0ex}[0pt]{0.95}\\

\hline

\end{tabular}
\end{center}
\end{table}

\renewcommand{\arraystretch}{1.0}

\end{document}